\documentclass[aps,prd,amssymb,floats,showpacs,twocolumn,floatfix,altaffilletter]{revtex4}

\usepackage{epsfig}
\usepackage{color}
\pagecolor{white}
\usepackage{latexsym}
\usepackage{hyperref,color}
\usepackage{amsmath}
\usepackage{wasysym}
\usepackage{graphicx}
\usepackage{dcolumn}
\usepackage{bm}
\newcommand{\vect}[1]{\boldsymbol{#1}}
\newcommand{\fquad}[2]{#1^T #2 \: #1}
\newcommand{\de}{\mathrm{d}}
\newcommand{\defin}{=}

\def\ber{\begin{eqnarray}}
\def\eer{\end{eqnarray}}
\def\beq{\begin{equation}}
\def\eeq{\end{equation}}

\begin{document}
\title{Measuring gravito-magnetic effects by multi ring-laser gyroscope}

\author{F. Bosi}
 \email{bosi@pi.infn.it}
\author{G. Cella}
 \email{giancarlo.cella@pi.infn.it}
\author{A. Di Virgilio}
 \email{angela.divirgilio@pi.infn.it}
\affiliation{
INFN Sez. di Pisa, Pisa, Italy}
\author{A.Ortolan}
 \email{antonello.ortolan@lnl.infn.it}
\affiliation{Laboratori Nazionali di Legnaro, INFN Legnaro (Padova), Italy}
\author{A. Porzio}
 \email{alberto.porzio@na.infn.it}
\author{S. Solimeno}
\email{solimeno@na.infn.it}
\affiliation{University of Naples and CNR-SPIN, Naples, Italy}
\author{M. Cerdonio}
 \email{cerdonio@pd.infn.it}
\author{J. P. Zendri}
\email{zendri@lnl.infn.it}
\affiliation{INFN Sez. di Padova, Padova, Italy}

\author{M. Allegrini} 
\email{maria.allegrini@df.unipi.it}
\author{J. Belfi}
\email{belfi@df.unipi.it} 
\author{N. Beverini}
\email{beverini@df.unipi.it}
\author{B. Bouhadef}
\email{bouhadef@df.unipi.it}
\author{G. Carelli}
\email{carelli@df.unipi.it}
\author{I. Ferrante}
\email{isodoro.ferrante@pi.infn.it}
\author{E. Maccioni}
\email{maccioni@df.unipi.it}
\author{R. Passaquieti}
\email{roberto.passaquieti@pi.infn.it}
\author{F. Stefani}
\email{fabio.stefani@df.unipi.it}
\affiliation{University of Pisa and CNISM, Pisa, Italy}

\author{M. L. Ruggiero}
 \email{matteo.ruggiero@polito.it}
\author{A. Tartaglia}
 \email{angelo.tartaglia@polito.it}
\affiliation{
Polit. of Torino and INFN, Torino, Italy}

\author{K. U. Schreiber}
 \email{schreiber@fs.wettzell.de}
\author{A. Gebauer}
 \email{gebauer@fs.wettzell.de}
\affiliation{
Technische Universitaet Muenchen,
Forschungseinrichtung Satellitengeodaesie \\
Fundamentalstation Wettzell, 93444 Bad K\"otzting, Germany}
\author{J-P. R. Wells}
 \email{jon-paul.wells@canterbury.ac.nz}
\affiliation{
Department of Physics and Astronomy, University of Canterbury,
Christchurch 8020, New Zealand}
\author{Draft}
\affiliation{Draft}

\pacs{42.15.Dp, 42.30.Sy, 42.55.Lt, 91.10.Nj}
\begin{abstract}
\index{abstract}
We propose an under-ground experiment to detect the general relativistic effects due to the curvature of space-time around the Earth (de Sitter effect) and to rotation of the planet (dragging of the inertial frames or Lense-Thirring effect). It is based on the comparison between the IERS value of the Earth rotation vector and corresponding measurements obtained by a tri-axial laser detector of rotation. The proposed detector consists of six large ring-lasers arranged along three orthogonal axes. 
 In about two years of data taking, the $1\%$ sensitivity required for the measurement of the Lense-Thirring drag can be reached  with square rings of $6$ $m$ side, assuming a shot noise limited sensitivity ($ 20 prad/s/\sqrt{Hz}$).  The multi-gyros system, composed of rings whose planes are perpendicular to one or the other of three orthogonal axes, can be built in several ways. Here, we consider cubic and octahedron structures. The symmetries of the proposed configurations provide mathematical relations that can be used to study the stability of the scale factors, the relative orientations or the ring-laser planes, very important to get rid of systematics in long-term measurements, which are required in order to determine the relativistic effects.
\end{abstract}


\maketitle

\section{\label{sec:a} Introduction}

 The general theory of relativity is the most
satisfactory description of gravitational phenomena. The theoretical
breakthrough came with Einstein's geometrical representation of gravity: as
different test masses fall in the same way in a gravitational field, gravity
must be a property of space and time rather than of the masses themselves.

Until now, almost all successful tests of general relativity (Shapiro time
delay \cite{shap}, light deflection by the sun \cite{defl}, perihelion shift
of the orbit of Mercury \cite{MTW}) have been probing the gravitational
field of the Sun, without considering its proper rotation. However, general
relativity predicts that the stationary field of a rotating body is
different from the static field produced by the same non-rotating mass. The
difference is known as gravito-magnetism and consists of a drag of space-time
due to the mass currents. The rotational frame-dragging effect is also known
as the Lense-Thirring (LT) \cite{LTH} effect.

A direct experimental evidence of the existence of the GM field has
been obtained so far by Ciufolini \cite{ciufo} and by Francis Everitt and the GP-B
group \cite{GPB1}.
The Lense-Thirring effect, averaged over several orbits, has been recently
verified by analysing the node orbital motion of two laser ranged freely
falling satellites (LAGEOS-1 and LAGEOS-2) which orbit the Earth. In the
measurement presented in Ref. \cite{ciufo} the two LAGEOS satellites were
used to confirm the LT effect with an accuracy of the order of 10\%.
However, the launch of a third properly designed satellite LARES will give
the opportunity to measure the LT effect with an accuracy of the order of
1\% (citare).

The possibility to  detect Lense-Thirring with ring lasers has been discussed in the past \cite{StedLT, scully} .
Recently it has been already pointed out that a multi-gyros system is able
to test locally the Lense-Thirring effect \cite{essay}: an array of six, $6$
$m$ side, square ring-lasers have enough sensitivity for this purpose.
 The rings must have different
orientation in space. In the present paper we concentrate the attention on
the symmetries of the rings arranged on the faces of a cube or along the
edges of an octahedron, extracting the relevant relations important for the
diagnostics of the system. At the end we summarize and sketch the proposed
experiment. For completeness we must mention that an experiment of the type
we are planning and preparing could also be made in principle using matter
waves instead of light. This possibility has been proved experimentally for
various types of particles such as electrons \cite{electrons}, neutrons \cite%
{neutrons}, Cooper pairs \cite{pairs}, Calcium atoms \cite{calcium}, superfluid He3 \cite{He3} and superfluid He4 \cite{He4} . 
Cold atoms interferometry, in particular, yields very high sensitivity and it is suitable for space experiments because of the apparatus small size. However, atoms interferometry experiments in space do not provide an independent measurement of the Earth angular velocity, are affected by the mass distribution of the Earth, and test the average of the relativistic effect rather than the local one. Eventually, the comparison between in-space and on-ground measurements could be very valuable.

\section{\label{sec:b1} Detection of Gravito-magnetic Effects}

\index{Detection of Gravito-magnetic Effects}

Gravito-magnetism (GM) is a general relativistic phenomenon related to the
presence of mass currents in the reference frame of a given observer. In the
case of celestial bodies, including the Earth, and excluding translational
motion with respect to the center of the body, gravito-magnetic effects are
due to the absolute rotation of the massive source with respect to distant
stars. When the Einstein equations in vacuum are applied to this kind of
symmetry and are linearised (weak field approximation) GM is accounted for
by the analogue of a magnetic field of a rotating spherical charge. In
practice at the lowest approximation level, a dipolar GM field is obtained,
with the dimensions of an angular velocity. Its explicit form in a
non-rotating reference frame centred on the source (in our
case the Earth center), is (see e.g. \cite{ciufoliniwheeler})

\begin{equation}
\boldsymbol{B}=
\frac{2G}{c^2 R^3} \left[ \boldsymbol{J}_\oplus-3 (\boldsymbol{J}_\oplus \cdot%
\boldsymbol{u}_r)\boldsymbol{u}_r\right]  \label{GM}
\end{equation}
\noindent

where $\boldsymbol{R}\equiv R \boldsymbol{u}_r$ is the position of the
laboratory with respect to the center of the Earth and $\boldsymbol{J}_\oplus$ is the angular momentum
of the Earth, whose modulus is of course given by the product of the moment of inertia of the planet multiplied by its angular velocity.


The effect produced by a field like (\ref{GM}) on a massive test body  moving with velocity $\bm v$ looks
like the one produced by a magnetic field on a moving charge: in fact, the geodesic equation in weak field approximation reads
\beq
\frac{d \bm v}{dt}=\bm G +\bm v \wedge \bm B \label{eq:geod1}
\eeq
where $\bm G =-GM/R^{2} \bm u_{r}$ is the Newtonian gravitational field, so that the effect can be described in terms of a gravito-electromagnetic Lorentz force,  where the Newtonian gravitational field plays the role of the gravito-electric field (GE).

Furthermore, the rotation of the source of the gravitational field affects a gyroscope orbiting around it, in such a way that it undergoes the so-called Lense-Thirring precession, or dragging of the inertial frames of which the gyroscope defines an axis\cite{schiff,ciufoliniwheeler,iorio2011}. This phenomenon shows up also when one considers  a freely falling body with local zero angular momentum (ZAMO: Zero
Angular Momentum Observer): it  will be seen as rotating by a distant observer at
rest with the fixed stars \cite{ZAMO}.

\subsection{ Mechanical gyroscopes}

\index{Mechanical gyroscopes} Gravito-magnetic effects can in principle be
measured applying different methodologies. The one that has most often been
considered is focused on the behaviour of a gyroscope, that can be either in
free fall (on board an orbiting satellite) or attached to the rotating Earth.
The axis of the gyroscope is affected in various ways by the presence of a
gravitational field. As for GM, a little mechanical gyroscope is the analogous
of a small dipolar magnet (a current loop), so that it behaves as magnetic
dipoles do when immersed in an external magnetic field.

When studying  the motion around the Earth of a gyroscope whose spin vector is $\bm S$ ,   one is led to the formula
\cite{MTW,ciufoliniwheeler}:
\beq
\frac{d\bm S}{dt}=\bm \Omega' \wedge \bm S \label{prec}
\eeq
In Appendix \ref{A1}  we work out
the explicit expression of $\bm \Omega'$  in general relativity and, more in general, in  metric theories of gravity, using the Parametrized
Post-Newtonian (PPN) formalism\cite{Will}: we  show that  it is related to   the gravito-magnetic components $g_{0i}$ of the metric tensor  and its expression is given by (see Eqs. (\ref{eq:omegaprime})-(\ref{eq:OmegaTh})) $\bm \Omega' = \bm \Omega_{G}+ \bm \Omega_{B}+  \bm \Omega_{W} +\bm\Omega_{T}$, so that we can distinguish  four contributions, namely the geodetic term $\bm \Omega_{G}$,  the Lense-Thirring term $\bm\Omega_{B}$, the preferred frame term $\bm \Omega_{W}$, the Thomas term $\bm \Omega_{T}$.  All  terms in $\bm \Omega'$
are called relativistic precessions, but properly speaking only the second is due to the intrinsic gravito-magnetic field
of the Earth, namely it is $\bm \Omega_{B}=-\frac{1}{2} \bm B$, and manifests the Lense-Thirring drag.

Ciufolini \cite{LARES} deduced the relativistic
precession of the whole orbital momentum of two LAGEOS satellites whose
plane of the orbit is dragged along by the rotating Earth. Again on Eq. 
(\ref{prec}) was based the GP-B experiment, whose core were four freely falling
spherical gyroscopes carried by a satellite in polar orbit around the Earth
\cite{GPB1}. While time goes on and the available data grow it is expected
that the Lense-Thirring drag will emerge from the behaviour of the unique (so far) double
pulsar system \cite{double}.

\subsection{ Using light as a probe}

A different experimental approach consists in using light as a probe. In
this case the main remark is that the propagation of light in the
gravitational field of a rotating body is not symmetric. The coordinated
time duration for a given space trajectory in the same sense as the rotation
of the central source is different from the one obtained when moving in the
opposite direction. This asymmetry would for instance be visible in the
Shapiro time delay of electromagnetic signals passing by the Sun (or
Jupiter) on opposite sides of the rotation axis of the star (or the planet)
\cite{TRT2000}\cite{TNR2005}.

This property of the propagation of light is the one which we wish to
exploit in our Earth-bound experiment using a set of ring lasers. In a
terrestrial laboratory, light circulating inside a laser cavity in opposite directions
is forced, using mirrors, to move along a closed path in space. What is
closed from the view point of the laboratory is not so for a
fixed-stars-bound observer, but the essential is that the two directions are
not equivalent and that the two times required for light to come back to the
active region are (slightly) different. As it happened already in the case
of the mechanical gyroscopes, here too the difference in the two times of
flight is made up of various contributions depending on the rotation of the
axes of the local reference frame with respect to distant stars, on the fact
that the local gravitational (Newtonian) potential is not null, and of
course on the GM drag (which is our main interest). What matters, however, is
that the final proper time difference (a scalar quantity) is invariant: it
does not depend on the choice of the reference frame or of the coordinates.

Performing the calculation in linear approximation for an instrument
with its normal contained in the local meridian plane (see the Appendix \ref{A1}
details) we find

\begin{eqnarray}
c\delta \tau&=&\frac{4A}{c}\Omega_{\oplus} \left[  \cos \left(\theta+\alpha \right)
-2\frac{GM}{c^{2}R}\sin \theta \sin \alpha \nonumber \right.\\
&+&\left. \frac{GI_{\oplus}}{c^{2}R^{3}}  \left(2 \cos \theta \cos \alpha+\sin\theta \sin \alpha \right)  \right] \,
 \label{eq:totale}
\end{eqnarray}
where  $A$ is the area encircled by the light beams,
 $\alpha $ is the angle between the local radial direction and the normal to the
plane of the instrument, measured in the meridian plane, and $\theta $ is the colatitude of the
laboratory; $\Omega _{\oplus}$ is the rotation rate of the Earth as measured in the local reference frame (which includes the local gravitational time delay).

Eq. (\ref{eq:totale}) can also be written in terms of the flux of an
effective angular velocity $\mathbf{\Omega }$ through the cross
section of the apparatus:
\begin{equation}
\delta \tau = \frac{4}{c^{2}}\mathbf{A} \cdot \mathbf{\Omega },  \label{eq:flusso}
\end{equation}
where   $\mathbf A=A\vect{u}_n$ is the area enclosed by the beams and oriented according to its normal vector $\vect{u}_n$.
In particular, it is $\bm \Omega=\bm \Omega_{\oplus}+\bm \Omega'$, and the term proportional to $\bm \Omega_{\oplus}$ is the purely kinematic Sagnac term, due to the rotation of the Earth, while $\bm \Omega' = \bm \Omega_{G}+ \bm \Omega_{B}+  \bm \Omega_{W} +\bm\Omega_{T}$ encodes the relativistic effects (see Appendix \ref{A1})

\begin{figure}[t]
\includegraphics[width=251pt]{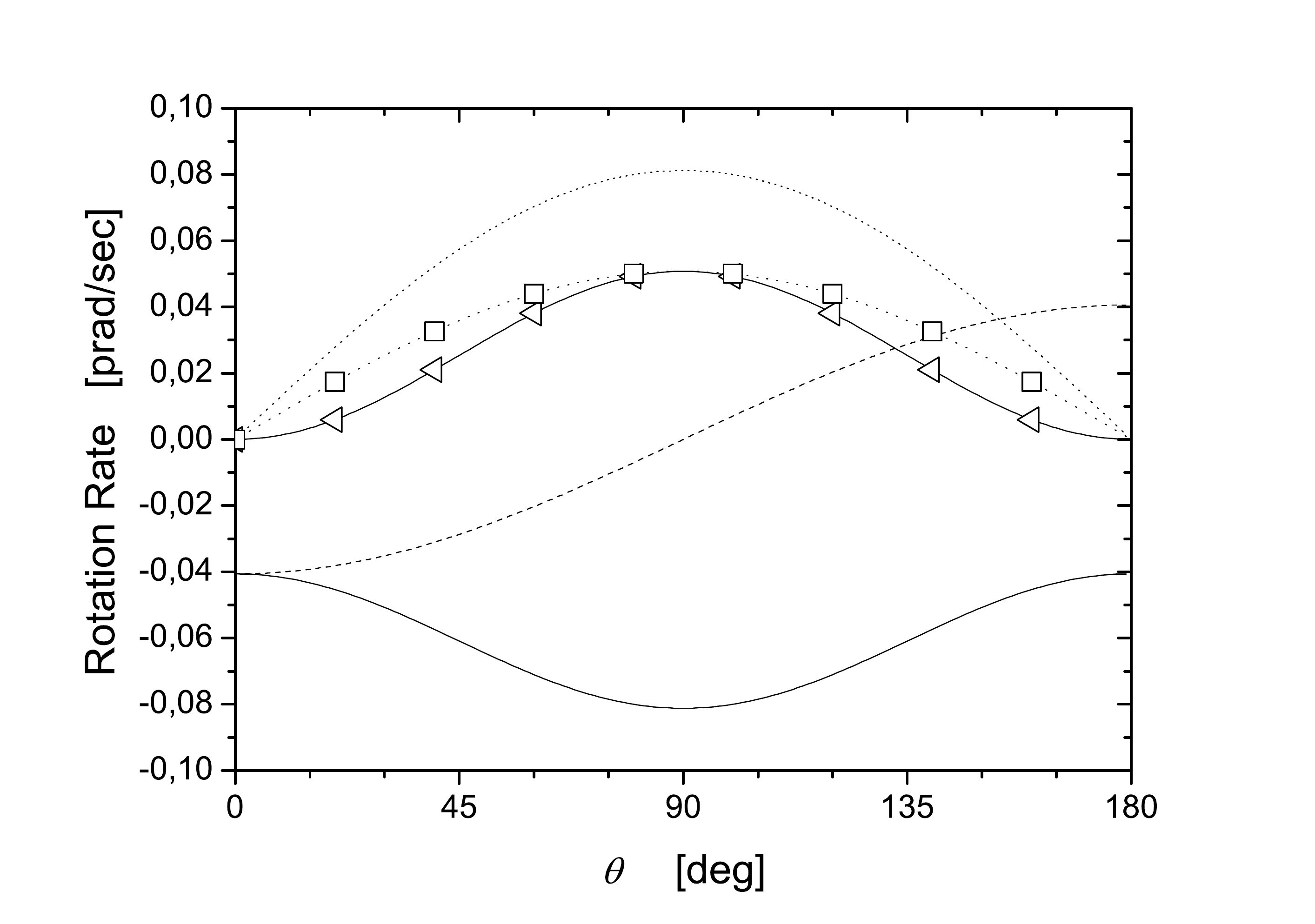}
\caption{ The amplitude of the relativistic effects on the surface of the Earth, according to the theory of general
relativity, in units of $prad/s$, as a function of the colatitude $\theta$. The continuous, dashed
and dotted lines correspond to $\vect{\Omega}^{\prime}=\vect{\Omega}_{G}+\vect{\Omega}_{B}$ projected along
the directions: i) parallel to $\vect{\Omega}_{\oplus}$ (i.e. ${\Omega}^\prime_\parallel$);
ii) $\vect{u_r}$ (local radial or zenithal direction); and iii) $\vect{u_\theta}$ (local North-South direction), respectively.
To evaluate the contribution to $\vect{\Omega}^\prime$ from
$\vect{\Omega}_{G}$, we have projected $\vect{\Omega}_{G}$ along  $\vect{\Omega}_{\oplus}$ (continuous line plus
triangles) and $\vect{u_\theta}$  (dotted lines plus squares). We note that the gravito-electric term has
only the $\vect{u_\theta}$ component and therefore along the radial direction we have
a pure gravito-magnetic term. }
\label{figpre}
\end{figure}

For a ring laser in an Earth-bound laboratory, the geodetic and Lense-Thirring terms are both of order $\sim 10^{-9}$
with respect to the Sagnac term, while the Thomas term is 3 orders of magnitude smaller.  As for the preferred frame term, the best estimates \cite{bell,damour96} show that this effect is about 2 orders of magnitude smaller than the geodetic and Lense-Thirring terms. Consequently,  to leading order, the relativistic contribution to the rotation measured by the ring laser turns out to be $ \bm \Omega' \simeq \bm \Omega_{G}+ \bm \Omega_{B}$, which we aimed at measuring in our experiment. In other words, the goal of our experiment will be the estimate of $\bm \Omega'$  (see Fig.~\ref{figpre})
which embodies the gravito-magnetic effects in a terrestrial laboratory.

In particular, the proposed experiment can also provide high precision tests of
metric theories of gravity  which are described in the framework of  (PPN) formalism. In fact, from Eqs. (\ref{eq:OmegaDS10}-\ref{eq:OmegaLT10}), we see that, on setting for the rotating Earth $\bm J=I_{\oplus} \bm \Omega_{\oplus}$, we obtain

\begin{eqnarray}
\bm \Omega_{G}&=&-(1+\gamma)\frac{GM}{ c^{2}R} \sin \vartheta   \Omega_{\oplus} \bm{u}_{\vartheta},
\label{eq:OmegaDS1b} \\
\bm \Omega_{B}&=&-\frac{1+\gamma+\frac{\alpha_1}{4}}{2} \frac{G I_{\oplus}}{c^2 R^3}\left[\bm \Omega_{\oplus}-3 \left(\bm \Omega_{\oplus} \cdot \bm{u}_{r} \right) \bm{u}_{r} \right]
\label{eq:OmegaLT1b} \
\end{eqnarray}

where  $\alpha_1$ and $\gamma$ are PPN parameters (e.g. $\alpha_1=0$ and $\gamma=1$ in  general relativity)
which account for the effect of preferred
reference frame and the amount of space curvature produced by a
unit rest mass, respectively.

As shown in Sect. \ref{con}, from a high precision measurement of the vector
$\vect{\Omega}'$ in the meridian plane, we should be able to place new constraints on the PPN
parameters $\alpha_1$ and $\gamma$.

\section { Theory of the measurement: combining together the response of several rings }
\subsection{ The response of a ring laser} \label{ssec:RLresponse}
 A ring laser converts time
differences into frequency differences. In fact, since
the emission is continuous, the right handed beam adjusts itself to give a
standing wave whose wavelength is an integer sub-multiple of the space length
of the loop $P$: $c\tau _{+}=P=N\lambda _{+}$. The same happens with the
left handed beam, but being the total time different, also the wavelength of
the corresponding standing wave will be different: $c\tau _{-}=N\lambda _{-}$.
The two modes of the ring can have different $N$, a situation usually called 'split mode', but the higher accuracy of the measurement has been
obtained so far with the two modes with equal $N$.
Considering the time of flight difference in terms of the wavelengths of the
two standing waves we see that:

\begin{equation}
c\delta \tau =N\left( \lambda _{+}-\lambda _{-}\right) =Nc%
\frac{f_{-}-f_{+}}{f^{2}}=P\lambda \frac{\delta f}{c}  \label{freque}
\end{equation}

The ring laser equation~\cite{geoff} relates the frequency splitting $\delta
f$ of the two optical beams inside the ring interferometer with the
experienced rotation rate of its mirrors
\begin{equation}
\delta f=\frac{4A}{\lambda P}\ \mathbf{u}_{n}\cdot \boldsymbol{\Omega }
 ,  \label{eq3.1}
\end{equation}
where $P$ is the perimeter and $\lambda $ is the laser wavelength. The
response $R$ of a ring laser to the rotation rate $\boldsymbol{\Omega }
 $, in units of rad/sec, is simply a rescaling of the frequency
splitting by the scale factor $S\equiv \frac{4A}{\lambda P}$,
i.e.
\begin{equation}
R\equiv \delta f/S=\mathbf{u}_{n}\cdot \boldsymbol{\Omega }.
\label{eq:dodb}
\end{equation}

The scale factor $S$ plays a crucial role in the accuracy of the
measurement of $\boldsymbol{\Omega }$ and to estimate the
relativistic effects the ratio $\frac{4A}{\lambda P}$ must be known and kept at $%
10^{-10}$ accuracy level for months. The requirements to keep the apparatus
in the optimal working conditions will be discussed in section \ref{sec:c}.

Since the
effective angular velocity as well as the gravito-magnetic one is of the
order of $10^{-9}\Omega _{\oplus }$, angles between vectors must be measured
at the corresponding accuracy level. Unfortunately, the absolute measurement
of $\boldsymbol{u}_{n}$ in the fixed stars reference system with the
accuracy of nano-radians can hardly be achieved. However, we can relax this
requirement by using $M\geq3$ ring lasers oriented along directions $\boldsymbol{u}^{\alpha }$
 ($\alpha =1\dots M$), where not all $\boldsymbol{u}^{\alpha }$
lie in the same plane. In fact, $\boldsymbol{\Omega }$ can be
completely measured by means of its projections on at least 3 independent
directions (e.g.defining a tri-dimensional Cartesian system) and the redundancies of the
measurement can be used as a monitor and control of the stability of
directions $\boldsymbol{u}^{\alpha }$. We further assume that ring lasers
have identical sensitivity and noise parameters. From an experimental point
of view this can be easily satisfied by building the devices with scale factors that differ less than $\%$.

In order to simplify the sensitivity calculations of the system one can
consider multi-axial configurations endowed with symmetries. As all the ring
laser normals $\boldsymbol{u}^{\alpha }$ are equivalent in space, symmetric
configurations should be more efficient in the rejection of spurious effects
and in the control and monitoring of the relative orientation of the ring
lasers.
The natural choice is to take advantage of space symmetries of regular polyhedra, setting one ring for each plane parallel to their faces. If we do not consider the degeneration between opposite faces, we have M=3 in the case of the cube, 4 for tetrahedron and octahedron, 6 for dodecahedron, and 10 for icosahedron. There is a peculiar geometry with $M=3$, obtained by arranging the rings along the edges of an octahedron, where the different rings can be nested together, sharing 2 by 2 the same mirrors. We will refer to it in the following by speaking of  ``octahedral configuration''. 
The $M=3$ is the minimum number of rings necessary to reconstruct the rotational vector, but a redundancy is very appropriate to enhance statistic and to have control tests on the geometric accuracy.\\
In general, by simple arguments, one can demonstrate that, for regular polyhedra configuration
\begin{equation}
\sum_{\alpha =1}^{M}\boldsymbol{u}_{\alpha }=\boldsymbol{0} \quad (M > 3)
\label{eqlin}
\end{equation}%
and that
\begin{equation}
\sum_{\alpha =1}^{M}(\boldsymbol{\Omega }\cdot \boldsymbol{u}%
_{\alpha })^{2}=\frac{M}{3}\ |\boldsymbol{\Omega }|^{2}\   \quad (M \geq 3).
\label{eqquad}
\end{equation}

As a consequence, one can study linear and quadratic combinations of ring
lasers responses $R_\alpha$ which are invariant under permutations of the
ring laser labels $\alpha$, i.e. $L=\sum_\alpha R_\alpha$ and $Q=\sum_\alpha
R_\alpha^2$.
For non-symmetric configurations we can generalize their definition as $
L=\sum_\alpha L_\alpha R_\alpha$ and $Q=\sum_\alpha Q_{\alpha\beta} R_\alpha
R_\beta$ , where $L_\alpha$ and $Q_{\alpha\beta}$ are suitable constants which
depend on $\vect{u}_\alpha$.
The interest in such linear or quadratic forms relies on their behaviours in
the presence of noise fluctuations or variations of the geometry of the
configuration. \newline
They also allow us to carry out analytical estimates of the overall
sensitivity of a tri-axial system of ring laser  to relativistic effective rotation rates.


\subsection{Requirements for the geometry of the configuration}
\index{requirements for the geometry of the configuration}

The response of each ring laser can be
conveniently written as
\begin{equation}
R_{\alpha }=\boldsymbol{\Omega }\cdot (\boldsymbol{u}_{\alpha
}+\delta \boldsymbol{u}_{\alpha })+\varepsilon _{\alpha }\ ,
\label{eqmis}
\end{equation}%
where $\delta \boldsymbol{u}_{\alpha }\equiv \delta S_{\alpha }\,\boldsymbol{%
u}_{\alpha }+\delta \boldsymbol{\vartheta }_{\alpha }\wedge \boldsymbol{u}%
_{\alpha }$ account for systematic errors in the scale factors and
orientations in space and $\varepsilon _{\alpha }$ represents the additive
noise that affects the rotation measurement $R_{\alpha }$, that we assume
averaged on the observation time $T\simeq 1\ day$. We assume as well that $%
\varepsilon _{\alpha }$ are Gaussian distributed random variables with
zero mean and variance $\sigma _{\Omega }^{2}$. Modulus $|\delta \boldsymbol{%
u}_{\alpha }|\simeq \delta S_{\alpha }$ and direction $\delta \boldsymbol{%
\vartheta }_{\alpha }\wedge \boldsymbol{u}_{\alpha }$ represent the
deviations from regular polygon geometry in the plane and from polyhedra
geometry in the space, due to scale factor fluctuations $\delta S_{\alpha }$
and infinitesimal rotations $\delta \boldsymbol{\vartheta }_{\alpha }$,
respectively. In what follows the crucial assumption is that systematic
errors (scale factors and relative alignment of $\boldsymbol{u}_{\alpha }$)
are negligible with respect to statistical errors, i.e. $|\boldsymbol{\Omega
}\cdot \delta \boldsymbol{u}_{\alpha }|<\sigma _{\Omega }$ or
equivalently $|\delta \boldsymbol{u}_{\alpha }|<\sigma _{\Omega }/\Omega
$, while the dihedral angles $\arccos(\boldsymbol{u}_{\alpha }\cdot
\boldsymbol{u}_{\beta })$ can nearly approximate a regular polyhedron
configuration.


Redundancy of responses, if $M>3$ rings are involved, can be used to control systematic
errors projected along the direction of $\vect{\Omega}$. In fact, the rigidity of the configuration imposes
some linear kinematic constraints among different estimates of the laboratory rotation.
In general, any linear combination of 3 responses $R_\alpha$ gives an
estimate of the local rotation $\vect{\Omega}$ and we can test the consistency among
different estimates by means of the ordinary least square fit.
A very simple linear constraint can be found for regular polyhedral configurations
\begin{equation}
L=\sum_{\alpha=1}^M R_\alpha
\end{equation}
and we will illustrate its statistical property as an example of the power of the method.

From the definition of $L$ immediately follows that it is Gaussian distributed with zero mean
and standard deviation $\sigma_L= \sqrt{M} \sigma_\Omega$. In addition, possible
misalignments $\delta \vect{\vartheta}_\alpha\wedge \vect{u}_\alpha$
or scale factor fluctuations $\delta S_\alpha \vect{u}_\alpha$ are amplified by a factor of
$\Omega$ in the mean value of $L$
\begin{eqnarray}
<L>&=&
\sum_{\alpha=1}^M \vect{\Omega}\cdot\delta\vect{u}_\alpha\\
&=& \Omega \sum_{\alpha=1}^M (\delta S_\alpha \vect{u}_\alpha+\delta\vect{\vartheta}_\alpha\wedge \vect{u}_\alpha)^{\parallel} \ ,
\end{eqnarray}
without affecting the corresponding variance $\sigma^2_L$.
Thus $<L>$ can be used as a ``null constraint" which
is minimum when the configuration geometry is a regular polyhedron, and so
the overall mean error parallel to $\vect{\Omega}$
can be monitored at $\sim \sqrt{M}\sigma_\Omega/\Omega \simeq 10^{-10}$ accuracy level.

\subsection{Estimate of the parallel component of the relativistic effective rotation vector}

An estimate of $\Omega^2$ for symmetric configurations readily follows from Eq. (\ref{eqquad})
\begin{eqnarray}
Q &=& \frac{3}{M}\sum_{\alpha=1}^M R^2_\alpha \label{EqQua} \\
&=&  \Omega^2 + \frac{6}{M} \sum_{\alpha=1}^M \varepsilon_\alpha \vect{\Omega} \cdot \vect{u}^\alpha +
\frac{3}{M}\sum_{\alpha=1}^M \varepsilon_\alpha^2 \ .
\end{eqnarray}
Its mean value and standard deviation read (see App. \ref{A2})
\begin{eqnarray}
<Q>&=& \Omega^2 + \frac{1}{3}\sigma_\Omega^2 \\
\sigma_Q &=& \sqrt{\frac{18}{M} \sigma_\Omega^4 +  \frac{12}{M}\Omega^2 \sigma^2_\Omega } \ .
\end{eqnarray}
In addition, one can demonstrate that $Q$ is non-central $\chi^2$ distributed with
$M$ degrees of freedom and non-centrality parameter $\Omega^2$. In order to estimate the relativistic effective rotation, we must subtract $\Omega_\oplus$ from the rotation rate estimated in the laboratory.
To this end we calculate the difference  $\Delta \equiv (Q-\Omega_\oplus^2)$ that,
in the limit of high SNR ($|\Omega|/\sigma_\Omega>>1$),
tends to be  Gaussian distributed with mean
\begin{equation}
<\Delta> \simeq 2\, \Omega_\oplus \ \Omega^{\prime}_\parallel
\end{equation}
and standard deviation
\begin{equation}
\sigma_\Delta  \simeq (2 \sqrt{3}/\sqrt{M})\, \Omega_\oplus \,  \sigma_\Omega \ ,
\label{sensQ}
\end{equation}
where we have neglected terms of the order of $\sigma_\Omega/\Omega$. The
$SNR = <\Delta>/\sigma_\Delta$
of the parallel component of relativistic effective rotation is increased by a factor
of $\sqrt{M/3}$ with respect to the sensitivity of each ring laser.

The advantage of this approach is that we compare scalar quantities (moduli of rotation vectors)
measured with respect to the local and distant stars reference systems. Its drawback
is the very poor sensitivity to the perpendicular
component $\Omega_\perp$ of the relativistic effective rotation. In fact, $\Omega^2-\Omega_\oplus^2 =
2 \vect{\Omega}^\prime \cdot\vect{\Omega}_\oplus $
$+ |\vect{\Omega}^\prime|^2$, and the ratio between the second term
(which is associated to the perpendicular component as $|\vect{\Omega}^\prime|^2= \Omega^{\prime 2}_\parallel+
\Omega^{\prime 2}_\perp $) and the first term is
$\sim GM/c^2R \simeq  10^{-10}$.

It is worth noticing that statistical fluctuations of $L$ (control of geometry by redundancy)
and $Q$ (measure of relativistic effects) are uncorrelated, and that they tend to be
independent in the limit of high SNR.

\subsection{Estimate of the components of the relativistic effective rotation  vector}

By arranging the response of ring lasers $R_\alpha$ as
M-tuples in a $M$-dimension vector space $\vect{R}=(R_1,R_2,R_3, \dots, R_M)$,
we can easily define projection operators that allows the estimate
of local meridian plane ${\cal M}$ and also the direction $\vect{w}$
of $\vect{\Omega}$ in the physical space. Moreover, the norm of projected random vectors are
described by remarkably simple statistics.
According to the definition of the
matrix product we have $\vect{R}= \vect{N} \vect{\Omega}+\vect{\varepsilon}$,
where $\vect{N}$ is a $M\times3$ matrix whose elements are $\vect{N}_{\alpha i} = (u_\alpha)_i$
and $\vect{\varepsilon}=(\varepsilon_1,\varepsilon_2,\varepsilon_3, \dots, \varepsilon_M)$.
Thus, the random vectors $\vect{R}$ can be projected on the
linear subspaces ${\cal P}_{{\cal M}}$ and ${\cal Q}_{{\cal M}}$ of dimensions 2 and $M-2$,
which represent respectively a plane in the physical space and its complementary space.
The physical symmetry of the rotating Earth imposes that
the relativistic effective rotation vectors and $\vect{\Omega}_\oplus$ lie in the same plane, i.e. the meridian plane,
and therefore the knowledge of the orientation of this plane is crucial if we want to measure not only the modulus
but the whole vector. We recall that a plane is
defined as the set of the points $s \vect{v} + t \vect{w}$, where $s$ and $t$ range over
all real numbers, $\vect{v}$ and $\vect{w}$  are given orthogonal unit vectors in the plane.

The parallelism of $\vect{v} \wedge \vect{w}$ with the normal to the meridian plane can be tested under
the hypothesis that the rotation signal is fully located in the ${\cal P}_{{\cal M}}$
subspace while the ${\cal Q}_{{\cal M}}$ subspace contains only noise.
The test can be easily performed over the norms of the two projections
$E_P(\vect{v},\vect{w})\equiv ||\vect{P}_{\vect{v},\vect{w}} \vect{R}||^2$ and $E_Q(\vect{v},\vect{w})\equiv
||\vect{Q}_{\vect{v},\vect{w}} \vect{R}||^2$, where we have introduced the symbol
$||\vect{R}||=(\sum_{\alpha=1}^M R_\alpha^2)^{1/2}$ to indicate the
L-2 norm in the M-dimensional Euclidean response space. The $M\times M$ projection matrices
$\vect{P}_{\vect{v,w}}$ and $\vect{Q}_{\vect{v,w}}$ can be written explicitly as functions of
the unit vectors $\vect{v}$ and $\vect{w}$
\begin{eqnarray}
\vect{P}_{\vect{v},\vect{w}} &=& \vect{N} \vect{V} (\vect{N}\vect{V})^T \\
\vect{Q}_{\vect{v},\vect{w}} &=& \vect{I} - \vect{N} \vect{V} (\vect{N}\vect{V})^T \ ,
\end{eqnarray}
where $\vect{V}$ is a $3 \times 2$ matrix with columns $\vect{v}$ and $\vect{w}$,
and $\vect{I}$ is the $M\times M$ identity matrix. As shown in appendix \ref{A2} , the probability distribution
of $E_P$ is non-central $\chi^2$ with 2 degrees of freedom and
non-centrality parameter $\Omega^2 $, while $E_Q$ should be $\chi^2$ distributed with M-2
degrees of freedom.

The best estimate of $ \vect{v}$ and $ \vect{w}$ is then obtained by
\begin{equation}
(\hat{v},\hat{w})=\arg\max_{\vect{v},\vect{w}} ||\vect{P}_{\vect{v},\vect{w}} \vect{R}||^2 \ ,
\label{eqsti}
\end{equation}
where the $\max_{\vect{u},\vect{w}}$ is taken over the unit sphere. The direction of
the Earth rotation axis can be estimated  as a particular case of Eq. (\ref{eq:dodb}).
In fact, the projectors $\vect{P}_{\vect{w}}$ and $\vect{Q}_{\vect{w}}$ can be obtained by
substituting the matrix  $\vect{V}$ for the $3 \times 1$ matrix $\vect{W}$
with columns $\vect{w}$. The difference lies in the dimension of the corresponding
subspaces, i.e. ${\cal P}_{{\cal W}}$ and ${\cal Q}_{{\cal W}}$ have dimension 1 and M-1 respectively.
It is worth noticing that the maximum of Eq. (\ref{eqsti}) can be computed by an analytical formula
both for the location of the meridian plane and the direction of
the Earth rotation axis. In fact, if we introduce in the local reference frame the (local) spherical
coordinate ($R$, $\Theta$ and $\Phi$) (we use capital letters to avoid confusion with Sect. II)
and parametrize the unit vectors $\vect{v}$ and  $\vect{w}$ with these angles, for instance
$\vect{w}= (\cos\Phi\sin\Theta,\sin\Phi\sin\Theta,\cos\Theta)$ and
$\vect{v}=(\cos\Phi\cos\Theta,\sin\Phi\cos\Theta,-\sin\Theta)$ , we have that the maximum
of $||\vect{P}_{\vect{v},\vect{w}} \vect{R}||^2$ and $||\vect{P}_{\vect{w}}\vect{R}||^2$ is achieved for
\begin{eqnarray}
\tan\widehat{\Theta}&=& \left(\frac{\vect{R}^T \vect{F} \vect{R}}{\vect{R}^T \vect{H}\vect{R}}\right)^{1/2}
\nonumber \\
\tan \widehat{\Phi}&=& \left(\frac{\vect{R}^T \vect{K} \vect{R}}{\vect{R}^T \vect{J}\vect{R}}\right)^{1/2}
\label{eqstitf}
\end{eqnarray}
where $\vect{F}$,  $\vect{H}$,  $\vect{K}$,  $\vect{J}$ are $M\times M$ symmetric matrices which are functions of
the $\vect{u}_\alpha$ alone.


In general, there are no analytical calculations for mean and variance of $\hat{\boldsymbol{v}},\hat{\boldsymbol{w}}$
and one must run Monte Carlo simulations to get their estimates. However, in the limit of high SNR
$E_P$ and $E_Q$ tend to be Gaussian distributed, as well as fluctuations of
$\hat{\boldsymbol{v}}$ and $\hat{\boldsymbol{w}}$ around their mean values.  The same reasoning holds true
also for the estimation of $\widehat{\Theta}$ and  $\widehat{\Phi}$.

\begin{figure}[h]
\includegraphics[width=241pt]{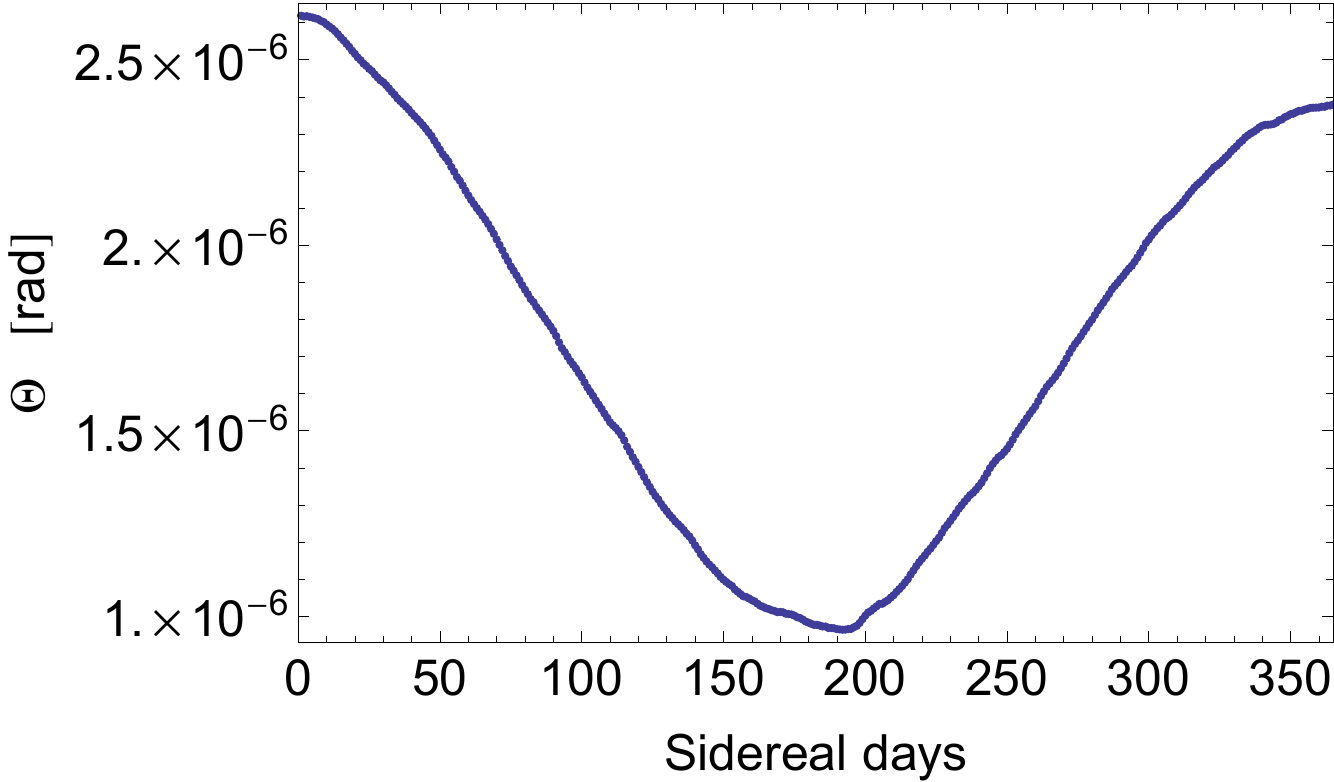}
\caption{ Change of the angle $\Theta$ due to polar motion as measured by the
ring laser responses in one year. }
\label{THETA}
\end{figure}

\begin{figure}[h]
\includegraphics[width=241pt]{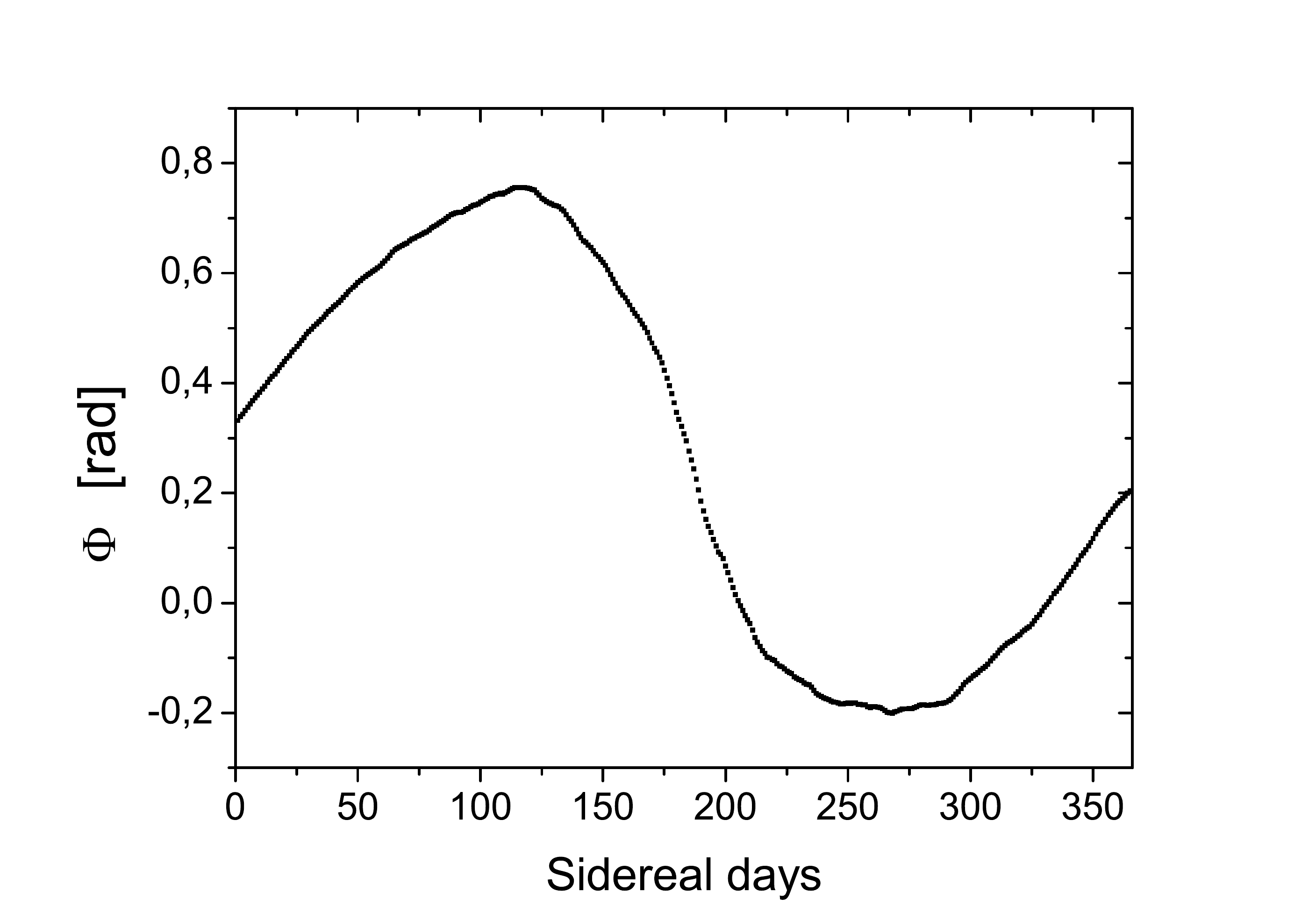}
\caption{ Change of the angle $\Phi$ due to polar motion as measured by the
ring laser responses in one year. Note the large variation  of  $\Phi$ which correspond to
a nearly complete precession cycle of the Earth axis in one year.
}
\label{PHI}
\end{figure}

\begin{figure}[!h]
\includegraphics[width=241pt]{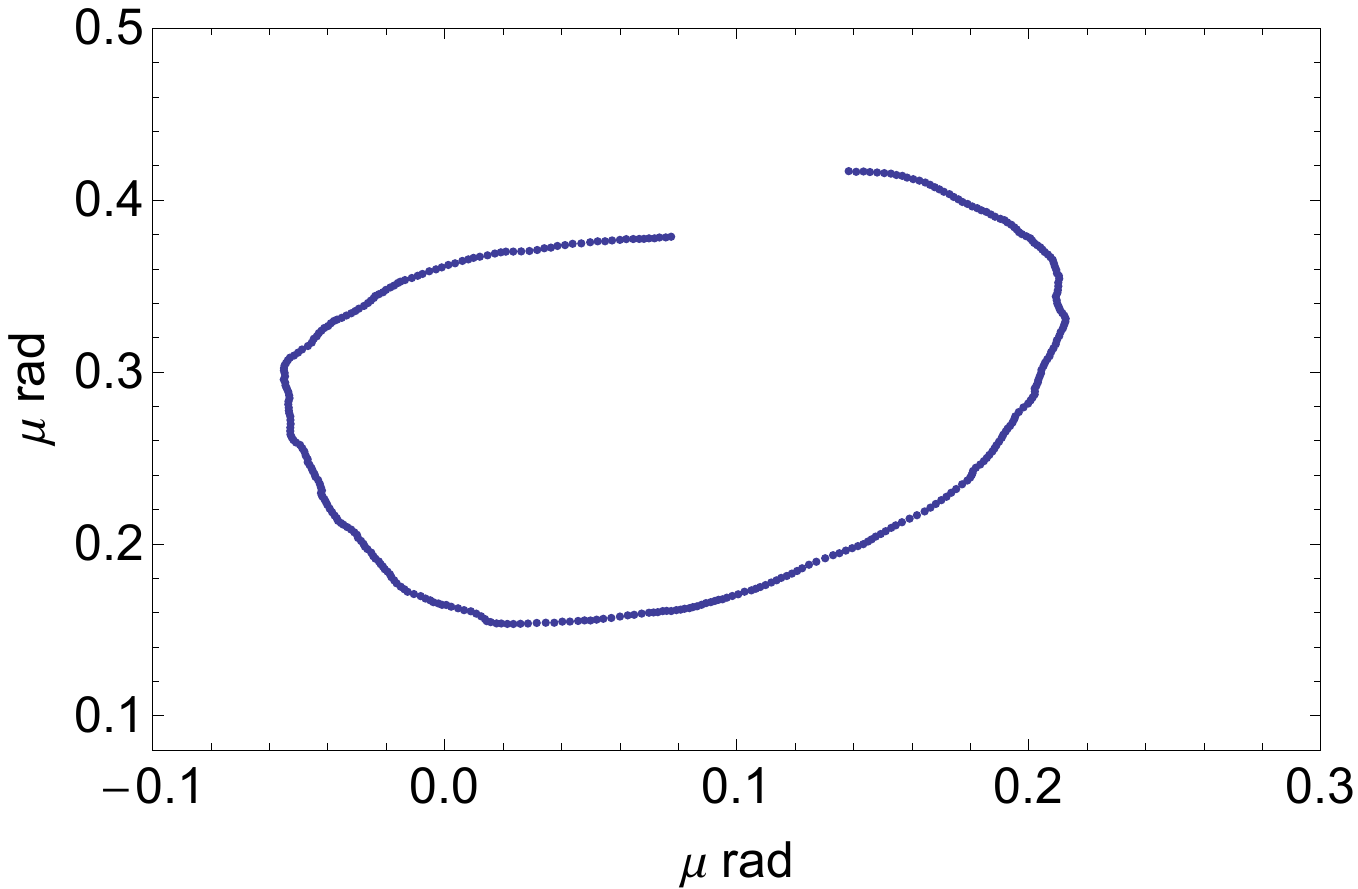}
\caption{ The estimated polar motion from the 6 ring laser responses. }
\label{RPOM}
\end{figure}

The validity of the proposed experimental configuration has been checked by
 a numerical simulation over a period of 1 year of the six responses of
the octahedral configuration oriented as in Fig.~\ref{nodi} of subsection \ref{guide}.
In order to simplify the calculations
we assume that the laboratory  colatitude is $\theta=\pi/4$ and that the normal
to the plane of a ring forms a $\pi/4$ angle with respect to the
Earth axis, and another normal is orthogonal to the former and forms again a $\pi/4$ angle
with the west-east direction.
This configuration is close to a possible experimental arrangement at the \emph{Gran Sasso
National Laboratories} (LNGS) within few degrees. The directions of the
unit vector $\vect{u}_\alpha$ in the local reference frame are
\begin{eqnarray}
\left\{
\begin{array}{l} \vect{u}_1=\vect{u}_4=\left(\frac{1}{2},\frac{1}{\sqrt{2}},\frac{1}{2}\right)\\
\vect{u}_2=\vect{u}_5=\left(-\frac{1}{2},\frac{1}{\sqrt{2}},-\frac{1}{2}\right)\\
\vect{u}_3=\vect{u}_6=\left(-\frac{1}{\sqrt{2}},0,\frac{1}{\sqrt{2}}\right)\\
\end{array}\right.
\label{versori}
\end{eqnarray}
and the rotation signal for the 6 rings are equal within a factor $\sqrt{2}$.
\begin{figure}[!h]
\includegraphics[width=241pt]{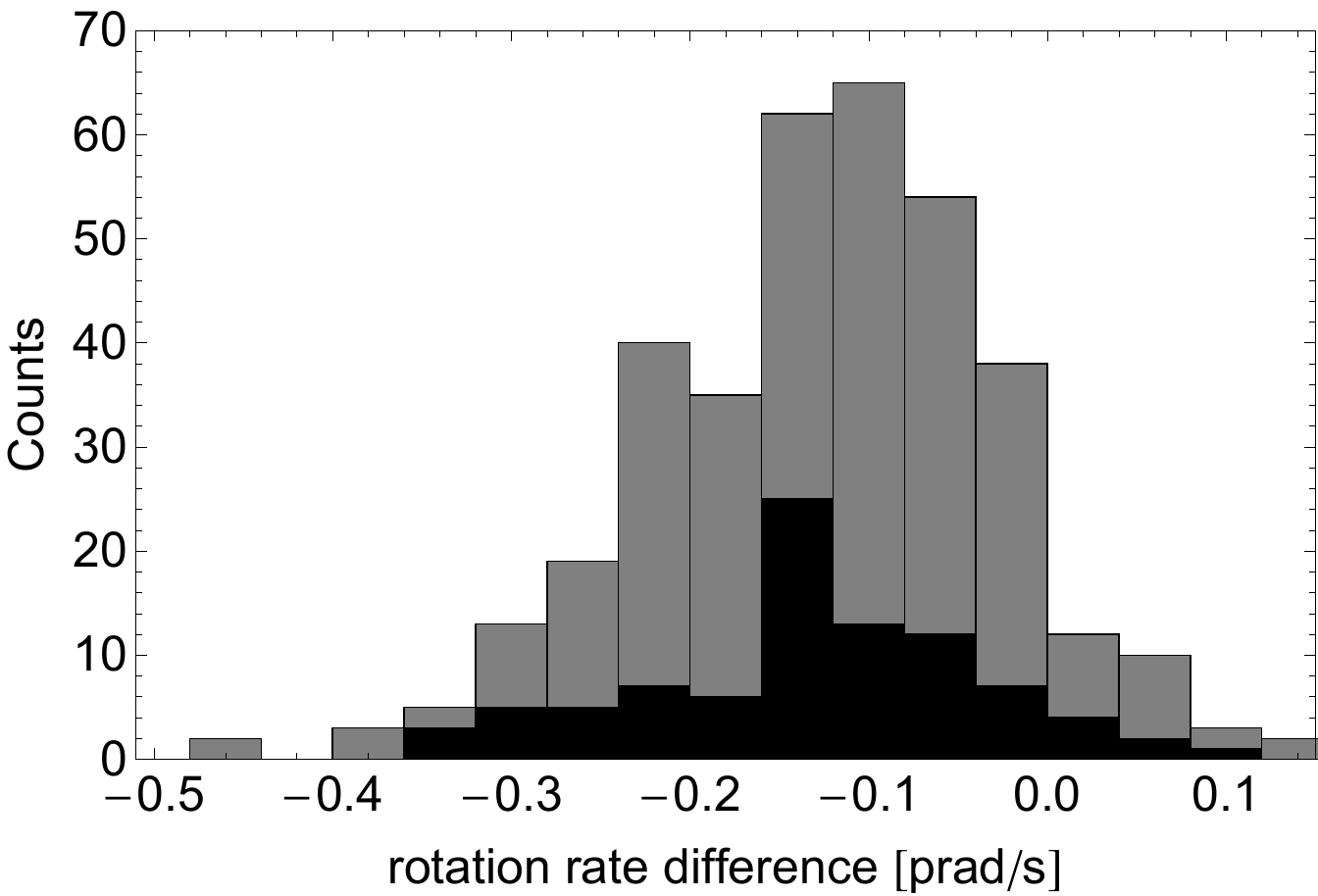}
\caption{ Histograms of the  difference $\Delta$
between $Q$ and $\Omega^2_\oplus$, normalized with the mean sidereal day, collected for 3 months (dark histogram) and one year (light histogram).
}
\label{Fhist}
\end{figure}
We assume one mean sidereal day $T_S=86164.0989 \ s$ of integration and a noise standard deviation
$\sigma_\Omega= 7 \times 10^{-2} \ prad/s$. The variance $\sigma_\Omega$ is extrapolated from present "G" sensitivity
at $10^{4}$ $s$ and scaling by a factor $5$, due  by the increase of the ring size and the power of the laser of a factor
$1.5$ and $10$, respectively. The relativistic rotation contributions
$\Omega^\prime_r= -2.8 \times 10^{-2} prad/s$ and $\Omega^\prime_\theta = -5.6 \times 10^{-2} prad/s$
have been added to the Earth rotation vector $\vect{\Omega}_\oplus$, as estimated by IERS \cite{IERS}.
The component of relativistic effects parallel to $ \vect{\Omega}_\oplus$ is  $\Omega^\prime_\parallel=
(\Omega^\prime_\theta+\Omega^\prime_r)/\sqrt{2}=5.9 \times 10^{-2} prad/s$.
Using Eq. (\ref{eq:dodb})
we calculated the responses of the 6 rings and then we injected the Gaussian noise.
In Fig.~\ref{Fhist} we show the histograms of $T_S \Delta/(2 \pi)$ accumulated for 90 and 366 sidereal days.
The corresponding mean values of the parallel component of relativistic effects
are $-6.0\times 10^{-2}\ prad/s$ and $-6.2\times 10^{-2}\ prad/s$ with standard deviations
$4.7 \times 10^{-3}\  prad/s$  and $2.6 \times 10^{-3}\ prad/s$, respectively.  Thus a $\sim 10$\% accuracy
can be achieved in 3 months by simply comparing the square modulus of rotation vectors.
In order to give a full estimate of the vector
$\vect{\Omega}^\prime$, we have also explicitly calculated day by day the angles $\widehat{\Theta}$ and
$\widehat{\Phi}$ for describing the orientation  of the meridian plane and the direction of the Earth rotation vector.
The results are summarized in Fig.~\ref{THETA} and \ref{PHI}, where we report the time evolution of these angles
and in Fig.~\ref{RPOM} where we show the corresponding annual polar motion.

By synchronizing the polar motion measured in the local reference system with the polar motion measured 
by IERS in the fixed star reference system, the two reference frames will coincide within the accuracy 
of the measurement of $\vect{\Omega}$ and $\vect{\Omega}_\oplus$, say 1 part of $10^{10}$.     
As a final remark, we point out that the full measurement of the vector $\vect{\Omega}^\prime$
allow us for the estimate of $\Omega^\prime_\perp\simeq 2\times 10^{-2}\ prad/s$ with a standard deviation
of the same order of magnitude of the estimate of $\Omega^\prime_\parallel$. This represent an increase
of the relativistic rotation signal of $\sim 30$ \%. However, the estimate of $\vect{\Omega}^\prime$
is crucial to separate the geodetic from Lense-Thirring contributions and/or to
measure the PPN parameters $\alpha_1$ and $\gamma$.

\subsection{\label{sec:b}  The Earth motion and feasibility of the experiment}

Since our goal is the estimate of the Lense-Thirring effect at few \% accuracy,
the independent measurement of $\vect{\Omega}_\oplus + \vect{\Omega}_{_{REL}}$, which
represents the rotation of the laboratory with respect to distant stars,
must be determined to $10^{-10} \Omega_\oplus$. Due to tidal forces  and to the exchange of angular
momentum between the solid Earth and geophysical fluids, the angular velocity of the Earth varies
in time, both in direction and modulus. Changes in modulus correspond to a variation
of the Length of the Day (LoD) of few milliseconds with respect to atomic clocks. The
direction of the rotation axis of the Earth varies with respect to both the fixed stars and the
Earth-fixed reference frames.
Nowadays, the best Earth rotation monitoring is provided by
the {\sl IERS 05C04} time series \cite{IERS}
which are routinely obtained using the geodetic space techniques VLBI (Very Long Baseline Interferometry),
SLR (Satellite Laser Ranging), GPS ( Global Positioning System) and
DORIS (Doppler Orbitography and Radiopositioning Integrated by Satellite).

\begin{figure}[th]
\includegraphics[width=241pt]{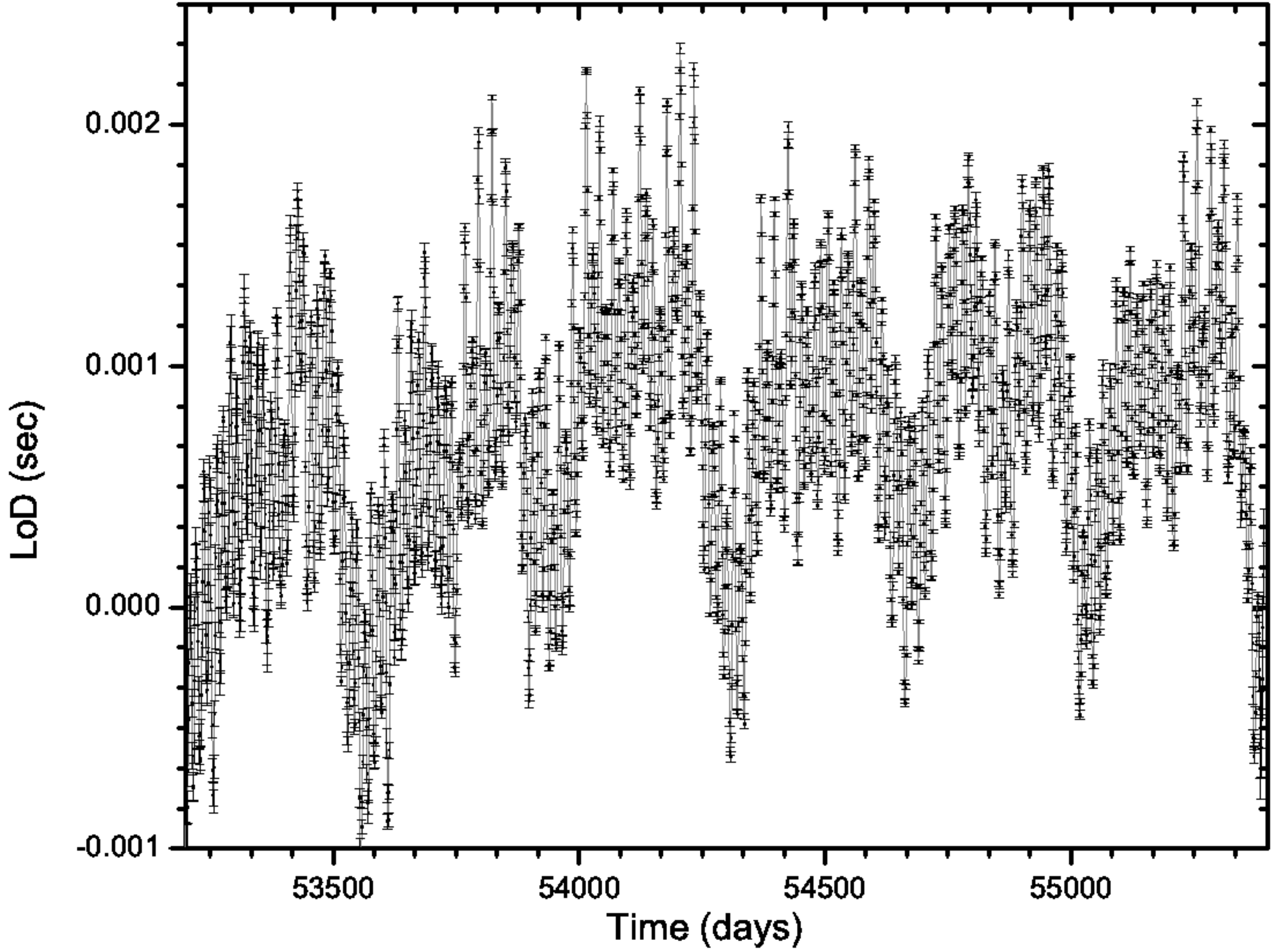}
\caption{ The change of the Length of the Day (LoD) over the last $6$ years from the {\sl IERS 05C04}
time series. Notice that estimated errors of LoD decreased in the last years to
a level which correspond to  $10^{-14} \ rad/sec$, i.e. 0.1 ppb $\Omega_\oplus$.}
\label{figlod}
\end{figure}

\begin{figure}[th]
\includegraphics[width=241pt]{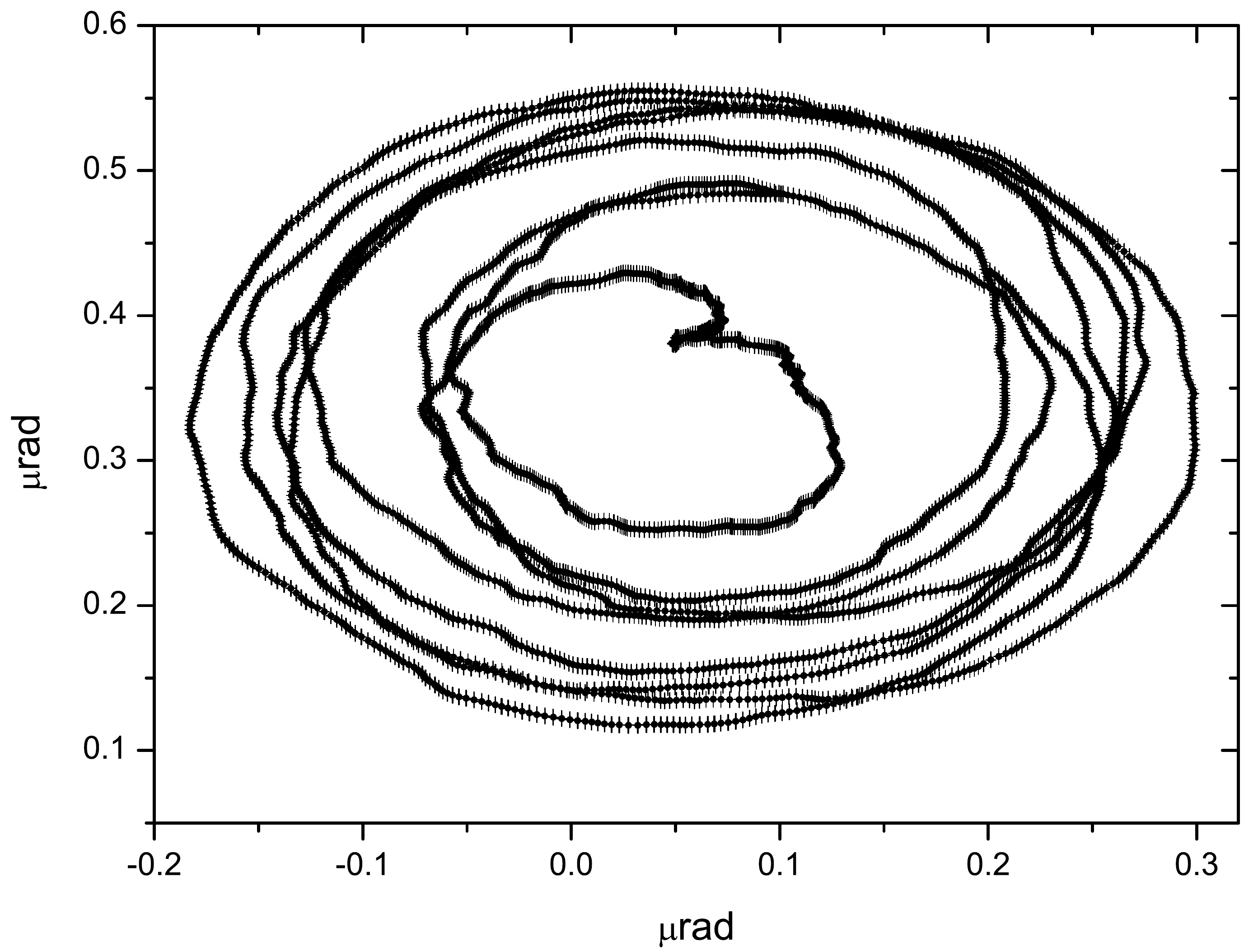}
\caption{ The change of the direction of the Earth rotation axis (i.e. pole position) over the last
6 years from the {\sl IERS 05C04} time series. Estimated errors are also plotted.}
\label{figpmo}
\end{figure}

In  Figs.~ \ref{figlod} and \ref{figpmo} we report the Length of the Day (LoD) and the pole position
with the corresponding errors of the last six years. It is worth to noticing that
the achieved precision is 0.001 ms in the LoD and $0.1 \ \mu arc sec$ in the pole position.

Further improvements are expected in the next few years and the overall errors
in LoD and pole position should decrease of a factor 10 that
is crucial for a 1\% measurement of the relativistic rotation terms.
However, the {\sl IERS 05C04} time series is already sufficient to get $|\vect{\Omega}_\oplus|$
with $3 \%$ accuracy.

For what concerns the differential rotation of the laboratory
with respect to the rotation estimated by IERS, it is expected to be sufficiently small
to contribute to $\Omega_\oplus$  only trough  $\vect{\Omega}_{\oplus\parallel}$.
However, $\vect{\Omega}_{_{REL}}$ is still largely unknown due to possible
micro-rotations of the crust of the Earth. This is one of the causes limiting the performances of G in Wettzell: the Earth crust motion caused by atmospheric changes. It is assumed that an underground facility is less sensitive to this kind of noise sources. It is as well important to keep the experiment  close to VLBI stations. The underground Gran Sasso Laboratories
is placed half way between two relatively close VLBI stations, Medicina \cite{medicina} and Matera \cite{matera} which
can provide estimates of the crustal motion of the Adriatic plate \cite{crumo}.
A significant contribution
to $\vect{\Omega}_{_{REL}}$  comes from the "diurnal polar motion" (periodic motion
of the Earth crust due to tides) and consists in periodic changes of
amplitude $\sim 10^{-7} \Omega_\oplus$. This effect has been already measured by
large ring laser gyroscopes \cite{PMwettzell},
and can be accurately modeled and then subtracted from ring lasers responses.

We conclude that  by means of available geodesics and geophysics  techniques, provided that
the experiment is located in an area with very low relative angular motion ($\Omega_{_{REL}}$),
 a suitable tri-axial detector of rotation can in principle detect $\Omega^\prime$ with $\%$ precision.

\section{\label{apparatus} The 'Real Apparatus', the present sensitivity of G in Wettzell}

\subsection*{\label{sec:c} Sensor properties}

A closer look at equation~\ref{eq3.1} reveals that there are three basic effects one has to carefully
account for. These are:
\begin{itemize}
\item scale factor stability ($4 A/\lambda P$)
\item orientation of the gyroscope with respect to the instantaneous axis of rotation of the Earth
\item instantaneous rate of rotation of the Earth -- Length of Day (LoD)
\end{itemize}
The scale factor for all practical purposes has to be held constant to much better than 1 part in
$10^{10}$. Otherwise the frame-dragging parameter cannot be determined unambiguously. For G, the
base of the gyroscope has been manufactured from Zerodur, a glass ceramic with a thermal expansion
coefficient of $\alpha < 5 \times 10^{-9}/^{o}C$. Furthermore the instrument is located in a thermally
insulated and sealed environment with typical temperature variations of less than 5 $mK$ per day. However, because
the underground laboratory is only at a depth of 5~m, there is still a peak to peak temperature
variation of about 1 degree per year, accounting for the change of seasons. Changes in the atmospheric pressure also affect the dimensions of the ring laser structure
by changing the compression of the Zerodur block
and cannot be neglected. Hence G is kept in a pressure stabilized enclosure. A feedback system
based on the determination of the current value of the optical frequency of the lasing mode of one sense of propagation
allows for active control of the pressure inside the steel vessel such that an overall geometric scale
factor stability of better than $10^{-10}$ is routinely obtained.  At the same time the design of the
instrument is made as symmetric as possible. So changes in area and perimeter are compensated
with a corresponding change in wavelength as long as no shear forces are present and the longitudinal
mode index stays the same.

\begin{figure}[ht]
\centering
\includegraphics[height=5.4cm]{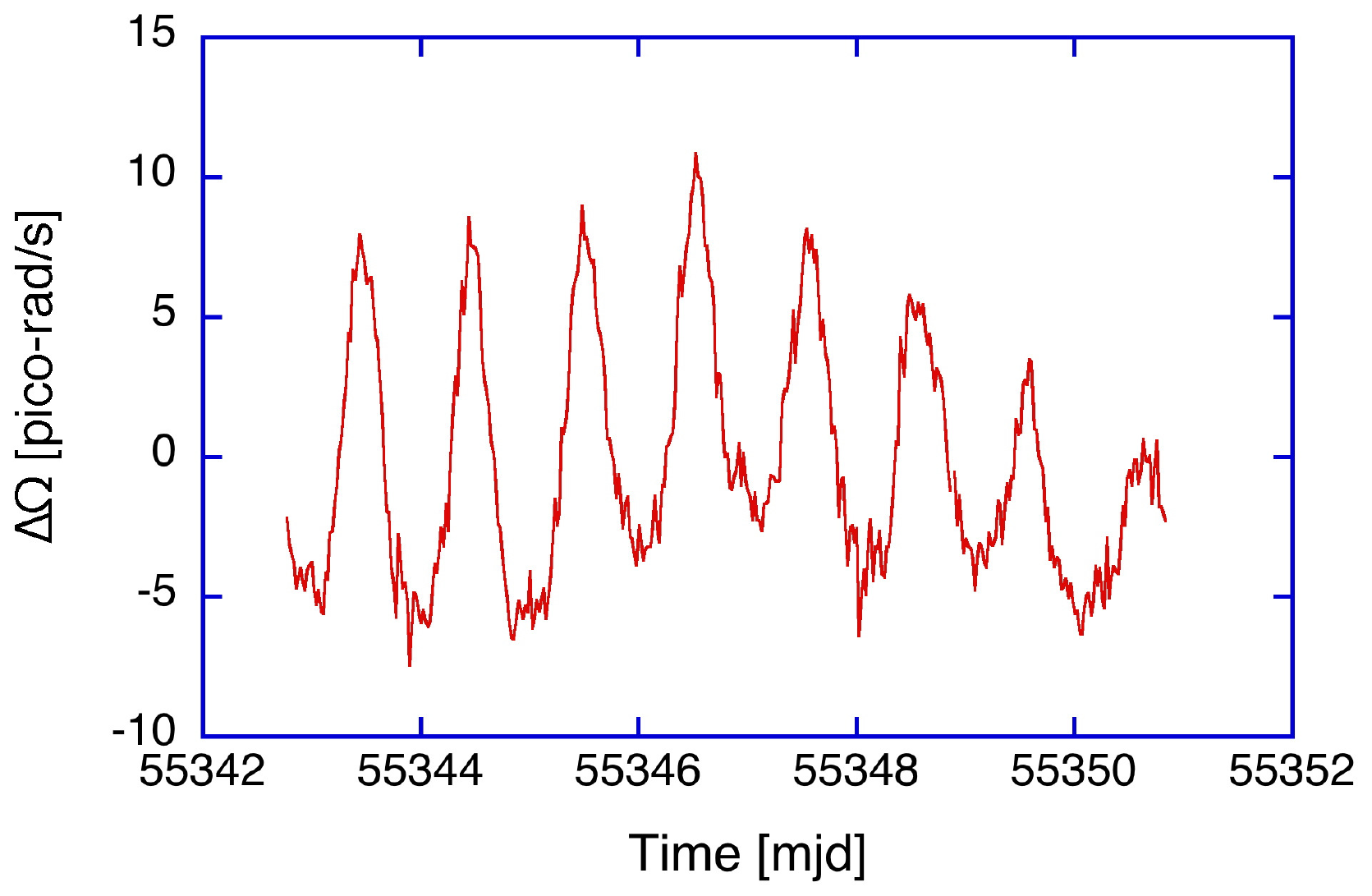}
\caption{Approximately eight days of raw G data taken with 30 minutes of integration time. One can clearly see the contributions from diurnal polar motion, solid Earth tides and local tilt.}
\label{G-raw}
\end{figure}
\begin{figure}[ht]
\centering
\includegraphics[height=4.cm]{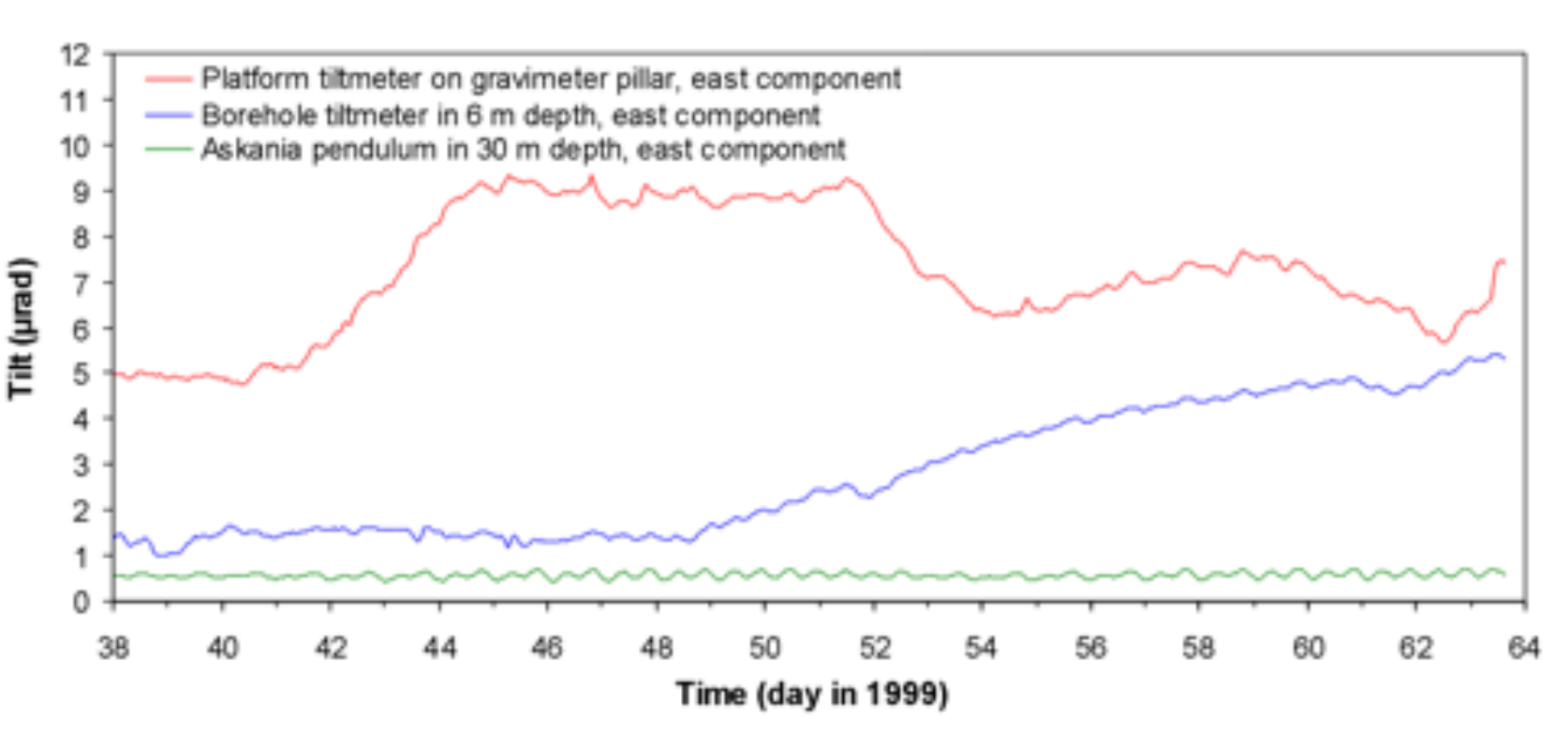}
\caption{Measurement of  local tilts as a function of depth in the Earth.}
\label{Tilt}
\end{figure}

A typical eight day long measurement sequence of rotation rate data from the G ring laser is shown in Fig.~\ref{G-raw}.
In order to demonstrate the obtained sensor sensitivity we have subtracted the mean Earth rotation
rate from the gyroscope data.
The y-axis gives the measured variation of the rate of rotation, while the x-axis shows the time expressed in the
form of the modified Julian date. Each data point was taken by integrating over 30 minutes of measurement data.
There are several distinct signal contributions in the data, which come from known geophysical
effects. The most prominent signal is caused by diurnal polar motion \cite{polmot}. The polar motion
data is superimposed by a tilt signal caused by the semi-diurnal and diurnal tides of the solid Earth,
distorting the otherwise sinusoidal diurnal frequencies slightly. At the
Geodetic Observatory in Wettzell the tilt effects of the solid Earth tides can be as large as 40 $nrad$ in amplitude.
In Fig.~\ref{G-raw} the diurnal signal is dominated by the polar motion \cite{bere}.
Less evident in Fig.~\ref{G-raw} are the effects from local tilt, which contains periodic signals of tidal origin as well as non-periodic signals.
The latter   are non-periodic and
usually change slowly over the run of several days. High resolution tiltmeters inside the pressure stabilizing
vessel of the G ring laser keep track of these local effects and the data is corrected for gravitational
attraction (atmosphere, sun and moon)l \cite{polmot}. Large non-periodic
local tilts occur most prominently
after abundant rainfall, indicating hydrological interactions with the rock and soil
beneath the ring laser monument.  Fig.~\ref{Tilt} shows the east component of three tiltmeters installed
i) on a gravimeter pillar at the surface, ii) in 6 $m$ depth, and iii)  in 30 $m$ depth.

While the tilmeter in 30m depths clearly shows the periodic signal of the solid earth tides, the tilt record of the instruments near to the surface
is dominated by large non-periodic signals hydrological, thermoelastic and barometric origin.
Several investigations have shown that the site and the installation depths of tiltmeters has a major impact
on environmental noise mainly coming from hydrology \cite{jent}, \cite{kump}, \cite{weis},  \cite{jahr}  has shown that
even in 100 $m$ depths effects caused by hydrological changes are detectable, but strongly reduced in comparison to a 50 $m$ deep installation.
First investigations related to topographic and temperature induced effects were carried out by \cite{harr1} and \cite{harr2}.
 Detailed investigations using the finite-element method have shown that these effects can amount to more than 10 $nrad$
(\cite{geba1}, \cite{geba2}), while the distance between the source and the location of observation can be several hundred meters.
Additionally, recent work using the G ring laser data reveals that effects caused by wind friction at the Earth surface yields to
high frequency rotations of large amplitudes.

The large seasonal temperature effect on the G ring laser as well
as the substantial local tilt signals and the rather high ambient noise level of our near soil surface
structures give reasonable hope of much better performances of a ring laser installation in a deep
underground laboratory such as the Gran Sasso Laboratories.

\begin{figure}[!ht]
\centering
\includegraphics[height=5.4cm]{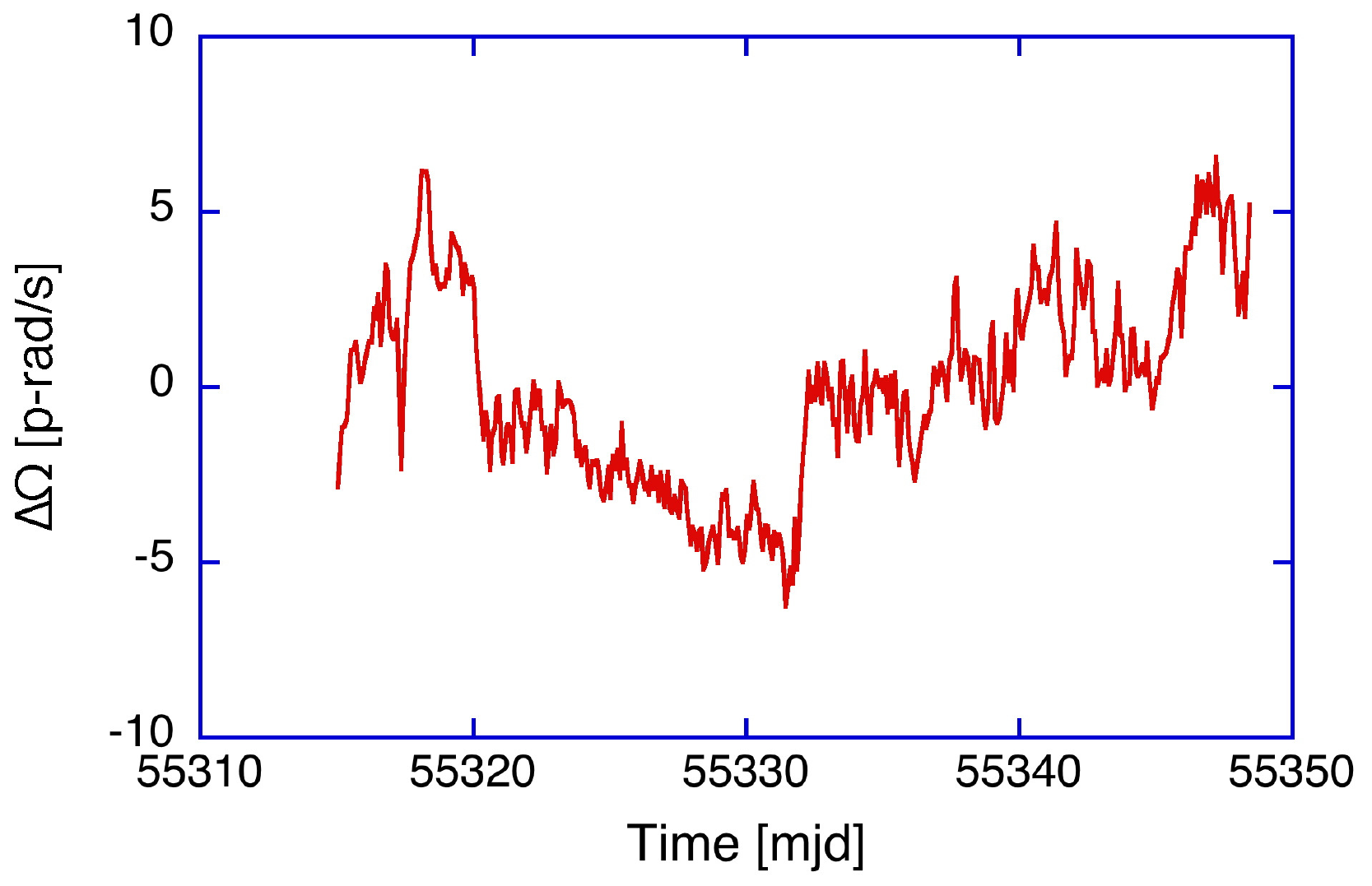}
\caption{ The rotation rate of the Earth measured with the G ring laser as a function of time. Averaging over 2 hours was applied to a corrected dataset, where all known geophysical signals have been removed.}
\label{avg2h}
\end{figure}

\begin{figure}
\includegraphics[scale=0.28]{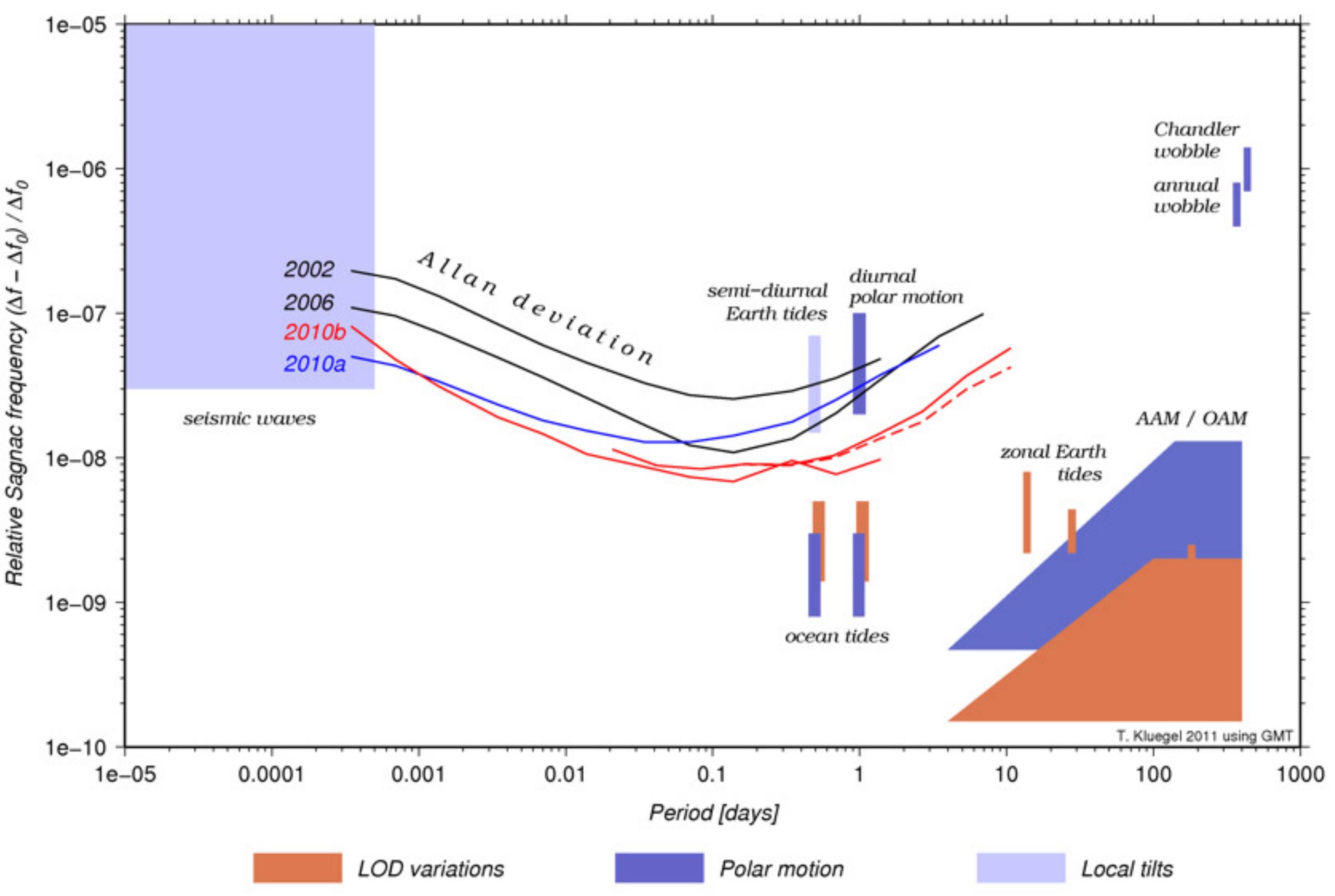}
 \caption{Resolution and stability of G, compared with Earth signals}
 \label{allan}
 \end{figure}

For the detection of fundamental physics signals one has to remove all known perturbation signals of the Earth from the
ring laser time-series. Furthermore we have applied 2 hours of averaging of the data in order
to reduce the effect from short period perturbations. Fig.~\ref{avg2h} shows an example.
In Fig.~\ref{allan} we show
the current sensitivity expressed in term of Allan deviation of the G, the expected
sensitivity of each ring laser at  Gran Sasso  Laboratories and the relevant geophysical signal.

In order to reduce the local orientation uncertainties, which remain after local tilts measured with the
high resolution tiltmeters have been removed,
averaging as indicated above was applied to a series of 30 days of data
collection, including the period shown in Fig.~\ref{G-raw}. It can be expected that a similar data set
from the Gran Sasso laboratory would become substantially smoother, since most of the
perturbations, caused by ambient atmosphere - topsoil interaction still contained in the data
of Fig.~\ref{avg2h} would no longer exist in the deep underground facility. Changing hydrologic conditions
presumably causing small local rotation and temperature variations, atmospheric pressure and wind loading
are among the sources for the systematic signatures in the residual data.

\section{\label{sec:c1} Configuration of a tri-axial detector }

\begin{figure}
\includegraphics[scale=.15]{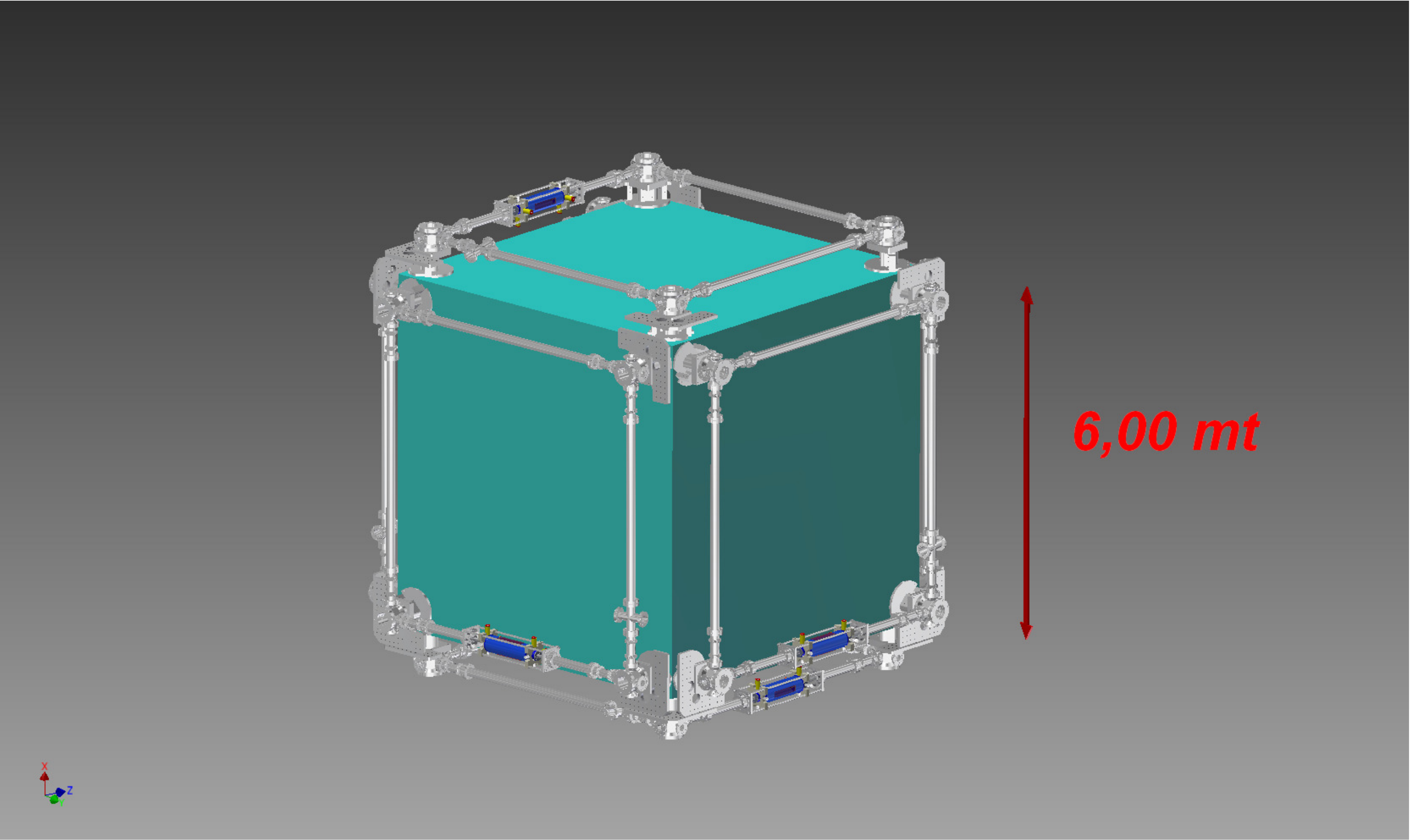}
\caption{Six rings arranged on the faces of a Cube, using the GEOSENSOR design, which has been successfully used so far for middle size rings, as our prototype G-Pisa}
\label{cube}
\end{figure}

\begin{figure}
\includegraphics[scale=.25]{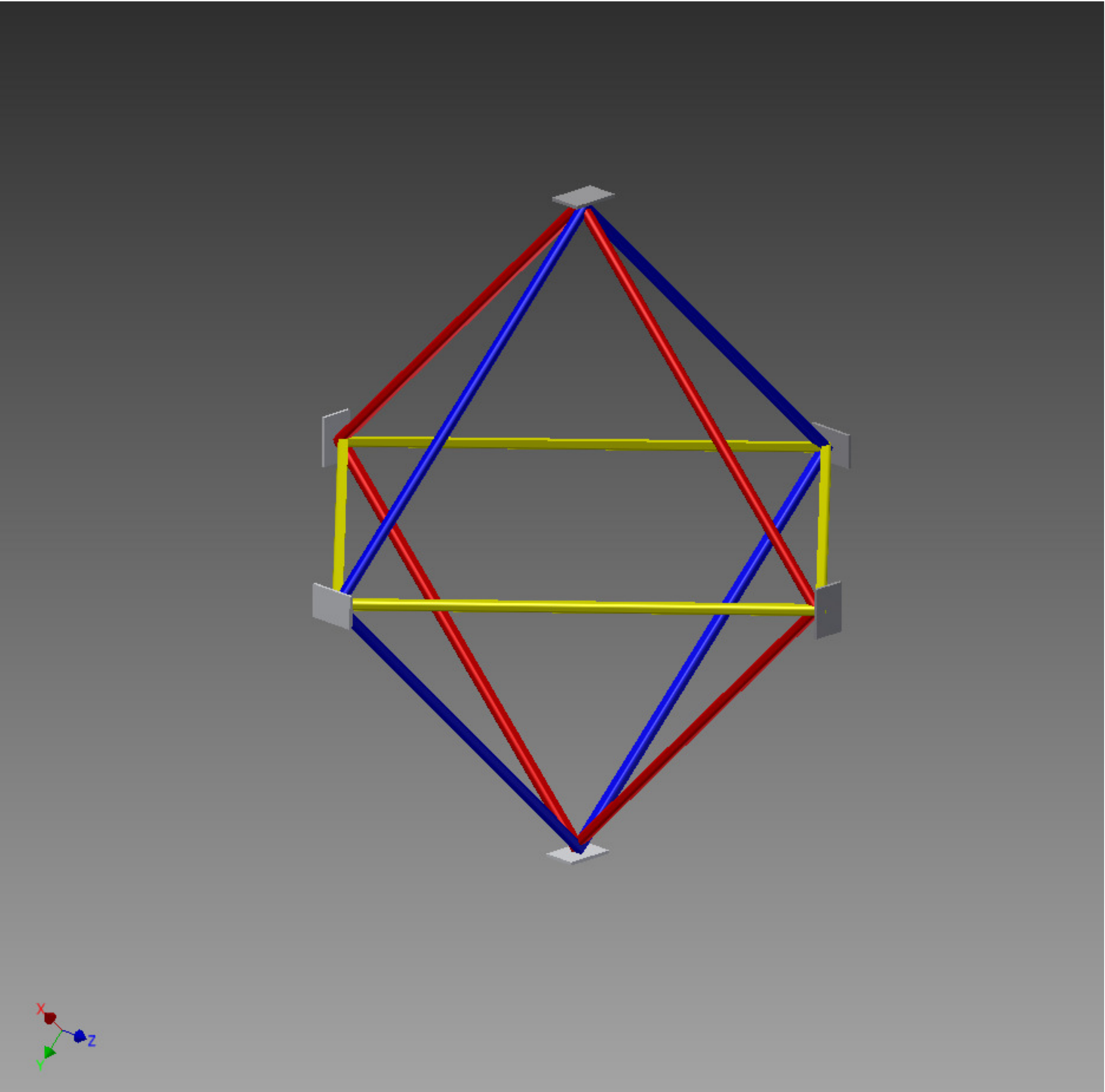}
\caption{Three rings are formed using $6$ mirrors located on the vertices of an octahedron}
\label{ott}
\end{figure}

\begin{figure}[!hbt]
\includegraphics[scale=.22]{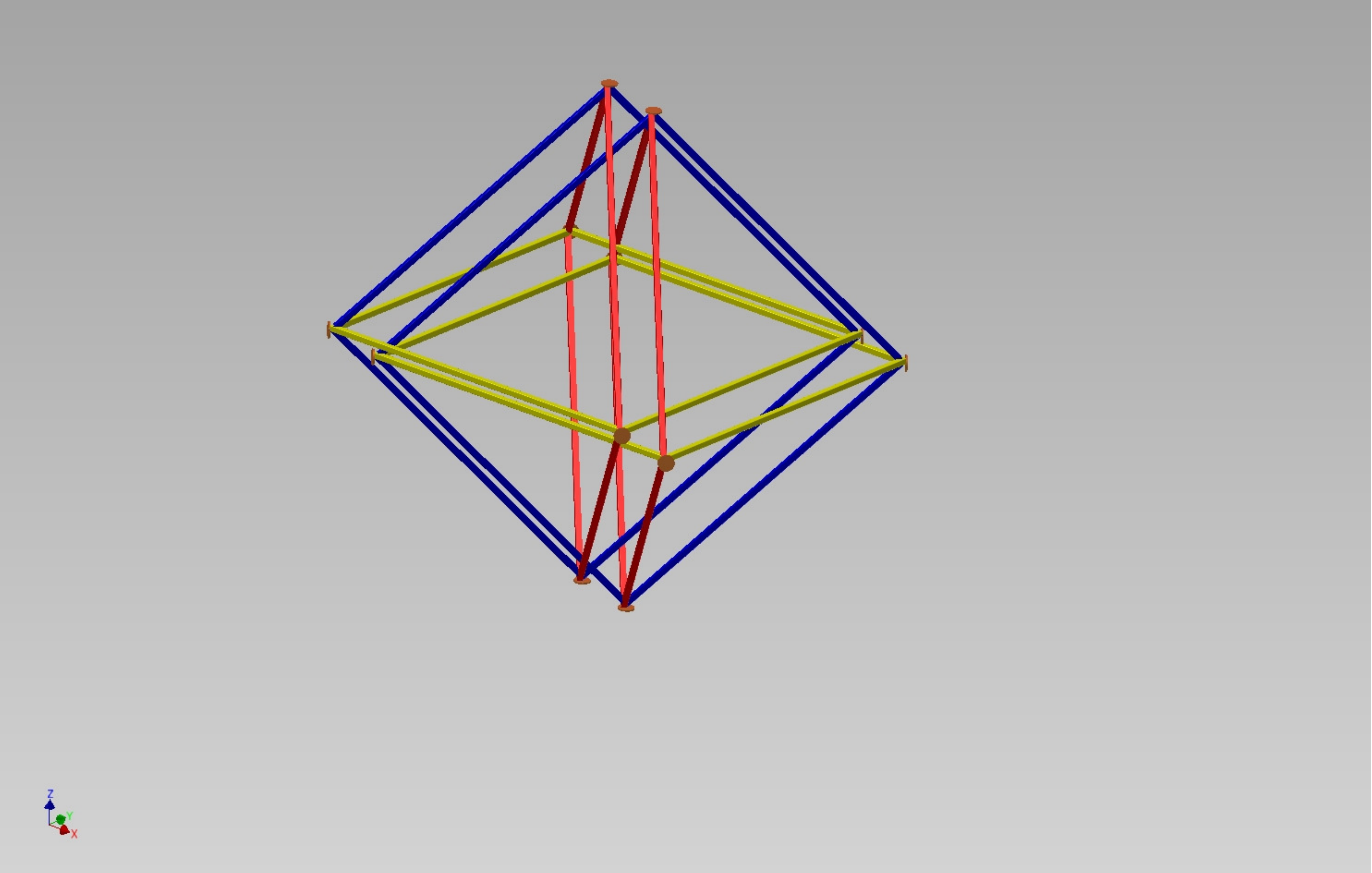}
\caption{Six rings, two by two parallel, with mirrors on the vertices of two octahedron, constructed very close one to the other in order to reduce the
dimension of the apparatus.}
\label{ott2}
\end{figure}

From now on, we will restrict  our analysis to $24$ $m$ perimeter rings, arranged in two configurations that are of some
experimental interest, i.e.  6 ring lasers
rigidly mounted on the faces of a cube, as shown in Fig.~\ref{cube}, and $3$ ring lasers oriented along the edges of an octahedron, see Fig.~\ref{ott}.
The cubic configuration requires $24$ mirrors forming
$6$ independent rings and the extension of the GEOSENSOR design is straightforward (see subsection \ref{guide} );  while the octahedral configuration require $6$ mirrors only to form $3$ orthogonal  rings.
By itself the configuration which uses a cube is redundant, each ring has a parallel companion, which can be used for the study of systematics. For the octahedron configuration the implementation of the GEOSENSOR design needs further development.
Redundancy  can be easily obtained constructing a second octahedron with planes parallel to the other one. The two structures should be built very close to each other, in order to keep as much as possible the whole apparatus compact; in this way $6$ rings are available, analogously to the cube configuration, see Fig.~\ref{ott2}.
This configuration  has the advantage that there are constraints in the relative angle between rings, since each mirror is in common between two rings, and
three linear Fabry-Perot cavities are available using the three diagonals of the rings. Those linear cavities have the capability of monitoring the relative angles between different rings, and as well the length of each diagonal.

\subsection {\label{s1} Ring-laser sensitivity}

The rotation sensitivity $\sigma^2_\Omega$, for noise fluctuations which are dominated by laser
shot noise over an integration time T, reads
\begin{equation}
\sigma^2_\Omega = \frac{c P}{4 A Q} \sqrt{\frac{h f}{W T}}\ ,
\end{equation}
where $Q$ is the quality factor of the optical cavity, $f=c/\lambda$ is the laser frequency,
$h$ is the Plank constant and $W$ is the power of the laser \cite{rlg-noise}. The limiting sensitivity can be
conveniently calculated scaling the parameters of the Wettzell "G" ring laser
\begin{eqnarray}
&\sigma_\Omega& = 2.9 10^{-13} \left(\frac{P}{16\ m} \right)  \left(\frac{16\ m^2} {A}\right) \left(\frac{3 \times 10^{12}}{Q}\right) \times \nonumber \\
&& \left(\sqrt{\frac{20\ nW}{W}} \right)  \left(\sqrt{\frac{10^5 \ s}{T}}\right) rad/s
\label{sensring}
\end{eqnarray}
In order to obtain in few weeks a 10\% accuracy level in the measurement of the relativistic effective rotation rates,
we must achieve the sensitivity goal of $\sigma_\Omega = 7 \times 10^{-14}\ rad/s$ (or equivalently a rotation
noise level $ 20\ prad/sec/Hz^{1/2}$ at a frequency of $1 \ day^{-1}$).
From Eq. (\ref{sensring}) we have that a system of $6$ rings with $P=24 \ m$, $Q=3 \times 10^{12}$ and $W=200\ nW$
can fulfill this requirement.

\subsection {\label{s2} Expected performances of not optimally oriented rings}

We assume that  the ring lasers are identical in the sense
described in Sect. \ref{sec:b} and that the dihedral angles $\arccos({\vect{u}_\alpha\cdot \vect{u}_\beta})$ are measured better than one part in $10^{10}$ in order
to estimate $\Omega$ independently from the reference frame. Note that only the stability of
dihedral angles can be monitored by means of the Earth signal itself only for short times (few days), 
while their measurements and controls must be performed independently in the laboratory.
For instance, assuming that the scale factors are controlled to
the $10^{-10}$ accuracy, the responses of two parallel rings is statistically different from noise 
when their parallelism is modified.

From an experimental point of view, to arrange in the Cartesian planes several rings  and keep the  configuration stable over the integration time $T\simeq 1\ day$
is a demanding task. However, we can relax such a demanding requirement by means of
data analysis procedures that account for slightly non-orthogonal dihedral angles.


For instance, we can use the measured dihedral angles to estimate directly $\Omega$. In fact,
 we can substitute the quadratic combination of ring laser responses in Eq. (\ref{EqQua}) with the
equivalent bilinear combination
\begin{equation}
Q=\sum_{\alpha=1,\beta=1}^M Q_{\alpha\beta} R_\alpha R_\beta
\end{equation}
where $Q_{\alpha\beta}$ are the elements of the $M\times M$ matrix $\vect{Q}=\vect{N}(\vect{N}\vect{N}^T)^{-2}\vect{N}^T$. The statistics of $Q$ is no longer non-central $\chi^2$; however, we can easily compute (see App. \ref{A2} for details)
its mean
\begin{equation}
<Q>= |\vect{\Omega}|^2 + M\sigma_\Omega^2
\end{equation}
and variance
\begin{equation}
\sigma^2_Q= 2 \sigma^4_\Omega \sum_{\alpha_\beta} Q^2_{\alpha_\beta} +
4 \sigma^2_\Omega \Omega^2 \sum_{\alpha_\beta}
Q^2_{\alpha_\beta} \vect{u}^\parallel_\alpha \vect{u}^\parallel_\beta
\end{equation}

In the limit of high SNR, fluctuations of $Q$ tend to be Gaussian distributed, and so we recover
the results in Eq. (\ref{sensQ})  for the overall sensitivity of the system. If we start with dihedral
angle close to $\pi/2$  (say 1 part in $10^5$), then sensitivity loss is very small since it
is of the same order.

\subsection{\label{guide}Guidelines of the Experimental Apparatus}

\begin{figure}
 \includegraphics[scale=0.25]{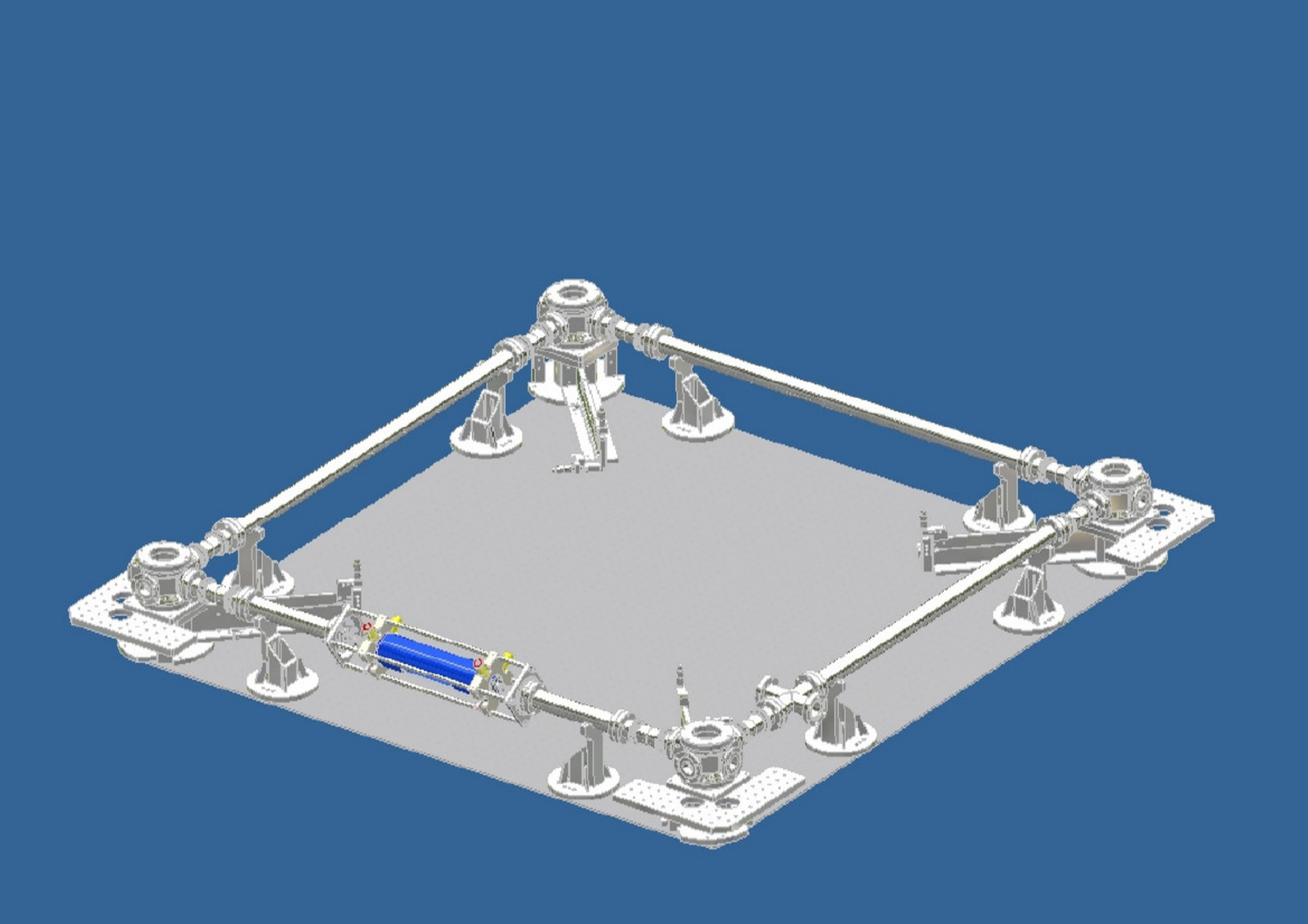}
 \caption{Drawing of G-Pisa, based on the GEOSENSOR design}
 \label{gpisa}
 \end{figure}

\begin{figure}
 \includegraphics[width=\columnwidth]{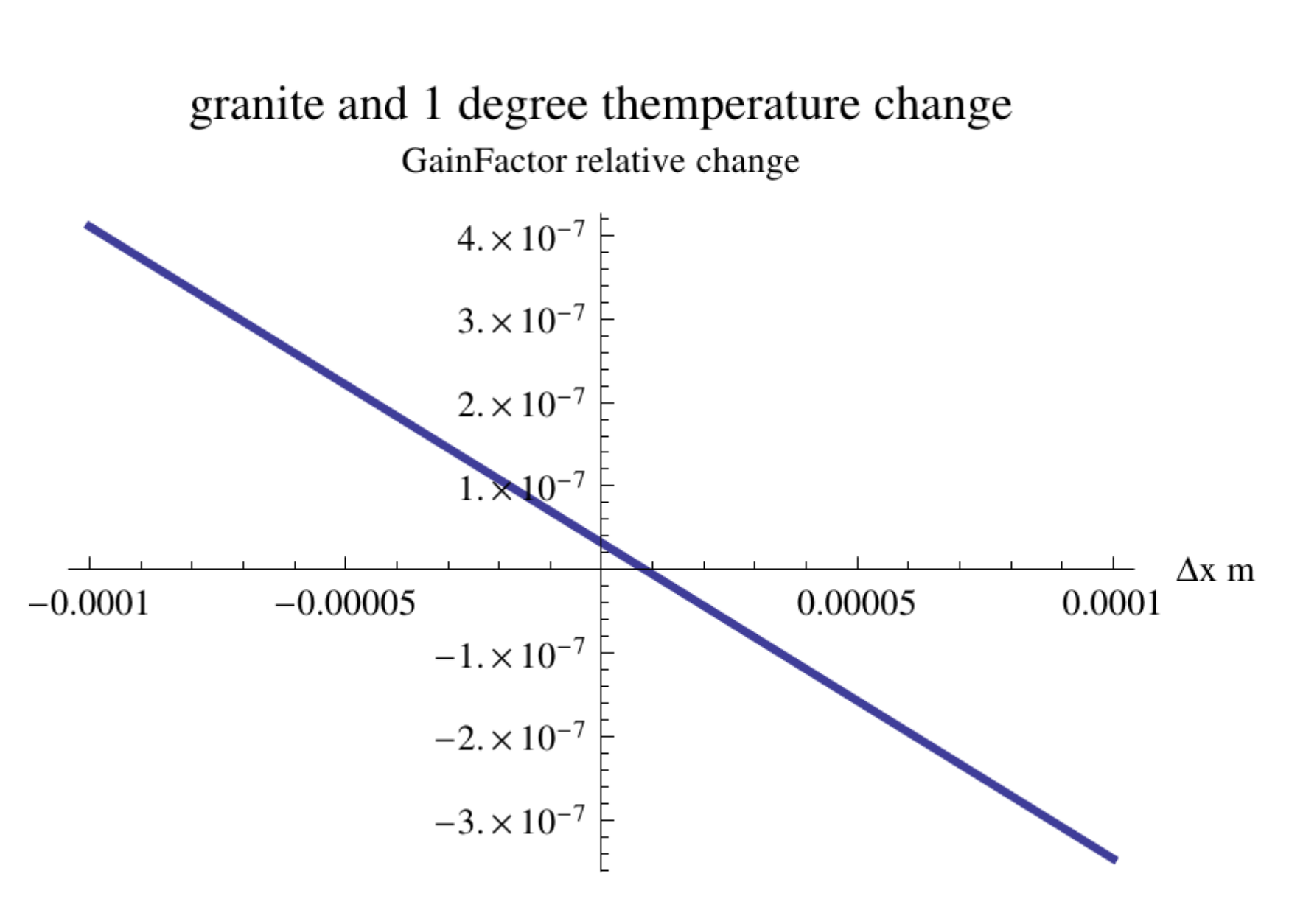}
 \caption{$M_{acc}$, for a rectangular ring, with sides $6\times 6.6$ $m$, in function of a misalignment of one of the four mirrors with respect to the ideal position, the perimeter control acts on four  mirrors, maximum thermal excursion of 1 degree and the support of the GEOSENSOR  has thermal expansion coefficient of $7 \times 10^{-6}$ $ K^{-1}$ (granite)  }
 \label{mmer}
 \end{figure}

\begin{figure}
 \includegraphics[scale=1]{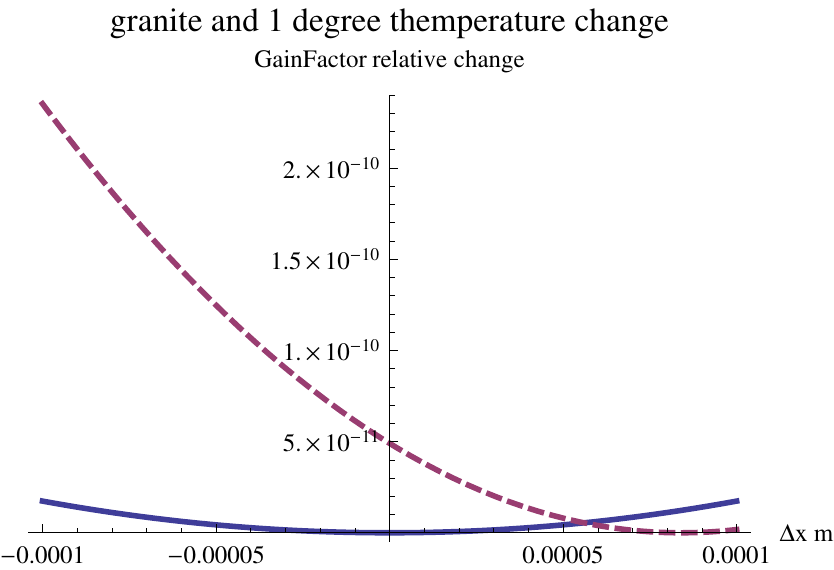}
 \caption{$M_{acc}$ in function of a misalignment of one of the four mirrors with respect to the ideal position, the perimeter control acts on four (thick line) or two (dashed line) mirrors, maximum thermal excursion of 1 degree and the material has $7 \times 10^{-6}$ $m/m K$ (granite)  }
 \label{twofour}
 \end{figure}

\begin{figure}
 \includegraphics[scale=1]{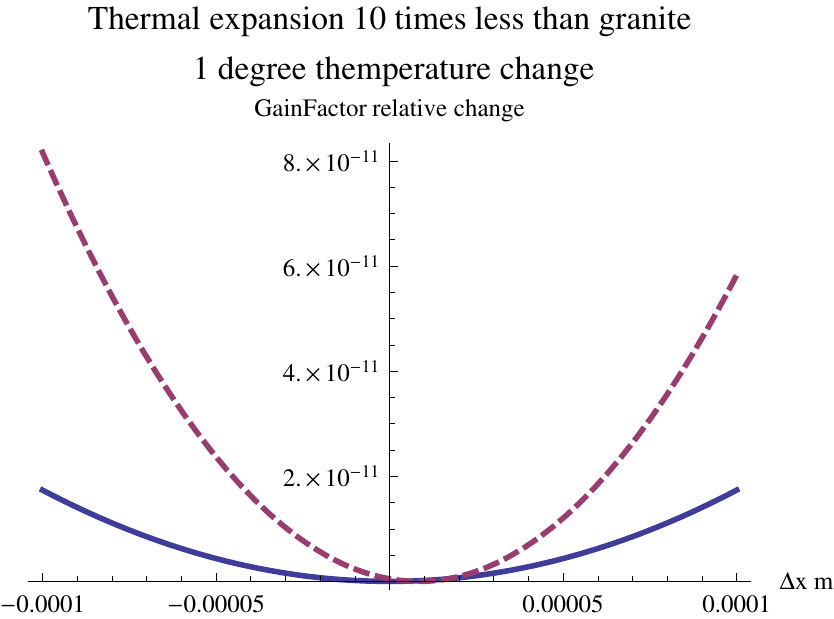}
 \caption{Same as Fig.~\ref{twofour}, but with expansion coefficient 10 times lower}
 \label{twofour1}
 \end{figure}

\begin{figure}
 \includegraphics[scale=0.5]{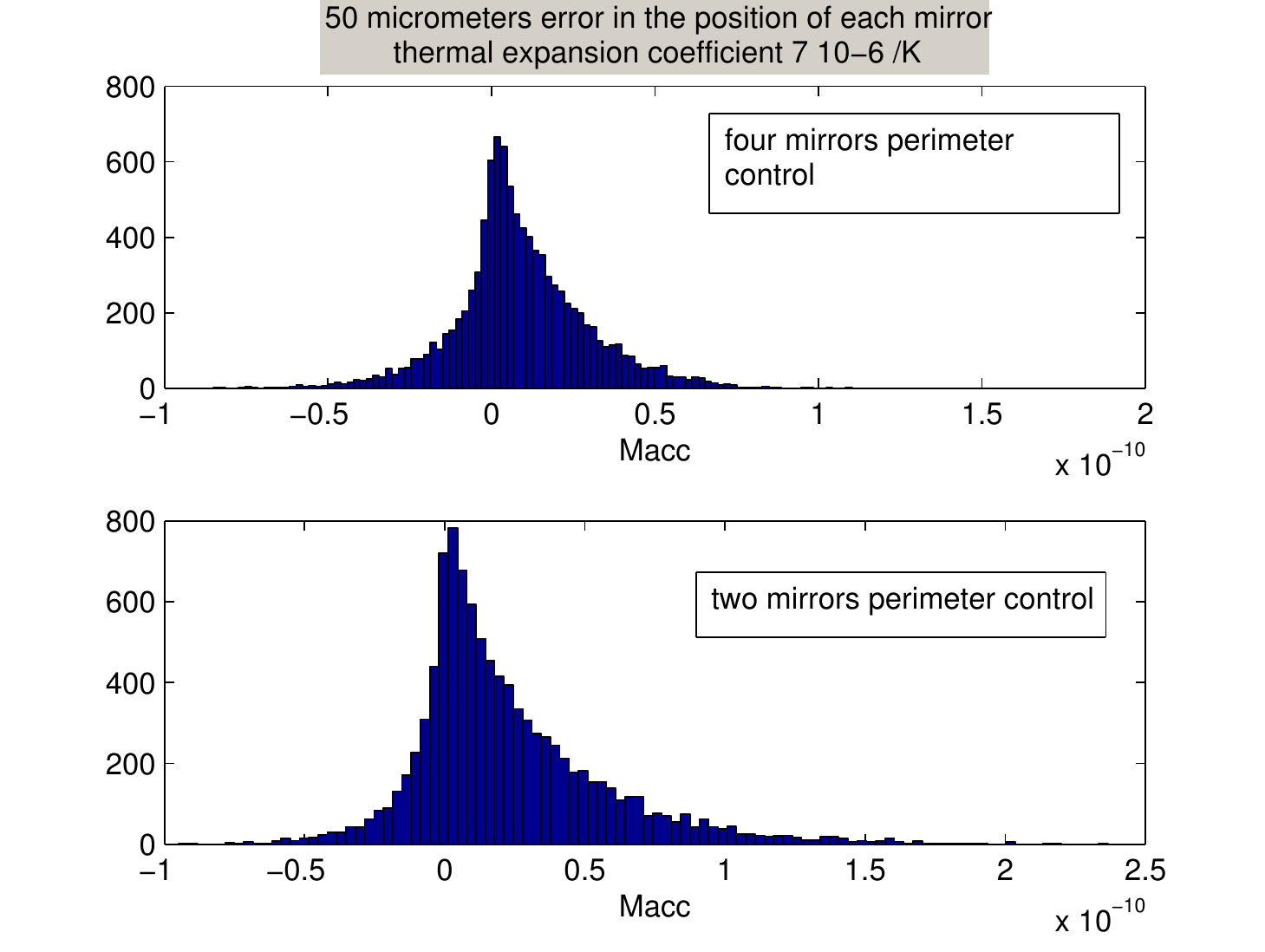}
 \caption{Histogram of $M_{acc}$ when the position of $3$ mirrors are within $\pm 50$ $\mu m$ close to the ideal position, 10000 points have been evaluated by randomly extracting the position error ($\pm 50\mu m$). The thermal expansion coefficient is $7\times 10^{-7} K^{-1}$, thermal excursion $1$ $K$, top histogram shows the case with $4$ mirrors control of the perimeter, bottom curve with $2$ mirrors control. }
 \label{precision}
 \end{figure}

\begin{figure}
 \includegraphics[scale=1]{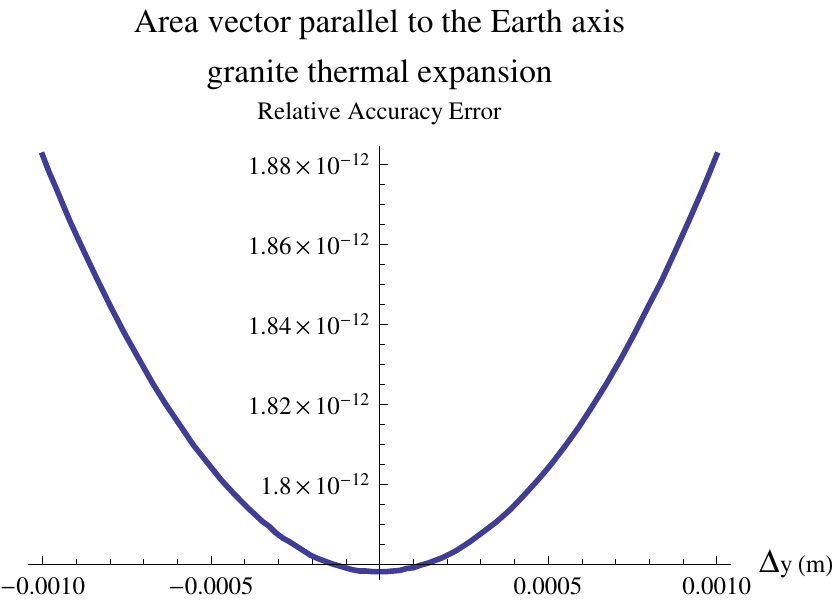}
 \caption{Accuracy change in percentage for vertical misalignments and area vector close to the parallel alignment to the Earth rotational axis. The ring geometry is not perfect in the plane, there is a misalignment of $100\mu m$, a maximum temperature change of $ 1 $ $K$, and the thermal expansion coefficient is $7\times 10^{-6}m/m /K$}
 \label{orientatiopp}
 \end{figure}

\begin{figure}
 \includegraphics[scale=1]{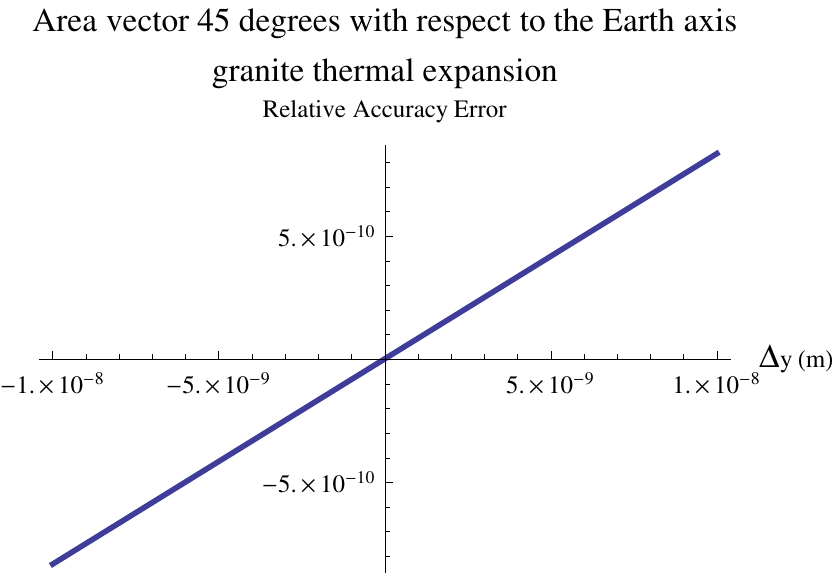}
 \caption{Accuracy change in percentage for vertical misalignments and area vector close to $45$ degrees with respect to the Earth rotational axis. The ring geometry is not perfect in the plane, there is a misalignment of $100\mu m$, a maximum temperature change of $ 1 $ $K$, and the thermal expansion coefficient is $7\times 10^{-6}m/m /K$}
 \label{orientationpp}
 \end{figure}

The best performing ring, so far, is G which is a four mirrors ring. This is one of the reasons why the present scheme
uses a square ring geometry. In principle a triangular ring, with $3$ mirrors could be preferable since the three mirrors
 are always inside a plane, and  the losses will be minimized as well, reducing the number of mirrors. It could be advantageous in principle,
 but a triangular ring is less sensitive. For instance, let us compare the performance of two rings inscribed in a circle
 of radius $r$; for a regular polygon with different number of sides $N$,
  the area is $A = N\, \frac{r^2}{2}\, \sin(\frac{2 \pi}{N})$  and the perimeter is
   $P = 2  N\, r\, \sin(\frac{\pi}{N})$; it is straight forward to demonstrate
    that the triangular ring has $0.7$ times the signal than the square one,
    which is equivalent to say that the triangular ring  needs $2$ times more
    time to reach the same level of accuracy as the square one.\\
 The ring-laser response is proportional to the Scale Factor "S". For a perfect
  square ring this proportionality factor is equivalent to $N$ the number of wavelength
   inside the ring: when the length of the ring changes, because of a change in the temperature,
   the laser changes its wavelength in order to keep $N$ constant. This is true as long as the
   perimeter change is below a wavelength, $632$ $nm$ in our case, and in this conditions the
   gain factor of the instrument guarantees a very high accuracy of the measurement.
   For example: if the laboratory has $\delta T = 1^o$ degree temperature excursion, the ring perimeter is $36$ $m$,
   in order to guarantee the operation of the ring-laser with a fixed number of wavelengths $N$, it is
    necessary to realize the whole apparatus using materials  with temperature expansion coefficient of the order of $10^{-8}$ $K^{-1}$.  This is the concept used for G in Wettzell:  a structure realised with material  as Zerodur, with a design which can be defined monolithic, i.e. relative motions of the mirrors are not allowed. G has a very high stability, but is rather expensive, and not very flexible with regards to changing the mirrors and align the laser cavity. Moreover the extension of this design to a large array of rings seems rather difficult. Later on, a more flexible and less expensive design has been realized, called GEOSENSOR, which so far has been employed especially for smaller size rings. This design allows a very good relative alignment of the mirrors, it is relatively easy to change mirrors and tools to move each mirrors along different degrees of freedom have been implemented. So far this kind of instruments have been done in steel.  
Fig.~\ref{gpisa} shows a drawing of G-Pisa, our prototype.
  The optical cavity vacuum chamber has a stainless steel modular structure: 4 towers, located at the corners of the square and containing the mirrors holders inside, are connected by pipes, in order to form a ring vacuum chamber with a total volume of about $5\cdot 10^{-3}$~m$^3$. The mirrors are rigidly fixed to the tower. The cavity alignment can be adjusted by moving the towers with respect to the slab through a lever system that allows $2$ degrees of freedom of movements. No window delimits the active region and the vacuum chamber is entirely filled with a mixture of He and a $50\%$ isotopic mixture of $^{20}$Ne and $^{22}$Ne. The total pressure of the gas mixture is set to 560 Pa with a partial pressure of Neon of $20~Pa$. The active region is a plasma produced in a capillary pyrex tube inserted at the middle of one of the ring sides by a radio frequency capacitively coupled discharge. In a non monolithic device, temperature changes could interrupt the continuous operation, and the perimeter is actively controlled by acting on the mirrors and using as reference a stabilized laser; very highly stabilised lasers are commercially available, for instance wavelength stabilization at the level of $2.5\times 10^{11}$ using iodine line can be obtained. G-Pisa is kept in continuous laser operation trough a perimeter stabilization servo system which acts along  the diagonal direction, for two opposite placed mirrors, through piezoelectric actuators \cite{AppPhys} . \\
  The GEOSENSOR design has other advantages as well: the mirrors are under vacuum and are not affected by the outside pressure changes, they can be very easily aligned and the cost is pretty much reduced compared with the monolithic design. The experience of G-Pisa has shown so far that it can work with different orientations. In fact G-Pisa has worked both horizontally and vertically oriented. It is in steel, inside the thermally stabilized room in the central area of Virgo, in order to improve thermal stability, it has been mounted on top of a granite table (thermal expansion coefficient about $5\times 10^{-6}$ $m/m K$).\\
  To find the guidelines of the mechanical project, we have used a simple program which consists in considering the ring as four points (the light spot on the mirrors) which can be moved from the ideal position, both inside the plane or outside the plane. The model takes into account thermal expansion and the perimeter is kept constant by acting diagonally on pairs of mirrors; the use of $2$ mirrors or $4$ mirrors for the feedback correction have been investigated; the thermal excursion is considered of $1^{o}$ degree. The scale factor $S$ in presence of misalignments is compared with $S_{0}$ ( scale factor  at the optimal configuration); this comparison is expressed as $M_{acc}=\frac{S_{0}-S}{S_{0}}$, which gives the accuracy limit induced by misalignments. The required level of accuracy of $1$ part of $10^{10}$ is $M_{acc}=10^{-10}$.\\ 
Fig.~\ref{mmer} shows $M_{acc}$ for a rectangular ring, with sides $6$ $m$ and $6.6$ $m$, in function of a misalignment of one of the four mirrors.
  \\ Fig.~\ref{mmer}  clearly shows that the Gain Factor changes a lot with small change of mirrors positions. The situation strongly improves by considering a perfect square ring. In fact, for a closed figure with a fixed number of sides, the area over perimeter ratio has a maximum when the polygon is regular one, as for example a 'perfect' square ring. Fig.~\ref{twofour} and \ref{twofour1} show $M_{acc}$ with $100$ $\mu m$ construction precision and two possible choices of the thermal expansion coefficient.
For instance, let us assume that each mirror position is in the ideal position within a quantity $\delta$ which depends on  the precision of the construction. Fig.~\ref{precision} shows $M_{acc}$ when $3$ out of the $4$ mirrors are positioned with an error, $10000$ points have been evaluated pseudo-randomly distributed between $\pm 50\mu m$ along each coordinate.
 In summary, if the thermal excursion is $1\ K$, the position of the mirrors is an ideal square within $\pm 50 \mu m$, the support has a thermal expansion coefficient below $7\times 10^{-7} K^{-1}$, $M_{acc}$ remains in the range necessary for the needed accuracy using four or two mirrors active control of the perimeter.
\\ Misalignments  which bring the light spots outside the plane of the ring do  not have appreciable effect on the gain factor, but they change the orientation of area vector $\vect{u}_{\alpha}$; in this case the effect for $M_{acc}$ depends on the relative angle between  the ring and the Earth rotational axis. Figs.~\ref{orientatiopp} and \ref{orientationpp} shows how the accuracy changes for two different ring orientations: parallel to the axis of the Earth and at $45^{o}$ degrees respectively. The first is almost insensitive, while the other is sensitive to nano-metric misalignments.

\begin{figure}
 \includegraphics[scale=1]{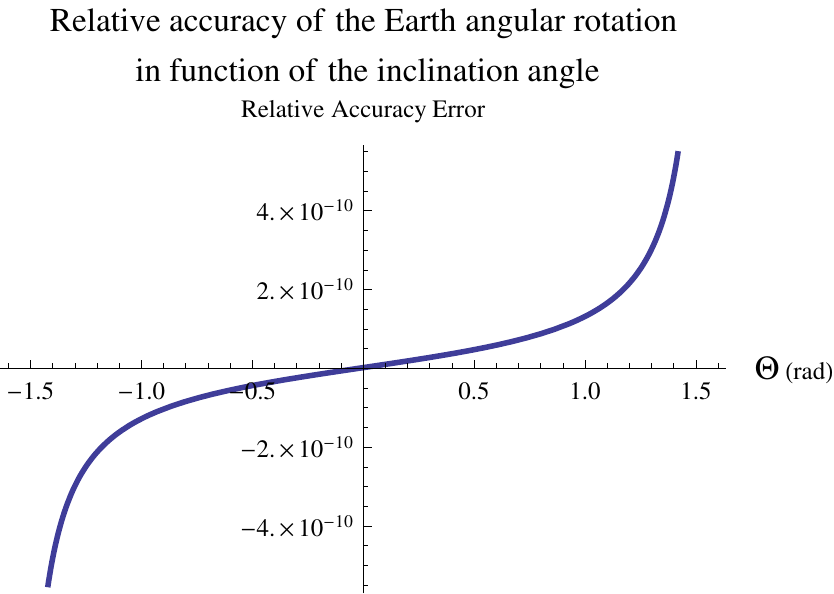}
 \caption{Relative limit of the accuracy in the measurement of the Earth angular rotation induced by a $nm$ change in the position of one of the mirrors with respect to its original position, in function of angle with the Earth rotation axis. The area vector of the ring lays in the meridian plane. The accuracy loss is zero when the Earth axis and the area vector are parallel, and is very high in the orthogonal alignment }
 \label{angle}
 \end{figure}

 Fig.~\ref{angle} shows $M_{acc}$ for a $nm$ vertical misalignment of one of the rings in function of the angle with respect to the Earth rotational axis.
\\In summary: The Gain Factor of each ring can be kept constant at the level of $1$ part in $10^{10}$, if the positions of the mirrors are constructed and kept within $+/- 50$ $\mu m$ error close to the ideal square ring; the relative position between mirrors can be rigidly constrained with granite, super-invar or similar low thermal expansion coefficient spacers, it is preferable to use all the four mirrors for the perimeter active stabilization, but two mirrors control could be acceptable as well if the structure has thermal coefficient better than granite. It is necessary to constantly monitor the relative angle between rings with $n rad$ precision (only the relative alignment matters). This can be accomplished looking at the modal structure of the FP cavities formed along the diagonals.\\ Moreover the orientation of each ring with respect the Earth rotation axis should be such to avoid alignment too sensitive to the relative angle (relative angle with the Earth rotation axis below $60^{o}$).
 Using the Earth angular velocity rotation, which is perfectly stable for few days, the whole apparatus can be calibrated at the beginning; the relative angle, or the area of each ring could be not perfectly planar or exactly $90^{0}$, but it is important to monitor the geometry of the structure during the whole measurement time (years). The mirror holders plays an important role, it can be advantageous to build them in Zerodur or similar material, in order to avoid displacements out of the plane. The mirrors holder should be designed  in order to provide the tools to align the cavities; in principle each mirrors should have $5$ degrees of freedom: $three$ translations and two tilts, the rotation around the axis orthogonal to the mirror itself does not play a role; but since the mirrors are spherical only three motions are fundamental: we may have one translation along the diagonal and two mirrors tilts or three translations.
\\Let us consider now an octahedral geometry, containing the three rings.

\begin{figure}
 \includegraphics[scale=.4]{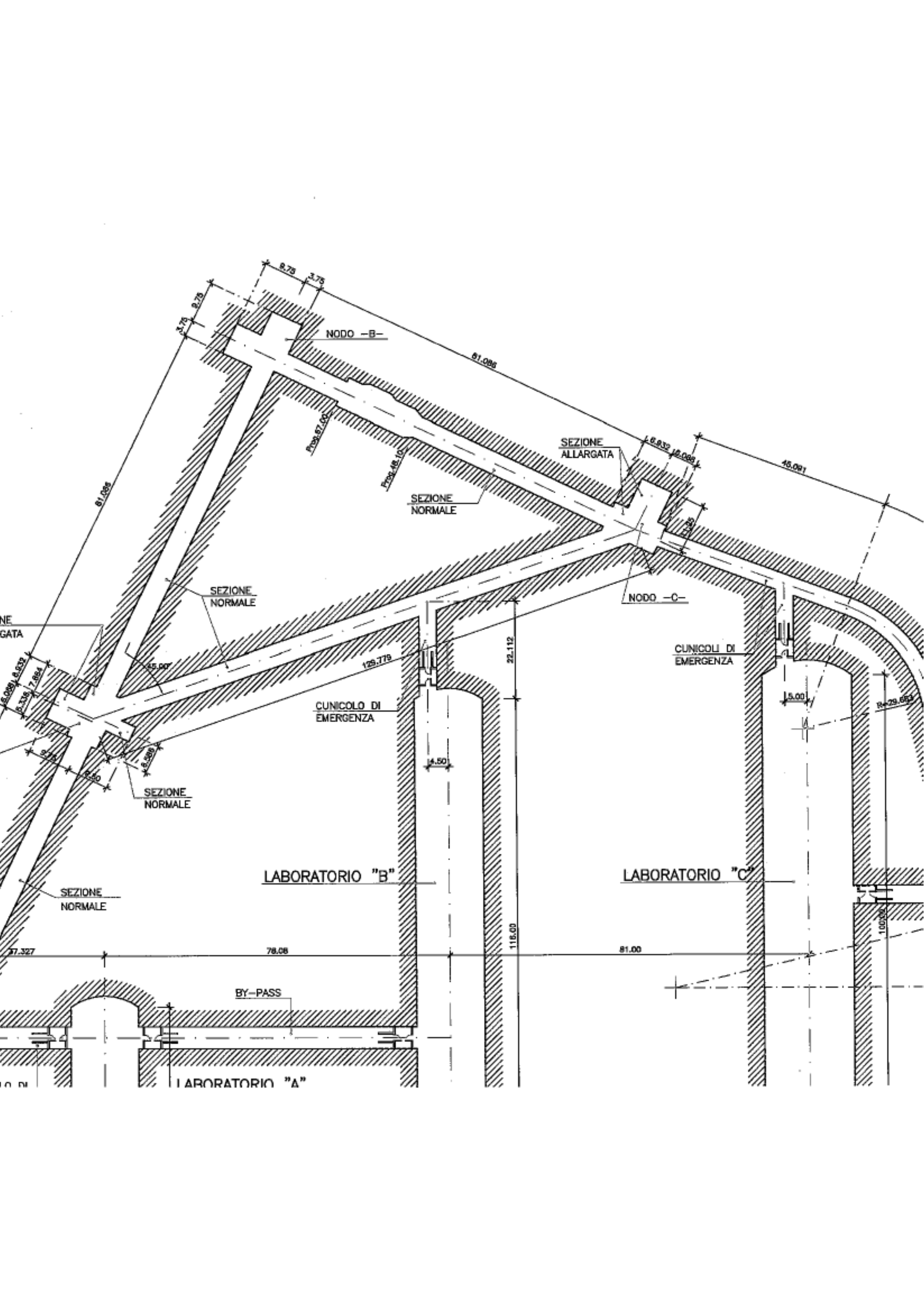}
 \caption{Plan of LNGS laboratory close to node B}
 \label{nodi}
 \end{figure}

\begin{figure}
 \includegraphics[scale=.15]{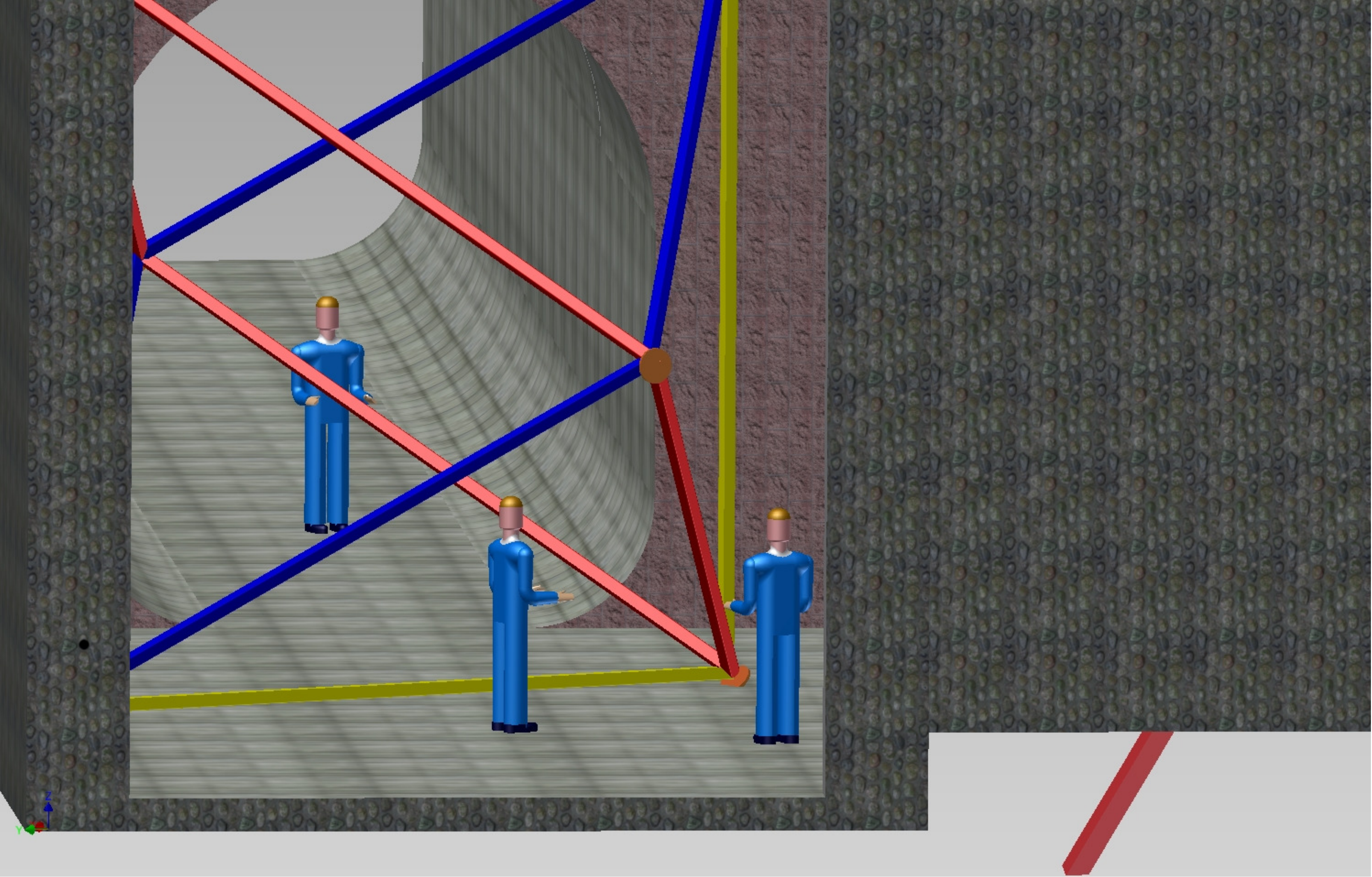}
 \caption{The ring laser system inside node B of LNGS, side view showing that passage between the two entrances}
 \label{man}
 \end{figure}

Fig.~\ref{angle} shows that the relative angle between the different rings must be monitored at the level of nrad, and that it  should be avoided to put one of the rings with an angle larger than $60^{0}$ with respect the Earth rotation axis.
 We have done the exercise to fit the octahedron, with rings of $24$ $m$ perimeter, inside the node B of LNGS, considering that this node is $8$ $m$ tall, and imposing the constrains discussed in Fig.~\ref{angle}. 
The exercise is done with the octahedron since it needs more space. Considering that the latitude of LNGS is $42^o$ $27''N$  two configurations are given: to have the octahedron straight up ($8.4$ $m$ tall) or laying on one side, one ring is respectively horizontally or vertically oriented and the other two symmetrically positioned with respect to the meridian plane. Let us consider the maximum size $9$ $m$: $8.48$ $m$, the diagonal of the octahedron, plus $0.6$ $m$ necessary to hold mirrors and optics in general necessary for the read out. 
This octahedron can be contained inside each of the big halls of LNGS, in both orientation, but inside node B, which is the most isolated room of LNGS, see Figs.~\ref{nodi} and \ref{man}, the only possibility is the shorter configuration, with the longer side parallel to the floor.

\begin{figure}
 \includegraphics[scale=0.2]{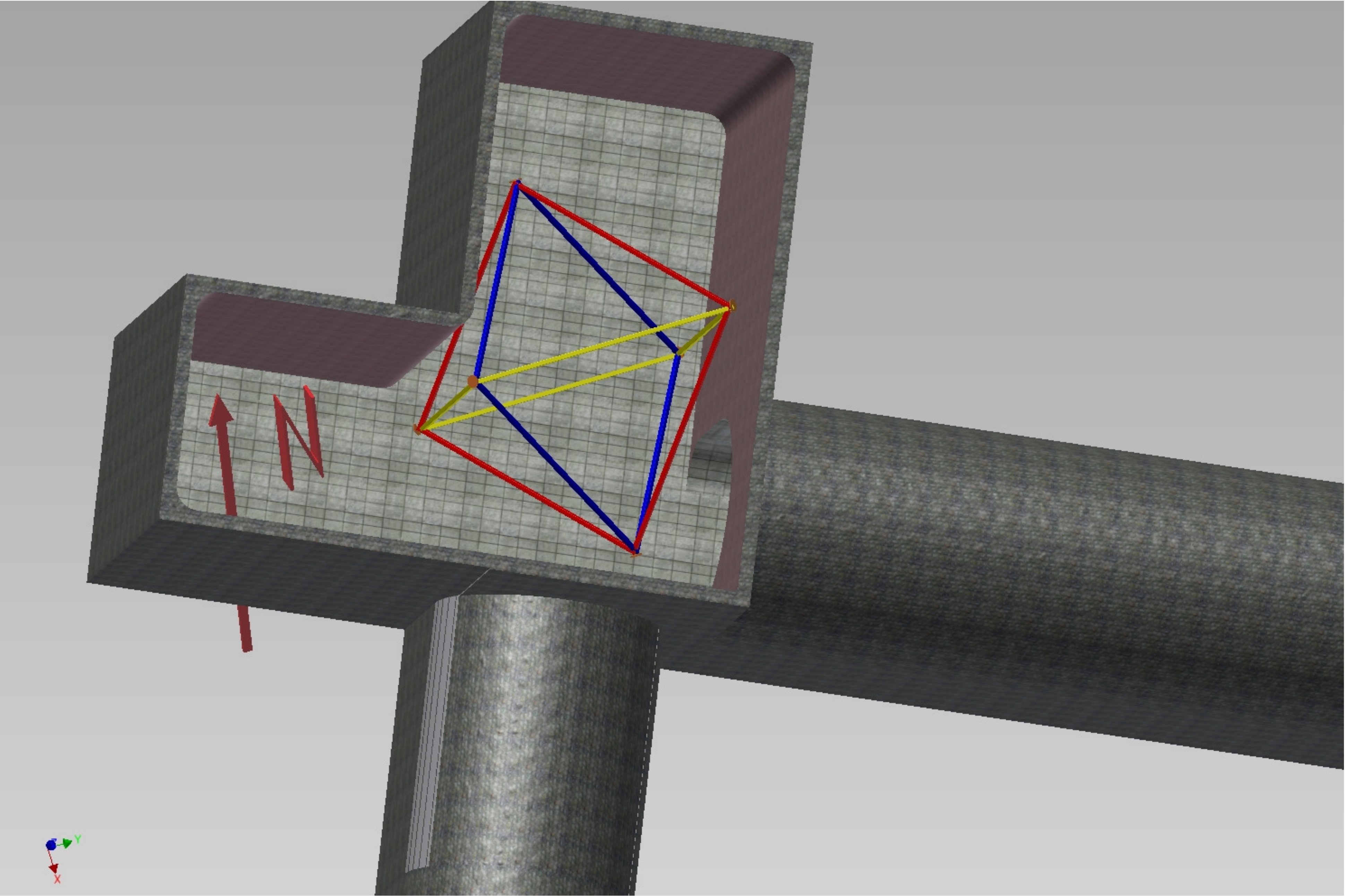}
 \caption{the octahedron inside node B  }
 \label{octahedron}
 \end{figure}

So, the octahedron, with rings of $24$ $m$ perimeter, can be contained inside node B, where the ceiling is $8$ $m$ tall, while for node C the structure should be scaled, probably no more than $20$ $m$ perimeter can be contained inside node C, since the ceiling there is $6$ $m$ tall. Fig.~\ref{octahedron} shows the octahedron inside node B.


\section{Diagnostics of dihedral angles and scale factors}

To reduce the influence of systematics in long--term measurements,
the control of the geometrical stability of ring laser system is of 
paramount importance. In particular, it is crucial to monitor the deviations 
from planarity of each ring laser and their mutual orientations.

\begin{figure}[!h]
\centerline{\includegraphics[width=7.5cm]{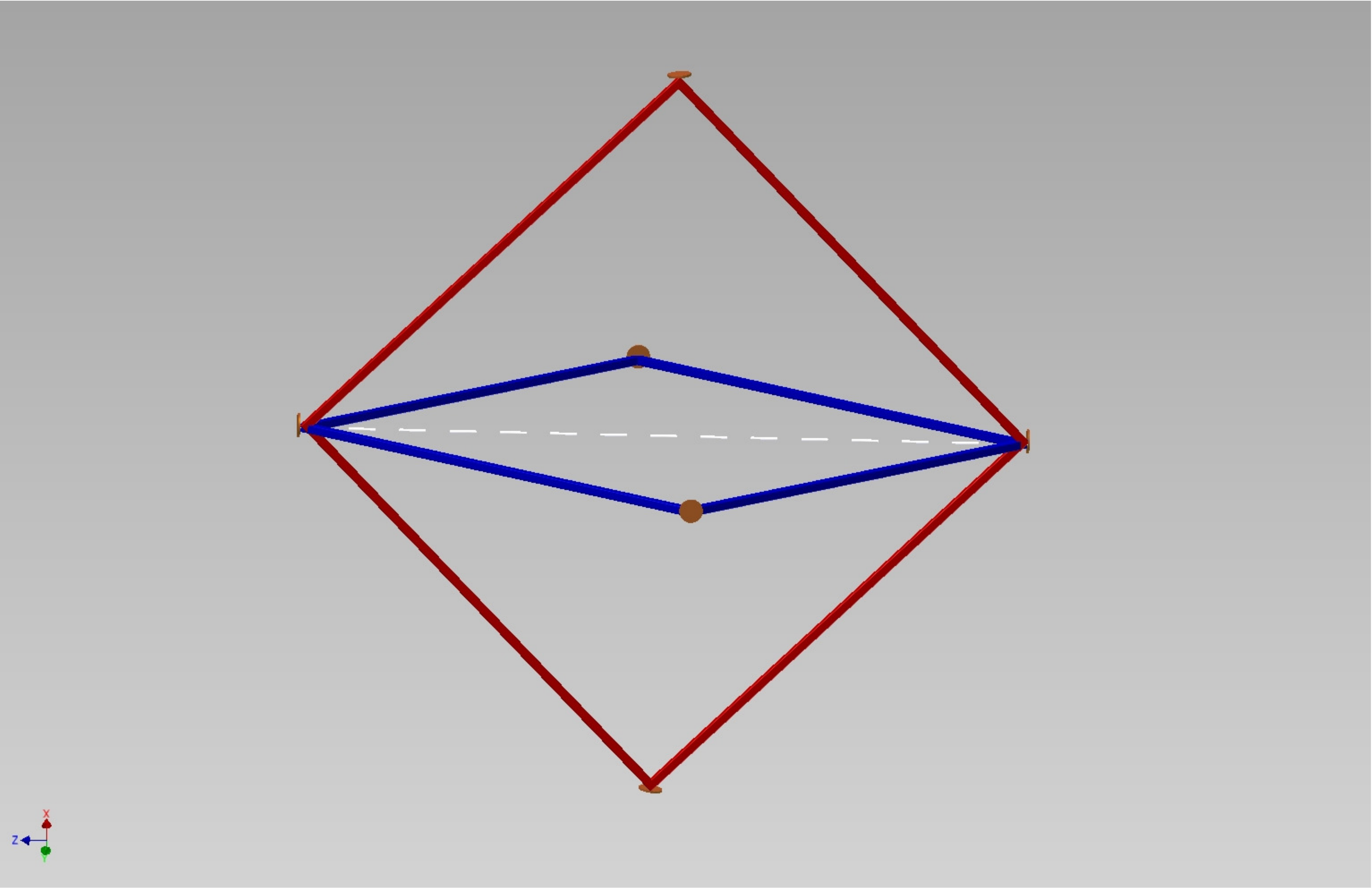}}
\caption{In a square ring configuration passive Fabry--Per\`{o}t cavities
are formed along the square diagonals (dashed line in the sketch). In the
case of an octahedron each of these three passive cavities is shared by two
rings.}
\label{diagFP}
\end{figure}

A square ring consists of four spherical mirrors with the same curvature
radius $R$, placed at the corners. Square geometry guarantees that opposite
mirrors are parallel so that they form two extra linear Fabry--Per\`{o}t
cavities (see Fig. \ref{diagFP}). As a consequence, each square ring is made
of three optical resonators: the ring itself and two linear ones oriented
along the diagonals. These latter can be used to monitor the
geometrical stability of the whole ring system. Deviations from a square
geometry result in tilting and/or displacements of the diagonal vectors,
which in turn change the cavity eigenmodes.

A linear symmetric FP cavity with spherical mirrors in $z_{M}=\pm \frac{1}{2}%
d=\pm \frac{1}{\sqrt{2}}L$ ($L$ being the square ring arm) and centers on
the z-axis, supports the Gaussian modes 
\begin{multline}
E_{\ell ,m}\left( x,y,z\right) =\frac{1}{w_{c}\left( z\right) }H_{\ell
}\left( \frac{\sqrt{2}x}{w_{c}\left( z\right) }\right) H_{m}\left( \frac{%
\sqrt{2}y}{w_{c}\left( z\right) }\right)  \\
\times \exp \left[ -ik\frac{x^{2}+y^{2}}{2q_{c}(z)}-ikz+i\left( \ell
+m+1\right) \arctan \left( \frac{2z}{b}\right) \right] \ ,
\label{Hermite-Gauss}
\end{multline}%
where $q_{c}(z)=z-ib=\left( \frac{1}{R_{c}(z)}-i\ \frac{\lambda }{\pi
w_{c}^{2}(z)}\right) ^{-1}$ and $b= \sqrt{d\left( 2R-d\right) }$ are the complex curvature of the Gaussian beam and the confocal parameter, respectively; 
here the curvature radius $R_{c}\left( z\right) $  and  the spot-size 
$w_{c}\left( z\right) $ read  
\begin{align*}
R_{c}\left( z\right) & =\frac{d^{2}-2dR-4z^{2}}{4z} \\
w_{c}^{2}\left( z\right) & =\frac{\lambda }{\pi }\frac{4z^{2}+2dR-d^{2}}{2%
\sqrt{d\left( 2R-d\right) }} \ .
\end{align*}%
The eigenmodes $E_{\ell ,m}\left( x,y,z\right) $ form a complete set which
can be used for representing a generic field confined between the two
generally misaligned mirrors of the cavity 
\begin{equation*}
E\left( x,y,z\right) =\sum_{\ell ,m}C_{\ell ,m}E_{\ell ,m}\left(
x,y,z\right) \ ,  
\end{equation*}%
where
\begin{equation*}
C_{\ell m}=\int dx\int dyE_{in}\left( x,y\right) E_{\ell ,m}\left(
x,y,z_{M}\right) 
\end{equation*}%
and $E_{in}\left( x,y\right) $ is the beam illuminating the input mirror $M_{1}$. %
If we suppose the mirror tilted by $\Theta _{x}$ and $\Theta _{y}$ and displaced by $X$ and 
$Y$ with respect to cavity axis $\hat{z}$, we have
\begin{equation*}
E_{in}\left(x,y\right) \propto e^{  -ik\left[ \frac{(x-X)^{2}}{%
2q_{c}\left( z_{M}\right) }+\Theta _{x}x+\frac{(y-Y)^{2}}{2q_{c}\left(
z_{M}\right) }+\Theta _{y}y\right] } \ .
\end{equation*}%

As an example, the relative intensities $\left\vert C_{\ell
m}\right\vert ^{2}/\left\vert C_{00}\right\vert ^{2}$ for the first modes 
$\ell +m=0,1,2$ and $X=Y=0$ are reported in Tab. \ref{ClmTable}. 
\begin{table}[!tp] \centering%
\begin{tabular}{|c|c|c|c|}
\hline
$\left\vert C_{\ell m}\right\vert ^{2}/\left\vert C_{00}\right\vert ^{2}$ & $%
\ell =0$ & $\ell =1$ & $\ell =2$ \\ \hline
$m=0$ & $1$ & $5.16\times 10^{7}\Theta _{x}^{2}$ & $1.33\times 10^{15}\Theta
_{x}^{4}$ \\ \hline
$m=1$ & $5.16\times 10^{7}\Theta _{y}^{2}$ & $2.66\times 10^{15}\Theta
_{x}^{2}\Theta _{y}^{2}$ &  \\ \hline
$m=2$ & $1.33\times 10^{15}\Theta _{y}^{4}$ &  &  \\ \hline
\end{tabular}%
\caption{Power coupled to the first cavity higher modes ($\ell + m = 0, 1,
2$) as a fraction of the external laser power for $q_{c}=q_{x}=q_{y}$ and
$X=Y=0$. The value are obtained for a ratio between the cavity length and
the mirror radius of curvature of 1.5}\label{ClmTable}%
\end{table}

It is clear that the cavity axes misalignment can be detected by
looking at the intensity pattern of the beam transmitted though the output
mirror $M_{2}$. A modal decomposition of such a pattern gives a suitable set
of coefficients $\left\vert C_{\ell m}\right\vert ^{2}$ which can be used
for estimating the position and angular misalignment of the cavity with
respect to the reference beam $E_{in}\left( x,y\right)$.

Supposing that at the begining ($t=0$) the cavity external laser is 
perfectly aligned to a symmetric cavity (if the two mirror show 
equal transmittivity then the cavity transmission is $1$) so that 
all the incoming power $P_{in}$ is coupled to the TEM00 mode. 
The measurement procedure we have devised is a
tunable laser, showing a linewidth narrower than the cavity linewidth, tuned
over a cavity \textit{FSR} in a time interval $\Delta t$ so that each mode
is spanned in a time $\tau =\frac{\Delta t}{\mathcal{F}}$, where $\mathcal{F}$
is the cavity finesse. The number of photons in the $\ell m$\ mode are 
given by (we are now assuming a rectangular line shape instead
of a Lorentzian profile)%
\begin{equation*}
n_{\ell m}=k\frac{P_{in}}{hv}\tau \left\vert C_{\ell m}\right\vert ^{2}\ .
\end{equation*}%
This number of photons must be higher than the noise equivalent number of
photons hitting the detector in the same time interval. The noise equivalent
power in $W/\sqrt{Hz}$, is connected to the equivalent number of photons by%
\begin{equation*}
n_{NEP}=\frac{NEP}{h\nu }\frac{\sqrt{B}}{\eta }\tau \ ,
\end{equation*}%
where $\eta $ and $B$ are the quantum efficiency and the detection bandwidth, respectively.

To overcome the photon noise, we have to satisfy the inequality 
\begin{equation*}
\frac{n_{NEP}}{n_{\ell m}}=\frac{n_{NEP}}{n_{00}}\frac{\left\vert
C_{00}\right\vert ^{2}}{\left\vert C_{\ell m}\right\vert ^{2}}<1
\end{equation*}%
In particular, looking at $C_{01}$ coefficient we have  
\begin{equation*}
n_{00}>\frac{1}{5.16\times 10^{7}\Theta ^{2}}\frac{NEP}{h\nu }\frac{\sqrt{B}%
}{\eta }\tau \ ;
\end{equation*}%
further, by assuming $n_{00}\simeq \frac{P_{in}}{hv}\tau $ we obtain%
\begin{equation*}
P_{in}>\frac{NEP\sqrt{B}}{5.16\times 10^{7}\Theta ^{2}\eta } \ . 
\end{equation*}%
For typical silicon detectors $NEP\sim 10^{-14}W/\sqrt{Hz}$, $B\sim 10^{6}Hz$%
, $\eta \sim 0.9$, so that 
\begin{equation*}
P_{in}>2.2\times 10^{-19}\Theta ^{-2} \ .
\end{equation*}%
For a tilt sensitivity of $\Theta \sim 10^{-9}$ the power required at the input
is%
\begin{equation*}
P_{in}>220\;mW \ .
\end{equation*}

\section{More about the importance of this measurement}
\subsection{Post Newtonian Parameters}
 The proposed experimental apparatus
is well suited for performing optical test of metric theory of gravitation.
We start from the statement that the vector $\vect{\Omega}^\prime$ should be  entirely
contained in the meridian plane if the preferred frames effect, determined by $\vect{W}$ (see Eq. (A9)),
can be neglected. Indeed the currently available best
estimates \cite{Will} suggest that this effect is about 2 orders
of magnitude smaller than the geodetic and
Lense-Thirring contributions. As a consequence, we expect that the measured
components of $\vect{\Omega}^\prime$ outside the meridian plane should be
compatible with noise. In this case, our results could be used to obtain new
constraints, independent from the available ones, on the
preferred frames parameters. In addition, we can write the PPN parameters $\alpha_1$
and $\gamma$ as a function of the $\vect{u}_r$ and $\vect{u}_\theta$ components of
$\vect{\Omega}^\prime$
\begin{eqnarray}
\alpha_1 = \left(-4 \widehat{\Omega^\prime_\theta } \csc \theta -8 \widehat{\Omega^\prime_r}\sec \theta\right] \nonumber \\
\gamma-1 = \left(\widehat{\Omega^\prime_\theta} \csc \theta - \widehat{\Omega^\prime_r}\sec \theta/2\right ) -2 \ ,
\label{eqppn}
\end{eqnarray}
where $\widehat{\Omega^\prime_{r,\theta}} \equiv \Omega^\prime_{r,\theta}/w$ and $w$  is the very
precisely measured constant $w\equiv 2 \pi GM/(c^2 R T_S) $
$\simeq 5.0747798 \times 10^{-14}\ rad/sec$; here we have used $GM=3.986004418\times 10^{14}\ m^3/s^2$,
$R=6.378137\times 10^6 \ m$ and  $T_S=86164.0989 \ s$.  Assuming one year of data taking with the same
ring laser parameters used for the simulations in Sect. IIID we have that
the standard deviation of $\widehat{\Omega^\prime_\theta}$ and $\widehat{\Omega^\prime_r }$ is
 $\widehat{\sigma}_{\Omega_{r,\theta}} \simeq 0.03$, and therefore upper
limits of some interest can be put on $\alpha_1$ and $\gamma$ at the Gran Sasso colatitude $\theta\simeq \pi/4$.



 \subsection{Interdisciplinary: Geodesy and Geophysics}
\label{geophys}
 Earth rotation rate and the orientation of the rotational axis of the Earth in space are the linking quantities between
the terrestrial (ITRF) and the celestial (ICRF) reference frames. Currently a set of quasars, forming an external set of markers,
 provide the only way of determining the rotational velocity and the variations of the orientation of the rotational axis of the Earth
with sufficient accuracy. As already mentioned,  $10~\mu s$ for the measurement of length of day (LOD) and $0.1~mas$ for the pole position
are routinely achieved by a network of VLBI radio telescopes as one of the services (IERS) of the International Association of Geodesy (IAG).
 The operation of such a network requires expensive equipment and a lot of maintenance effort.
Huge amounts of data are recorded in each measurement session, which require physical transport over large distances for the
correlation in the analysis centers. Data latency and the fact that there is no continuous measurement coverage
are suggesting the investigation of alternative methods for the precise estimation of Earth rotation. Furthermore it is desirable to
develop an independent measurement technique, in order to identify intra-technique biases if they exist.
 Ring lasers are possible candidates for such an alternative measurement technique. They measure the earth rotation
locally and  within much shorter time intervals. Such gyros are widely used in aircraft navigation
and can measure rotations absolute, i.e. independent of an external reference frame. Therefore also
local contributions to earth rotation  are contained in the measurements. The effects of earth tides, strain,
 crust deformation, seismic events, polar motion are contained in the  ring laser measurements due to their
 contribution to earth rotation or  due to variations in the orientation of the respective ring laser. However, the demands on
such instruments are extremely high and cannot be met by existing commercial devices. They can be summarized as:
\begin{itemize}
 \item sensitivity to rotation $0.01~prad/s$ at about 1 hour of integration
 \item sensor stability of 1 part in $10^{10}$ over several month to years (Chandler Wobble)
 \item resolution in sensor orientation $\approx 1~nrad$. This corresponds to polar motion of around $1 cm$ at the pole.
 \end{itemize}
This means that a reasonable improvement in sensor sensitivity and stability is still required in order
 to make ring lasers viable tools to be applied to space geodesy. The design of the G ring laser is
one way of approaching these demands and it not too far away from reaching this goal \cite{schre}.
Operating several such ring laser gyroscopes in geophysical independent  regions simultaneously offers a
unique possibility to distinguish global from local (monumentation related) signal contributions through their
independent data streams.

\section*{\label{con} Discussions and Conclusions }
\index{discussions and conclusions}
The feasibility of the experiment for the measurement of
relativistic effective rotation rates appears to rest only on a tri-axial dynamical sensor
of local rotation of enough sensitivity. Despite the fact that large ring lasers as G are very
stable platforms and with the provision of tight feedback systems to stabilize the scale factor (cold cavity, as well
as the active cavity), currently ring laser gyroscopes are not able
to determine the DC part of the Earth rotation rate with a sensitivity compatible with the requirements
for detection of the Lense-Thirring effect. While the contribution of the varying Earth rotation itself
presumably can be removed with sufficient accuracy from the C04 series of VLBI measurements, there
remains the problem of determining the actual null-shift offsets from the laser functions in the
ring laser gyroscope. Since the gravito-magnetic effect is small and constant, a good discrimination
against laser biases, such as for example `Fresnel drag' inside the laser cavity must be achieved.
Therefore it will be advantageous to add one or several ring laser cavities in addition to the triad
structure for sufficient redundancy. We also intend to operate at least the G ring laser structure
in parallel to the here proposed structure in order to discriminate local perturbation signals from
regional and global ones. A second large ring laser located at the Cashmere facility in Christchurch,
New Zealand, will be used in the data analysis process, provided it can be run with sufficient resolution
and stability.

\appendix

\section{\label{A1} Ring Laser Measurements in the Laboratory Frame}
\index{Ring Laser Measurements in the Laboratory Frame}
In this Appendix we evaluate the response to the gravitational field of a ring laser in an Earth bound laboratory and, to know the space-time metric in the laboratory frame we shall use the construction of the ``proper reference frame'' as described in Ref. \cite{ciufoliniwheeler,MTW}.


As we discussed in Section III, a ring laser converts a time difference into a frequency difference (see e.g. Eq. (\ref{freque})).  It is possible to show that (see e.g. \cite{kajari}) in a stationary metric in the form\footnote{Greek and Latin indices
denote space-time and spatial components, respectively;
letters in boldface indicate spatial vectors, while letters in italic indicate four-vectors and four-tensors;
summation and differentiation conventions are assumed. In this Appendix,  if not otherwise stated, we use units such that $G=c=1$.} $g_{\mu\nu}=g_{\mu\nu}(x^{i})$ an observer at rest at $x^{i}=x^{i}_{0}$ measures  the proper-time difference $\delta \tau=\tau_{+}-\tau_{-}$ between the right handed beam propagation time ($\tau_{+}$) and the left handed one ($\tau_{-}$):
\beq
\delta \tau=-2 \sqrt{g_{00}(x^{i}_{0})}  \oint_{\mathcal S} \frac{g_{0i}}{g_{00}}  ds^{i}=-2 \sqrt{g_{00}(x^{i}_{0})}   \oint_{ S} \bm H \cdot d\bm s, \label{eq:deltataulocal1}
\eeq
where $\mathcal{S}$ is the spatial trajectory of the beams, whose tangent vector is $d\mathbf s$, and we set  $H_{i}=\frac{g_{0i}}{g_{00}}.$

In order to evaluate the proper-time difference (\ref{eq:deltataulocal1}), we need to know the  space-time metric in our laboratory, that is to say the gravitational field nearby the world-line of the observer which performs measurements with the ring laser. To this end, we consider an observer in arbitrary motion in a given
background space-time, and write the corresponding local metric in a neighborhood of its world-line
(see e.g. \cite{MTW})
\begin{eqnarray}
g_{(0)(0)}&=&1+2 \bm{\mathcal{A}} \cdot \bm x+O(x^{2}), \label{eq:G00a} \\
g_{(0)(i)}&=&\Omega_{(i)(k)}x^{(k)}+O(x^{2}),  \label{eq:G0i} \\
g_{(i)(j)}&=&\eta_{(i)(j)}+O(x^{2}). \label{eq:Gij}
\end{eqnarray}
It is worth pointing out that the Eqs. (\ref{eq:G00a})-(\ref{eq:Gij}) hold only near the world-line of the observer,
where quadratic displacements terms are negligible.
Here we suppose that the observer carries an orthonormal tetrad (parentheses refer to tetrad indices)
$ e_{(\alpha)}$, whose four-vector $ e_{(0)}$  coincides with his four-velocity $\mathcal{U}$, while
the four-vectors $ e_{(i)}$ define the basis of the spatial vectors in the tangent space along its
world-line. By construction we have $e_{(\alpha)} e_{(\beta)}=\eta_{(\alpha)(\beta)}$, where $\eta_{(\alpha)(\beta)}$ is the Minkowski tensor. The metric
components (\ref{eq:G00a})-(\ref{eq:Gij}) are expressed in  coordinates that are associated to the
given tetrad, namely the space coordinates $x^{(i)}$ and the observer's proper time $x^{(0)}$.
In the above equations, $\bm{\mathcal{A}}$ is the spatial projection
of the observer's four-acceleration, while  the tensor $\Omega_{(i)(k)}$ is related to the parallel
transport of the basis four-vectors along the observer's world-line: $ \nabla_{\mathcal{U}}
e_{(\alpha)}=- e_{(\beta)}\Omega^{(\beta)}_{\ (\alpha)}$. In particular, if $\Omega_{(i)(j)}$ were zero,
the tetrad would be Fermi-Walker transported. Let us remark that the metric (\ref{eq:G00a})-(\ref{eq:Gij}) is
Minkowskian along the observer's world-line ($x^{(i)}=0$); it is everywhere flat iff $\bm{\mathcal{A}}=0$,
i.e. the observer is in geodesic motion and the tetrad is non rotating (i.e. it does not rotate with
respect to an inertial-guidance gyroscope). In the latter case, the first corrections to the flat
space-time metric are $O(x^{2})$ \cite{MTW}.


In order to explicitly write the local metric, which through its  gravito-magnetic ($g_{0i}$)  and gravito-electric ($g_{00}$) components enables us to evaluate the proper-time difference (\ref{eq:deltataulocal1}),  we must choose a suitable tetrad by taking into account the motion of the Earth-bound laboratory in the background space-time metric. To this end, we consider the following PPN background metric which describes the gravitational field of the
rotating Earth (see e.g. \cite{Will}):
\begin{eqnarray}
&ds^{2}&=(1-2U(R))dT^{2}-\left(1+2\gamma U(R)\right)\delta_{ij}dX^{i}dX^{j}+ \nonumber \\
&2&\!\!\!\!\!\!\left[\frac{\left(1+\gamma+\alpha_{1}/4\right)}{R^{3}}{\left(\bm J_{\oplus}
\wedge \bm R\right)_{i}}-\alpha_{1}U(R) W_{i}\right]dX^{i}dT , \nonumber \\
&\ &
\label{eq:metricappn}
\end{eqnarray}
where $-U(R)$ is the Newtonian potential, $\bm J_{\oplus}$ is the angular momentum of the Earth,
$W_{i}$ is the velocity of the reference frame in which the Earth is at rest with respect to mean
rest-frame  of the Universe; $\gamma$ and $\alpha_{1}$ are post-Newtonian parameters that measure,
respectively, the effect of spatial curvature and the effect of preferred frames.
The background metric (\ref{eq:metricappn}) is referred to an Earth Fixed Inertial (ECI) frame,
where Cartesian geocentric coordinates are used, such that $\bm R$ is the position vector
and $R \doteq \sqrt{\sum_{i} X^{2}_{i}}=\sqrt{X^{2}+Y^{2}+Z^{2}}$. Then, we choose a laboratory
tetrad which is related to the background coordinate basis of (\ref{eq:metricappn}) by a pure Lorentz
boost,
together with a re-normalization of the basis vectors: in other words the local laboratory  axes have the same orientations as those in the background ECI frame, and they could be physically realized by three orthonormal telescopes, always pointing toward the same distant stars.

In this case,  one can  show that the
gravito-magnetic contribution in the local metric reads \cite{MTW,ciufoliniwheeler,tourrenc,ashby90}
$\Omega_{(i)(k)}x^{(k)}=-\left(\bm{\Omega}^\prime \wedge \bm x \right)_{(i)}$, where
the total relativistic contribution $\bm  \Omega'$ is the sum of four terms, with the  dimensions of angular rotation rates
\beq
\bm \Omega' = \bm \Omega_{G}+ \bm \Omega_{B}+  \bm \Omega_{W} +\bm\Omega_{T} \label{eq:omegaprime}
\eeq
defined by
\begin{eqnarray}\bm \Omega_{G}&=&
-\left( 1+\gamma \right) \bm \nabla U(R) \wedge \bm V, \label{eq:OmegaDS} \\
\bm \Omega_{B}&=&-\frac{1+\gamma +\alpha_{1}/4}{2} \left(\frac{\bm J_{\oplus}}{R^{3}}-\frac{3 \bm
 J_{\oplus} \cdot \bm R}{R^{5}}\bm R \right),\label{eq:OmegaLT} \\
\bm \Omega_{W} & = & \alpha_{1}\frac{{1}}{4} \bm \nabla U(R) \wedge \bm W, \label{eq:Omegaw} \\
\bm \Omega_{T}&=&-\frac{1}{2} \bm V \wedge  \frac{d \bm V}{dT}. \label{eq:OmegaTh}
\end{eqnarray}
The vector $\bm  \Omega'$ represents the precession rate that  an inertial-guidance gyroscope, co-moving with the laboratory, would have with respect to the  \textit{ideal} laboratory  spatial axes (see e.g. \cite{MTW,ciufoliniwheeler}) which are always oriented as those of the ECI frame; if the spin vector of the gyroscope is $\bm S$, its precession is hence defined by
\beq
\frac{d\bm S}{dt}= \bm \Omega' \wedge \bm S \label{precapp}
\eeq
Differently speaking, we may say that the local spatial basis vectors  are not Fermi-Walker transported along the laboratory world-line.
 In particular the total precession rate is made of four contributions:
 i) the geodetic or de Sitter precession $\bm \Omega_{G}$ is due to the motion of the laboratory
in the curved space-time around the Earth; ii) the Lense-Thirring precession $\bm\Omega_{B}$ is due to
the angular momentum of the Earth; iii)  $\bm \Omega_{W}$ is due to the preferred frames effect;
and iv) the Thomas precession $\bm \Omega_{T}$ is  related to the angular defect due to the Lorentz boost.

It is worth noticing that  for a laboratory  bounded to the Earth
\beq
\bm{\mathcal A}\simeq\frac{d\bm V}{dT}-\bm{\nabla}U(R), \label{eq:Alab}
\eeq
and the
 acceleration $\bm{\mathcal A}$ can not be eliminated.
However, for a geodetic motion (e.g. a free fall satellite) $\bm{\mathcal A}\equiv 0$ and
by substituting in Eqs. (\ref{eq:OmegaDS}) and (\ref{eq:OmegaTh}), we obtain
\beq
\bm \Omega_{G}= -\left( \frac 1 2 +\gamma \right) \bm\nabla U(R) \wedge \bm V,
\label{eq:omegageo22}
\eeq
and
\beq
\bm \Omega_{T}=\frac 1 2 \bm{\mathcal A} \wedge \bm V =0.
\label{eq:omegath22}
\eeq
Eq. (\ref{eq:omegageo22}) gives the geodetic precession for a free fall gyroscope, while Thomas precession (\ref{eq:omegath22}) is zero.

All  terms in (\ref{eq:OmegaDS})-(\ref{eq:OmegaTh}) must be evaluated along the laboratory world-line (hence, they are constant in the local frame),
whose position and velocity in the background frame are $\bm R$ and $\bm V$, respectively.
However, if we consider an \textit{actual} laboratory fixed on the Earth surface, the spatial axes of the
corresponding tetrad rotate with respect to the coordinate basis of the metric (\ref{eq:metricappn}),
and we must take into account in the gravito-magnetic term (\ref{eq:G0i})
the contribution of the additional rotation vector $\bm \Omega_{\oplus} $, which corresponds to the Earth
rotation rate, as measured in the local frame\footnote{For an Earth-bounded laboratory, it is
$\bm \Omega_{\oplus} \simeq \left[1+U(R)+\frac{1}{2}\Omega^{2}_{0} R  \sin^{2} \vartheta \right]\bm \Omega_{0}$ 
where $R$ is the terrestrial radius,  $\vartheta$ is the colatitude angle of the laboratory and $\bm \Omega_{0}$ is the terrestrial rotation rate, as measured in an asymptotically flat inertial frame.}.

As a consequence,  it is possible to show that, up to linear displacements from the world-line, the relevant local gravito-magnetic potential turns out to be
\beq
g_{(0)(i)}= \left(\bm{\Omega } \wedge \bm x \right)_{(i)},
\label{eq:Omegaik2}
\eeq
where $\bm \Omega=-\bm \Omega_{\oplus}-\bm \Omega'$, while the gravito-electric $g_{(0)(0)}$ one remains the same.

Now, we are able to evaluate the proper-time difference
\beq
\delta \tau=-2 \sqrt{g_{00}(x^{i}_{0})}   \oint_{ S} \bm H \cdot d\bm s. \label{eq:deltataulocal11}
\eeq
Without loss of generality, we suppose that the observer is at rest in the origin of the coordinates, so that, according to (\ref{eq:G00a}), $g_{00}(x^{i}_{0}) = 1$. As a consequence, we have
\beq
\delta \tau=-2  \oint_{ S}  \frac{\left(\bm \Omega  \wedge \bm x\right)}{\left(1+2  \bm{\mathcal{A}} \cdot \bm x \right)} \cdot d\bm s. \label{eq:deltataulocal2}
\eeq
Now, on taking into account the expression of acceleration of the laboratory frame (\ref{eq:Alab}) and evaluating the magnitude of the various terms, the leading contribution to (\ref{eq:deltataulocal2}) con be written, applying Stokes theorem
\beq
\delta \tau = -2 \int_{A} \left[{\bm \nabla \wedge  \left(\bm \Omega \wedge \bm x \right) }{} \right] \cdot d\mathbf A, \label{eq:deltataulocal3}
\eeq
where   $\mathbf A=A\vect{u}_n$ is the area enclosed by the beams and oriented according to its normal vector $\vect{u}_n$. On evaluating the curl, taking into account that $\bm \Omega$ is constant, we eventually obtain
\beq
\delta \tau = -4 \int_{A} \bm \Omega \cdot d\mathbf A= -4 \bm \Omega \cdot \mathbf A. \label{eq:deltataulocal4}
\eeq
On substituting $\bm \Omega=-\bm \Omega_{\oplus}-\bm \Omega'$ in (\ref{eq:deltataulocal4}), we see that the proper-time delay can be written in the form
\beq
\delta \tau=4 \bm \Omega_{\oplus} \cdot \mathbf A+ 4\bm \Omega' \cdot \mathbf A,\label{eq:ringlocal1}
\eeq
where $4 \bm \Omega_{\oplus} \cdot \mathbf A$ is the purely kinematic Sagnac term, due to the rotation of the Earth, while $ 4\bm \Omega' \cdot \mathbf A$ is the gravitational correction due to the contributions (\ref{eq:OmegaDS})-(\ref{eq:OmegaTh}).

According to Section \ref{ssec:RLresponse}, from Eq. (\ref{eq:deltataulocal4}), it is then possible to write the ring laser equation in the form
\begin{equation}
\delta f=\frac{4A}{\lambda P}\ \mathbf{u}_{n}\cdot \boldsymbol{\Omega }.
  \label{eq:ringlaserapp}
\end{equation}

To further clarify Eqs. (\ref{eq:OmegaDS})-(\ref{eq:OmegaTh}) it is useful to use an
orthonormal spherical basis $\bm{u}_{r}, \bm{u}_{\vartheta}, \bm{u}_{\varphi}$ in the ECI frame, such
that the $\vartheta=\pi/2$ plane coincides with the equatorial plane. As a consequence, the position
vector of the laboratory with respect to the center of the Earth is
 $\bm{R}=R \bm{u}_{r}$ and the kinematic constraint $\bm V= \bm\Omega_{\oplus} \wedge \bm R$ holds,
i.e. $\bm V= \Omega_{\oplus} R \sin \theta  \bm u_{\varphi}$.

\begin{widetext}
Thus, the components of $\bm \Omega'$ in physical units read
\begin{eqnarray}
\bm \Omega_{G}&=& -\left( 1+\gamma \right)\frac{GM}{c^{2}R} \sin \vartheta   \Omega_{\oplus} \bm{u}_{\vartheta},
\label{eq:OmegaDS10} \\
\bm \Omega_{B}&=&-\frac{1+\gamma +\alpha_{1}/4}{2} \frac{G}{c^{2}R^{3}}\left[\bm J_{\oplus}-3 \left(\bm J_{\oplus} \cdot \bm{u}_{r} \right) \bm{u}_{r} \right],
\label{eq:OmegaLT10} \\
\bm \Omega_{W} & = & -\frac{\alpha_{1}}{4} \frac{GM}{c^{2}R^{2}} \bm u_{r}   \wedge \bm W, \label{eq:Omegaw10} \\
\bm \Omega_{T}&=&-\frac{1}{2c^{2}} \Omega^{2}_{\oplus} R^{2} \sin^{2} \vartheta \bm{\Omega}_{\oplus},
\label{eq:OmegaTh10}
\end{eqnarray}
\end{widetext}
Moreover, we assume
the general relativistic values of the PPN parameters, $\gamma=1, \ \alpha_1=0$, and use for
the Newtonian potential of the Earth its monopole approximation, i.e. $U(R)=GM/R$.
Thus, the components  (\ref{eq:OmegaDS10})-(\ref{eq:OmegaLT10}) read
\begin{eqnarray}
\bm \Omega_{G}&=& -2\frac{GM}{c^{2}R} \sin \vartheta   \Omega_{\oplus} \bm{u}_{\vartheta},
\label{eq:OmegaDS1} \\
\bm \Omega_{B}&=&-\frac{G}{c^{2}R^{3}}\left[\bm J_{\oplus}-3 \left(\bm J_{\oplus} \cdot \bm{u}_{r} \right) \bm{u}_{r} \right],
\label{eq:OmegaLT1} \\
\bm \Omega_{W} & = & 0, \label{eq:Omegaw1} \\
\bm \Omega_{T}&=&-\frac{1}{2c^{2}} \Omega^{2}_{\oplus} R^{2} \sin^{2} \vartheta \bm{\Omega}_{\oplus},
\label{eq:OmegaTh1}
\end{eqnarray}

and, to leading order,  the total rotation rate which enters the Eq. (\ref{eq:ringlocal1}) is
\begin{eqnarray}
\bm{\Omega}&=& -\bm{\Omega}_{\oplus} +2\frac{GM}{c^{2}R} \sin \vartheta \Omega_{\oplus} \bm{u}_{\theta}+
\frac{G}{c^{2}R^{3}}\left[\bm J_{\oplus}-3 \left(\bm J_{\oplus} \cdot \bm{u}_{r} \right) \bm{u}_{r} \right] \nonumber \\
& & \label{eq:Omegalocal1}
\end{eqnarray}


\begin{widetext}
If we denote by $\alpha$ the angle between the radial direction $\bm u_{r}$ and the normal vector $\bm u_{n}$, on setting $\bm u_{n}=\cos \alpha \bm u_{r}+ \sin \alpha \bm u_{\theta}$ in (\ref{eq:ringlocal1}), and using (\ref{eq:Omegalocal1}), we may express the proper-time delay in the form
\beq
\delta \tau=\frac{4A}{c^{2}}\left[  \Omega_{\oplus} \cos \left(\theta+\alpha \right)
-2\frac{GM}{c^{2}R}\Omega_{\oplus}\sin \theta \sin \alpha +\frac{GI_{\oplus}}{c^{2}R^{3}}\Omega_{\oplus}  \left(2 \cos \theta \cos \alpha+\sin\theta \sin \alpha \right)  \right] \label{eq:delayexp1}
\eeq
where we have written $\bm J_{\oplus}=I_{\oplus} \bm \Omega_{\oplus}$, in term of the $I_{\oplus}$,  the moment of inertia of the Earth.
\end{widetext}

\section{\label{A2} Probability Distribution of Quadratic Forms}
\index{Probability Distribution of Quadratic Forms}
The statistics of quadratic forms of Gaussian random vectors  $\vect{x}$
are well known in the literature. In particular, if $\vect{x}$ is a multivariate Gaussian
random vector with mean $\vect{s}$ and covariance matrix $\vect{\Sigma}$,
the mean and the variance of a quadratic form $Q=\vect{x}^T\vect{Q}\vect{x}$ are given by
\begin{eqnarray}
<Q>&\equiv& <\vect{x}^T \vect{Q} \vect{x}> = Tr(\vect{Q}\vect{\Sigma}) + \vect{s}^T \vect{Q} \vect{s} \nonumber \\
\sigma^2_Q&\equiv &<(\vect{x}^T \vect{Q} \vect{x})^2>-<Q>^2 \nonumber \\
&=& 2 Tr(\vect{Q}\vect{\Sigma} \vect{Q}\vect{\Sigma} )
+ 4  \vect{s}^T \vect{Q} \vect{\Sigma} \vect{Q} \vect{s}
\label{eqA1}
\end{eqnarray}
where $\vect{Q}$ is a square symmetric matrix, $\ ^T$
and  $Tr$ are the transpose and trace operators, respectively. The statistics of $Q$
in general is not known, unless $\vect{Q}\vect{\Sigma}$ is an idempotent matrix \cite{QuadForm}.
In the case were $\vect{x}$ represents the response of ring lasers in a regular
polyhedral configuration $\vect{Q}= \vect{I} $, with no common noise source
and the same sensitivity $\vect{\Sigma}=\sigma^2 \vect{I}$, where $\vect{I}$ is
the identity matrix, the above formulas greatly simplifies
\begin{eqnarray}
<Q> &=& M \sigma^2 + E \\
\sigma^2_Q &=&2 M \sigma^4 + 4 E \sigma^2 \ ,
\end{eqnarray}
where $E=\vect{s}^t\vect{s}=||\vect{s}||^2$ is the signal energy. In this case also the statistics
of $Q$ readily follows. In fact, starting from its definition we have
\begin{equation}
P_Q(Q) \equiv  \int P(\vect{x}) \delta(Q-\vect{x}^T \vect{x} ) \: \de\vect{x}
\end{equation}
where $P(\vect{x}) = \exp[(\vect{x-s})^T(\vect{x-s})/(2\sigma^2)]/(2 \pi\sigma^2)^{M/2}$ is the
Gaussian probability density of one sample of the random vector $\vect{x}$. We can use
the integral representation of the Dirac's $\delta$-function
\begin{equation}
\delta(Q-\vect{x}^T \vect{x}) = \int_{-\infty}^{+\infty} e^{i \omega (Q- \vect{x}^T \vect{x})}
\: \de \omega
\end{equation}
and write
\begin{eqnarray}
&P_Q(Q)& =  \int_{-\infty}^{+\infty} \:  \de \omega e^{ i \omega Q} \frac{1}{(2 \pi\sigma^2)^{M/2}} \nonumber \\
&\ & \int  \exp\left[i \omega \: \vect{x}^T \vect{x} - \frac{1}{2\sigma^2}(\vect{x-s})^T(\vect{x-s})\right] \: \de\vect{x} \nonumber
\end{eqnarray}
By re-arranging the exponent, the last integral can be recast as a M-dimensional Gaussian integral
and calculated explicitly
\begin{widetext}
\begin{equation}
\frac{1}{(2 \pi\sigma^2)^{M/2}} \int  \exp\left[\left(i \omega -\frac{1}{2\sigma^2}\right) \vect{x}^T \vect{x} + \frac{1}{\sigma^2}\vect{s}^T\vect{x}  - \frac{E}{\sigma^2}  \right] \: \de\vect{x} =
\frac{\exp[i \omega \sigma^2 E  /(1- 2 i \omega\sigma^2) ]}{(1- 2 i \omega\sigma^2)^{M/2} }
\end{equation}
\end{widetext}
where in the last expression one recognizes the moment generating functions of a non-central
$\chi^2$ distributions with $M$ degrees of freedom and non-centrality parameter $E$.
The probability density function of $Q$ can be found using the tables of Fourier Transform pairs
\begin{eqnarray}
&P_Q(Q)& = \int_{-\infty}^{+\infty} \de \omega \: e^{ i \omega Q} \left\{\frac{\exp[i \omega \sigma^2 E/(1- 2 i \omega\sigma^2) ]}{(1- 2 i \omega\sigma^2)^{M/2} } \right\}\nonumber \\
&=&\!\!\!\! \frac{1}{2} \exp[-(Q+E)/(2 \sigma^2)] \left(\frac{Q}{E} \right)^{\frac{M-2}{4}} I_{M/2-1}(\sqrt{Q E}/\sigma^2)) \nonumber
\end{eqnarray}
where $I_k(x)$ are the modified Bessel functions of order $k$.

\section{\label{A3} Probability Distribution of Projectors}
\index{Probability Distribution of Projectors}

The norm of complementary projection operators $\vect{P}$ and $\vect{Q}$
acting on Gaussian random vectors $\vect{x}$ are described by
remarkably simple statistics.
In fact, starting from the definition of $E_P=||\vect{P}\vect{x}||^2$ and $E_Q=||\vect{Q}\vect{x}||^2$
we have that the joint probability density $P(E_P,E_Q)$ reads
\begin{equation}
        P(E_P,E_Q) = \int P(\vect{x}) \delta(E_P-\fquad{\vect{x}}{\vect{P}})
        \delta(E_Q-\fquad{\vect{x}}{\vect{Q}})\: \de\vect{x}
\end{equation}
where $P(\vect{x})$ is the probability density of one sample of the random vector
$\vect{x}$. The two Dirac $\delta$-functions can be written using their Fourier transforms,
\begin{equation}
        P(E_P,E_Q) = \int P(\vect{x}) e^{\left[ u E_P + v E_Q - \vect{x}^T
        (u\vect{P}+v\vect{Q}) \vect{x} \right]}\: \de u\:\de v\:\de \vect{x}
\end{equation}
where the integrals in $\de u$ and $\de v$ are performed along the imaginary axis (i.e. $u=i \omega_1$
and $v=i \omega_2$ are purely imaginary complex numbers).
Now assume the noise is Gaussian distributed, uncorrelated between different detectors and
with identical variance $\sigma^2$ in every detector, namely
\begin{equation}
        P(\vect{x}) = \frac{1}{(2\pi\sigma^2)^{\frac{M}{2}}} \exp\left(-\frac{1}{2\sigma^2}
        (\vect{x}-\vect{s})^T(\vect{x}-\vect{s})\right)
\end{equation}
where $\vect{s}\equiv (\vect{\Omega}\cdot \vect{u}_1, \dots \vect{\Omega}\cdot \vect{u}_{_M})$
is the rotation signal in vectorial form.
Then,
\begin{eqnarray}
        P(E_P,E_Q) &=& \left(\frac{\alpha}{\pi}\right)^{\frac{M}{2}} \int
        \exp\left[-\alpha (\vect{x}-\vect{s})^T(\vect{x}-\vect{s})\right]\times \nonumber\\
        &\times&\: \exp\left[-\vect{x}^T(u\vect{P}+v\vect{Q}) \vect{x} \right]\: \de\vect{x}\cdot \nonumber\\
        &\times&\: e^{u E_P} e^{v E_Q}\: \de u\:\de v,
\end{eqnarray}
where $\alpha \defin 1/2\sigma_\Omega^2$. Writing $\vect{x}$ as $\vect{s}+\vect{\varepsilon}$ and switching
the integration variable to $\vect{\varepsilon}$ yields
\begin{widetext}
\begin{eqnarray}
        P(E_P,E_Q) &=& \left(\frac{\alpha}{\pi}\right)^{\frac{M}{2}} \times \\
        &\times& \int \exp\left[-\vect{n}^T(\alpha\vect{I} + u\vect{P}+v\vect{Q})\vect{n}
        -2\vect{n}^T(u\vect{P}+v\vect{Q})\vect{s}\right] \de\vect{\varepsilon}
        \exp\left[ -\vect{s}^T(u\vect{P}+v\vect{Q})\vect{s} \right]
        e^{uE_P} e^{vE_Q} \ \de u\:\de v.\nonumber
        \label{jointprefinal}
\end{eqnarray}
\end{widetext}
The integration in $\de\vect{n}$ can be done by noting that
it is a standard $M$-dimensional Gaussian integral with the linear term,
and in general, for any $M\times M$ symmetric matrix $\vect{A}$ and $M$-vector $\vect{b}$,
\begin{equation}
        \int \exp\left( -\vect{n}^T \vect{A} \:\vect{n} + \vect{b}^T\:\vect{n}\right) \de\vect{n}
        = \frac{\pi^{M/2}}{\sqrt{\det(\vect{A})}} \exp\left(\frac{\fquad{\vect{b}}{\vect{A}^{-1}}}{4}\right).
        \label{gaussint}
\end{equation}
In our case,
\begin{eqnarray}
        \vect{A} &=& \alpha\vect{I} + u\vect{P}+v\vect{Q} \nonumber\\
        \vect{b} &=& 2(u\vect{P}+v\vect{Q})\vect{s}.
\end{eqnarray}
Now we exploit the properties of $\vect{P}$ and $\vect{Q}$.
Using their complementarity, we can write
\begin{equation}
        \vect{A} = (\alpha+u)\vect{P}+(\alpha+v)\vect{Q}
\end{equation}
and from the fact that they are orthogonal and idempotent we also have
\begin{equation}
        \vect{A}^{-1} = (\alpha+u)^{-1}\vect{P}+(\alpha+v)^{-1}\vect{Q},
\end{equation}
hence
\begin{equation}
        \fquad{\vect{b}}{\vect{A}^{-1}} = 4\left(\frac{u^2}{\alpha+u}\fquad{\vect{s}}{\vect{P}}
        + \frac{v^2}{\alpha+v}\fquad{\vect{s}}{\vect{Q}}\right).
        \label{gaussint1}
\end{equation}
Furthermore, as $\vect{P}$ and $\vect{Q}$ are projection matrices, their eigenvalues are
$\{0,1\}$ with multiplicities respectively $\{M-2,2\}$ for $\vect{P}$ and $\{2,M-2\}$
for $\vect{Q}$. Then, writing $\vect{A}$ in diagonal form is trivial and leads to
\begin{equation}
        \det(\vect{A}) = (\alpha+u)^2(\alpha+v)^{M-2},
        \label{gaussint2}
\end{equation}
determinants being independent from the basis.
By using \ref{gaussint1} and \ref{gaussint2} in \ref{gaussint} one can see that the
Gaussian integral splits into the product of factors involving either $u$ or $v$.
By further substituting in \ref{jointprefinal}, the remaining integrals separate and the
probability density remarkably factorizes as
\begin{equation}
    P(E_P,E_Q) = P(E_P)\:P(E_Q)
\end{equation}
with
\begin{eqnarray}
        P(E_P) &=& \int \frac{1}{1+2\sigma^2u} \exp\left( \frac{-s_p\:u}{1+2\sigma^2u} \right) e^{u E_P}\: \de u \nonumber\\
        P(E_Q) &=& \int \left(\frac{1}{1+2\sigma^2v}\right)^{\frac{M}{2}-1} \exp\left( \frac{-s_q\:v}{1+2\sigma^2v} \right) e^{v E_Q}\: \de v \nonumber\\
\end{eqnarray}
and $s_p \defin \fquad{\vect{s}}{\vect{P}}$, $s_q \defin \fquad{\vect{s}}{\vect{Q}}$.
The transformed functions is the moment generating functions of two non-central
$\chi^2$ distributions, with 2 and $M-2$ degrees of freedom respectively, and whose
non-centrality parameters are $\fquad{\vect{s}}{\vect{P}}$ and
$\fquad{\vect{s}}{\vect{Q}}$ respectively. Thus,
\begin{eqnarray}
        P(E_P) &=& \frac{1}{2}\: \exp\left(-\frac{E_P+s_p}{2\sigma^2}\right)\: I_0\left(\frac{\sqrt{E_P s_p}}{\sigma^2}\right)\nonumber\\
        P(E_Q) &=& \frac{1}{2}\: \exp\left(-\frac{E_Q+s_q}{2\sigma^2}\right)\: \left(\frac{E_Q}{s_q}\right)^{\frac{M}{4}-1}\: I_{\frac{M}{2}-2}\left(\frac{\sqrt{E_Q s_q}}{\sigma^2}\right) \nonumber \\
\end{eqnarray}
where $I_n(x)$ is the modified Bessel function of the first kind.
Some interesting conclusions can be drawn about the virtual channels $E_P$ and $E_Q$,
which make them interesting for the identification of meridian plane and the estimate
of $\vect{\Omega}$.
\begin{enumerate}
\item $E_P$ is distributed as a non-central $\chi^2$ with 2 degrees of freedom
and non-centrality parameter equal to $\fquad{\vect{s}}{\vect{P}}$, i.e.~the
magnitude of the signal projection in the P subspace.
\item $E_Q$ is distributed as a non-central $\chi^2$ with $M-2$ degrees of freedom
and non-centrality parameter equal to $\fquad{\vect{s}}{\vect{Q}}$, i.e.~the
magnitude of the signal projection in the ${\cal Q}$ subspace.
\item $E_P$ and $E_Q$ are \emph{statistically independent} processes.
\item In the limit of high SNR, $E_P$ and $E_Q$ are Gaussian distributed
with means  $<E_P> =\fquad{\vect{s}}{\vect{P}}$, $<E_Q> =\fquad{\vect{s}}{\vect{Q}}$
and variances $\sigma^2_{E_P}= 4 \sigma^2_\Omega \fquad{\vect{s}}{\vect{P}}$,
$\sigma^2_{E_Q}= (M-2)\sigma^2_\Omega \fquad{\vect{s}}{\vect{Q}}$, respectively.
\end{enumerate}


\begin{thebibliography}{99}

\bibitem{einstein}, Will, C.M. Theory and Experiment in Gravitational Physics, Cambridge University Press (1993)
\bibitem{LARES} Ciufolini, I. The 1995-99 measurements of the Lense-Thirring effect using laser-ranged satellites, Class. Quantum Grav. 17 2369 (2000)
\bibitem{GPB1} Everitt, C.W.F. \textit{et al.}. Gravity Probe B: Final Results of a Space Experiment to Test General Relativity, \textit{Phys. Rev. Lett.}, in press (2011).

\bibitem{double} Bourgay M., \textit{et al.}, An increased estimate of the merger rate of double neutron stars from observations of a highly relativistic system,
\textit{ Nature} \textbf{426}, 531-533 (2003)

\bibitem{shap} I.I. Shapiro et al., Phys.Rev. Lett., \textbf{26}, (1971) 1132

\bibitem{defl} Straumann, N., \textit{General Relativity and Relativistic Astrophysics}, Springer-Verlag, Berlin (1991)

\bibitem{MTW} Misner, C.W., Thorne, K.S., Wheeler, J.A., \textit{Gravitation}%
, Freeman, S. Francisco (1973)

\bibitem{LTH} H. Thirring, Phys. Z., \textbf{19}, (1918) 204

\bibitem{ciufo} I. Ciufolini, E. Pavlis, Nature \textbf{431}, 958 (2004); I.
Ciufolini et al, 71-104, Space Sci Rev (2009) 148.

\bibitem{essay} Di Virgilio, A. et al., A laser gyroscope system to detect the Gravito-Magnetic effect on Earth, \textit{Int. J. Mod. Phys. D}, \textbf{19} 2331-2343 (2010)

\bibitem{NZ} Triangular ring laser, $45$ $m$ perimeter, Univ. of Christchurch, NZ

\bibitem{electrons} Hasselbach F and Nicklaus M, Sagnac experiment with
electrons: Observation of the rotational phase shift of electron waves in
vacuum, \textit{Phys. Rev. A}, \textbf{48} 143 (1993)

\bibitem{neutrons} Werner S A, Staudenmann J L and Colella R, Effect of
Earth's Rotation on the Quantum Mechanical Phase of the Neutron, \textit{%
Phys. Rev. Lett.}, \textbf{42} 1103 (1979)

\bibitem{pairs} Zimmermann J E and Mercereau J E, Compton Wavelength of
Superconducting Electrons, \textit{Phys. Rev. Lett.}, \textbf{14} 887 (1965)

\bibitem{calcium} Riehle F, Kirsters Th, Witte A, Helmcke J and Borde\' Ch
J, Optical Ramsey spectroscopy in a rotating frame: Sagnac effect in a
matter-wave interferometer, \textit{Phys. Rev. Lett.}, \textbf{67} 177 (1991)

\bibitem{He3} R.W. Simmonds et al Nature 412 (2001) pag 55 

\bibitem{He4} E. Hoskinsons et al Phys. Rev. B74 (2006) 100509(R)

\bibitem{ZAMO} Raine, D. and Thomas, E., \textit{Black Holes - An Introduction}, World Scientific, Singapore (2009)

\bibitem{GPB} C.W.F. Everitt et al, 53-69, Space Sci Rev (2009) 148.

\bibitem{TRT2000} A. Tartaglia, Clas. Q. Grav., \textbf{17}, (2000)
2381-2384.

\bibitem{TNR2005} A. Tartaglia, A. Nagar, M.L. Ruggiero, Phys. Rev. D,
\textbf{71}, (2005).

\bibitem{geoff} G.E. Stedman, \textit{Ring-laser tests of fundamental
physics and geophysics}, Rep.\ Prog.\ Phys.\ \textbf{60}, 615--688 (1997).

\bibitem{ulli09} K.U. Schreiber, T. Kl\"ugel, A. Velikoseltsev, W.
Schl\"uter, G.E. Stedman and J-P. R. Wells; \textit{The large ring laser G
for continuous Earth rotation monitoring}; Pure and Applied Geophysics,
\textbf{166}, No 8-9, 1485-1498, (2009). 

\bibitem{tides} K. U. Schreiber, G. E. Stedman and T. Kl\"ugel; \textit{%
Earth tide and tilt detection by a ring laser gyroscope}, J.\ Geophys.\
Res.\ \textbf{108} (B)2, (2003) 


\bibitem{polmot} K. U. Schreiber, A. Velikoseltsev,M. Rothacher, T.
Kl\"ugel, G. E. Stedman, D. L. Wiltshire; \textit{Direct measurement of
diurnal polar motion by ring laser gyroscopes} J. Geophys. Res. Vol. {\bf 109}, No.
B6, B06405, (2004) 

\bibitem{Benni} B. Pritsch, K.U. Schreiber, A. Velikoseltsev and J-P. R.
Wells; \textit{Scale factor corrections in large ring lasers}; Applied
Physics Letters, \textbf{91}, No. 6, 061115, (2007), 

\bibitem{rlg-noise} K.U. Schreiber, J-P. R. Wells and G.E. Stedman; \textit{%
Noise processes in large ring lasers}; General Relativity and Gravitation,
\textbf{40}, No. 5, 935-943, (2008) 

\bibitem{IERS} Time series of the daily estimate of Earth rotation vector
can be downloaded from 
\verb|http://data.iers.org/products/176/11165/orig/| \verb|eopc04.62-now|

\bibitem{medicina} See e.g. \verb|http://www.med.ira.inaf.it/index_EN.htm|

\bibitem{matera} See e.g. \verb|http://www.asi.it/en/flash_en/observing/| \verb|space_geodesy_center|

\bibitem{crumo} E. Mantovani, D. Albarello, C. Tarnburelli, M. Viti, Annals
of Geophysics \textbf{38}, 67 (1995); R. Haas, E. Gueguen, H. Scherneck, A.
Nothnagel, and J. Campbell, Earth Planets Space, \textbf{52}, 759-764,
(2000).

\bibitem{diupm } K. U. Schreiber, \textit{et al.} Journal of  Geophysical Research,
\textbf{109}, B06405 (2004).

\bibitem{geoff1} G.E. Stedman, R.B. Hurst and K.U. Schreiber; \textit{On the
potential of large ring lasers}, Optics Communications, \textbf{279}, No. 1,
124-129, (2007). 

\bibitem{scully} Scully, M.O., Zubairy, M.S., Haugan, M.P., Proposed optical
test of metric gravitation theories, \textit{Phys. Rev. A} \textbf{24}, 2009
(1981).

\bibitem{PMwettzell} Schreiber, K.U. \textit{et al.},
\textit{J. Geophys. Res.} \textbf{109}, B06405 (2004).

\bibitem{bere} Brzezinski, A.,  Contribution to the theory of polar motion for an elastic earth with liquid core,
Manuscr. Geod., {\bf 11}, 226-241 (1986).

 \bibitem{geba1} Gebauer, A.; Kroner, C.; Jahr, T.,  The influence of topographic and lithologic features on horizontal deformations, Geophys. J. Int., {\bf 177}, 586-602 (2009).

\bibitem{geba2} Gebauer, A.; Steffen, H.; Kroner, C.; Jahr, T.,  Finite element modelling of atmosphere loading effects on strain, tilt and displacement at multi-sensor stations, Geophys. J. Int., {\bf 181} (3), (2010). 

\bibitem{harr1} Harrison, J.C., Cavity and topographic effects in tilt and strain measurements, JGR, {\bf 81/2}, 319-328 (1976).

\bibitem{harr2}  Harrison, J.C., Herbst, K.,  Thermoelastic strains and tilts revisited. Geophysical Research Letters, {\bf 4/11} (1976).

\bibitem{jent} Jentzsch G.,Liebing M.,Weise A., Deep Boreholes for High Resolution Tilt Recording. Bulletin Inf. Marees Terrestres, {\bf 115},  8498-8506 (1993).

\bibitem{jahr} Jahr, T., G. Jentzsch, A. Weise, Natural and man-made induced hydrological signals, detected by high resolution tilt observations at the Geodynamic Observatory Moxa/Germany, Geodynamics, {\bf 48}, 126-131, (2009). 

\bibitem{kump} Kumpel, H.-J.,  Verformung in der Umgebung von Brunnen. Habilitationsschrift, Univ. Kiel, 198p (1989).

\bibitem{weis}  Weise, A., Jentzsch, G., Kiviniemi, A., Kaariainen, J.,  Comparison of long period tilt measurements: results from two clinometric stations Metsahovi and Lohja. Finland J. Geodyn. {\bf 27}, 237-257 (1999).

\bibitem{schre} Schreiber, K. U., Klugel, T., Velikoseltsev, A., Schlater,
W., Stedman, G. E., and  Wells, J. R.,  The Large Ring Laser G for Continuous Earth Rotation Monitoring. Pure and Applied
Geophysics, {\bf 166}(8), 301 498 (2009). 


\bibitem{ciufoliniwheeler} Ciufolini, I., Wheeler, J.A., Gravitation and
Inertia, Princeton University Press, Princeton (1995).

\bibitem{chow} Chow, W.W., Gea-Banacloche, J., Pedrotti, L.M, Sanders, V.E.,
Schleich, W., Scully, M.O., The ring laser gyro, \textit{Rev. Mod. Phys. }
\textbf{57}, 61 (1985).

\bibitem{Will} Will, C.M., \textit{Living Rev. Relativity} \textbf{9}, 3
(2006), \verb|http://www.livingreviews.org/lrr-2006-3|

\bibitem{tourrenc} Angonin-Willaime, M.C., Ovido, X., Tourrenc, Ph.,
Gravitational perturbations on local experiments in a satellite: the
dragging of inertial frame in the HYPER project, \textit{Class. Quantum Grav.%
} \textbf{36}, 411 (2004).

\bibitem{ashby90} Ashby, N., Shahid-Saless, B., Geodetic precession or
dragging of inertial frames?, \textit{Phys. Rev. D} \textbf{42}, 1118 (1990).

\bibitem{QuadForm} Mathai A.M., Provost S.B., \textit{Quadratic Forms in Random
Variables: Theory and Applications}, Marcel Dekker, New York (1992).


\bibitem{AppPhys}
J. Belfi, N. Beverini, F. Bosi, G. Carelli, A. Di Virgilio, E. Maccioni, A. Ortolan and Fabio Stefani,
  Perimeter actively stabilized ring laser gyroscope as nano-rotational motion sensor,Accepted \textit{Applied Physics B}  (2011).
  \bibitem{StedLT}
  G E Stedman, K U Schreiber and H R Bilger, On the detectability of the Lense-Thirring field from
rotating laboratory masses using ring laser gyroscope
interferometers, Class. Quantum Grav. {\bf 20} (2003). 


\bibitem{schiff} Schiff, L. I., Motion of a gyroscope according to Einstein's theory of gravitation, \textit{Proc. Nat. Acad. Sci.} \textbf{46}, 871 (1960).

\bibitem{kajari}  Kajari, E., Buser, M., Feiler, C., Schleich W.P., Rotation in relativity and the propagation of light, \textit{Riv.  Nuovo Cimento} \textbf{32}, 339 (2009).


\bibitem{bell} Bell, J.F., Camilo, F., Damour, T., A  Tighter Test of Local Lorentz Invariance using PSR J2317+1439, \textit{Astrophys. J.} \textbf{464}, 857 (1996).

\bibitem{damour96} Damour, T., Vokrouhlick\'y, D., Testing for gravitationally preferred directions using the lunar orbit, \textit{Phys. Rev. D} \textbf{53}, 6740 (1996).

\bibitem{iorio2011} Iorio, L., Lichtenegger, H.I.M., Ruggiero, M.L., Corda, C., Phenomenology of the Lense-Thirring effect in the Solar System, \textit{Astrophys. Space Sci.} \textbf{331}, 351 (2011). 


\end{thebibliography}
\end{document}